\documentclass{article}

\usepackage{arxiv}

\usepackage[T1]{fontenc}    
\usepackage{hyperref}       
\usepackage{url}            
\usepackage{booktabs}       
\usepackage{amsfonts}       
\usepackage{nicefrac}       
\usepackage{microtype}      
\usepackage{todonotes}
\usepackage{adjustbox}
\usepackage{pdflscape}
\usepackage{rotating}
\usepackage{amsmath}
\usepackage{multirow}
\usepackage{graphicx}
\graphicspath{{figures/}}

\usepackage[compress]{cite}

\usepackage{caption}
\usepackage{subcaption}
\usepackage{pifont}
\newcommand{\cmark}{\ding{51}}%
\newcommand{\xmark}{\ding{55}}%

\title{Resistive Neural Hardware Accelerators}

\author{Kamilya Smagulova$^{1,*}$, Mohammed E. Fouda$^{2,*}$, Fadi Kurdahi$^{2}$, Khaled Salama$^{1}$, and Ahmed Eltawil$^{1,2}$ \\
$^{1}$ Division of CEMSE,  King Abdullah University of Science and Technology, Thuwal 23955, Saudi Arabia.\\
$^{2}$ Electrical Engineering and Computer Science Dept., University of California–Irvine, Irvine, CA 92697 USA\\
$^{*}$Contributed equally to this work.\\
(e-mail: foudam@uci.edu)}

\begin{document}
\maketitle

\begin{abstract}

Deep Neural Networks (DNNs), as a subset of Machine Learning (ML) techniques, entail that   real-world data can be learned and that decisions can be made in real time. However, their wide adoption is hindered by a number of software and hardware limitations. The existing general-purpose hardware platforms used to accelerate DNNs are facing new challenges associated with the growing amount of data and are exponentially increasing  the complexity of computations.  An emerging  non-volatile  memory  (NVM) devices and processing-in-memory (PIM)  paradigm is creating a new hardware architecture generation with increased computing and storage capabilities.  In particular, the shift towards ReRAM-based  in-memory  computing  has   great potential in the implementation  of  area  and  power  efficient  inference  and in   training     large-scale  neural  network  architectures. These can accelerate  the  process  of  the IoT-enabled  AI  technologies entering  our  daily  life. In this survey, we review the state-of-the-art ReRAM-based DNN many-core accelerators, and   their superiority compared to CMOS counterparts was shown. The review covers different aspects of hardware and software realization of DNN accelerators, their present limitations, and future prospectives. 
In particular,   comparison of the accelerators   shows the need for the introduction of   new performance metrics and benchmarking standards.  
In addition, the major concerns regarding the efficient design of   accelerators include a lack of accuracy in simulation tools for software and hardware co-design.

\end{abstract}


\keywords{ Hardware acceleration \and In-memory computing \and Compute-in-memory \and processing-in-memory \and ReRAM \and Neural Networks \and nonidealities }

\section{Introduction}




The striking features of neural networks are the abilities for generalization and adaptive learning, which make them a powerful tool for solving abstract and complex problems. Currently, they are represented in a fairly wide range of fields such as agriculture \cite{ren2020survey,bannerjee2018artificial}, robotics \cite{karoly2020deep}, medicine \cite{briganti2020artificial}, renewable energy \cite{shamshirband2019survey,marugan2018survey}, and ecology and climate change \cite{barlow2020technological}. 
There is   active research in the fields of biometric recognition \cite{jacob2019capsule,labati2019deep}, navigation \cite{amer2021deep}, object-detection \cite{rohan2019convolutional}, and 
machine translation \cite{wu2016google}. Moreover, neural networks are also involved in the fight against the recent pandemic outbreak, e.g.,  estimation of its spread and quarantine measures \cite{dandekar2020neural} and the prediction of COVID-19 based on chest X-ray images
\cite{hussain2020coronavirus}. At such a fast pace, neural networks can bring the real-time virtual and augmented reality future closer, particularly   autonomous car driving  \cite{ahmad2020design,alhaija2018augmented}.
The rapid progress of state-of-the-art machine learning   (ML) algorithms and deep neural networks (DNNs) requires the utilization of power and memory-intensive computational resources. However, von-Neumann architectures for general-purpose computing have limited throughput. The problem is known as the "von Neumann bottleneck" and is caused by the need to carry data between processing and memory units of modern computers. The constant growth of data volume   aggravates the problem even more. Most proposed solutions attempt to mitigate   shortcomings by increasing the storage capacity of the main memory and improving the bandwidth of   interconnections \cite{efnusheva2017survey}.

The “classical” von Neuman architecture   consists of a Central Processing Unit (CPU), Random Access Memory (RAM), and an I/O interface.  The current market of memory devices,  sometimes also referred to as   RAM, can be divided into types of volatile and non-volatile memory. The charge-storage-based memory, such as dynamic random-access memory (DRAM), static random-access memory (SRAM), and Flash, is the most established  memory technology. However, they are facing challenges as CMOS technology approaches its physical limit. Moreover, DRAM and SRAM are volatile,   and the stored data are lost when power is turned off.   Contrarily, flash memory is non-volatile and can retain the data after power is removed. However, they suffer from large operation latency \cite{fariprimary}.
\par
The limited scalability down to 10 nm, low operation speed, and energy inefficiency of   traditional memory devices has provoked active research in the area of emerging nonvolatile memory (NVM) devices \cite{zahoor2020resistive}. The new devices function based on a change in resistance rather than charge. The resistance-switching phenomenon has been observed in different types of materials. The most popular types of devices include phase-change memory (PCM), spin-transfer torque random-access memory (STT-RAM), and resistive random-access memory (ReRAM). Moreover,  the emerging technologies have great potential not only in storing more data but also are capable of acting as efficient processing-in-memory (PIM) architectures. Among different NVM technologies,  ReRAM has better characteristics than PCM, including a higher density and scalability than STT-RAM \cite{banerjee2020challenges} and better write and read performance (<10 ns)   \cite{zahoor2020resistive}. It has a simple structure with the smallest $4F^{2}$ planar cell size, a wide $R_{on}$/$R_{off}$ ratio, and a high speed, compatible with existing CMOS technology \cite{meena2014overview}.

ReRAM  devices are widely considered as a promising technology for  the  implementation of multiply-accumulate (MAC)---the key operation in ML and NNs. The development of efficient  ReRAM-based hardware accelerators can significantly expedite the progress of intelligent hardware architectures. This includes processing of the enormous genomic sequence information \cite{onimaru2020predicting}, \cite{abbas20214mcpred} and the development of cryptocurrency and e-commerce \cite{pan2020study}, \cite{yazdinejad2020cryptocurrency}, \cite{li2020bitcoin}. In addition, ReRAM technology can be beneficial in biomedical applications, which require solutions with an extremely low power consumption. To date, several application-specific ReRAM-based accelerator designs have been demonstrated. One of them is DNA sequence alignment on the ReRAM-based platform, which provides  an acceleration of up to a 53 × and 1896 × times   \cite{huangfu2018radar} and a 7× better power efficiency \cite{gupta2019rapid}. Another application is  the cost-effective and high-accuracy ReRAM-based health monitoring system  \cite{liu2020monitoring}. ReRAM encryption engines for IoT is  also gaining popularity \cite{li2019reram,korenda2018secret}. In particular, a low cost weight obfuscation scheme for security enhancement was proposed in \cite{wang2021low}. Existing ReRAM accelerators have different designs and architectures, including  standalone macros, co-processors, and many-core processors. This paper reviews several state-of-the-art many-core and multi-node ReRAM-based PIM neural network accelerators and their pipeline designs,  identifies their limitations, and proposes potential directions for improvement.

This paper is organized as follows: Section \ref{sec:SotA} overviews the existing resistive accelerators with a brief explanation of each architecture. Section \ref{sec:SotA_comparison} compares their performance using several evaluation metrics including area and power distribution and efficiency in algorithm-to-hardware mapping. The comparison is drawn at device, design, and system levels. Section \ref{sec:SotA_RRAM_macro} summarizes the recent progress in   DNN implementation using the fabricated ReRAM macros. 
Section \ref{sec:SotA_potential} identifies gaps and potential directions in the deployment of DNNs using many-core ReRAM accelerators. 
Finally, Section \ref{sec:conclusion} presents conclusions.

\section{State-of-the-art Resistive Accelerators}
\label{sec:SotA}

During the last decade, numerous digital heterogeneous multi-core architectures have demonstrated their potential in efficient acceleration and  the training of large-scale neural networks. For instance, DaDianNao overcame  the memory-wall bottleneck problem of general-purpose computers and provided a 450x speedup and 150x energy efficiency \cite{chen2016diannao}. DaDianNao's architecture 
was comprised of 16 interconnected tiles, each including eDRAM buffers for storing neurons and synapses and a Neural Functional Unit (NFU) for parallel arithmetic computing.  This configuration became an inspiration for the ReRAM-based multi-core neural network accelerators. Unlike digital counterparts, ReRAM crossbar arrays allow for the integration of synaptic weight storage and matrix-vector multiplication (MVM) in ReRAM crossbar arrays. Therefore, ReRAM-based accelerators can provide more energy and area efficiency. Detailed descriptions of the most common accelerator designs and their operation are provided below.


\subsection {ISAAC}
\begin{figure}[!h]
     \centering
     \begin{subfigure}[t]{0.45\textwidth}
         \centering
         \includegraphics[width=\textwidth]{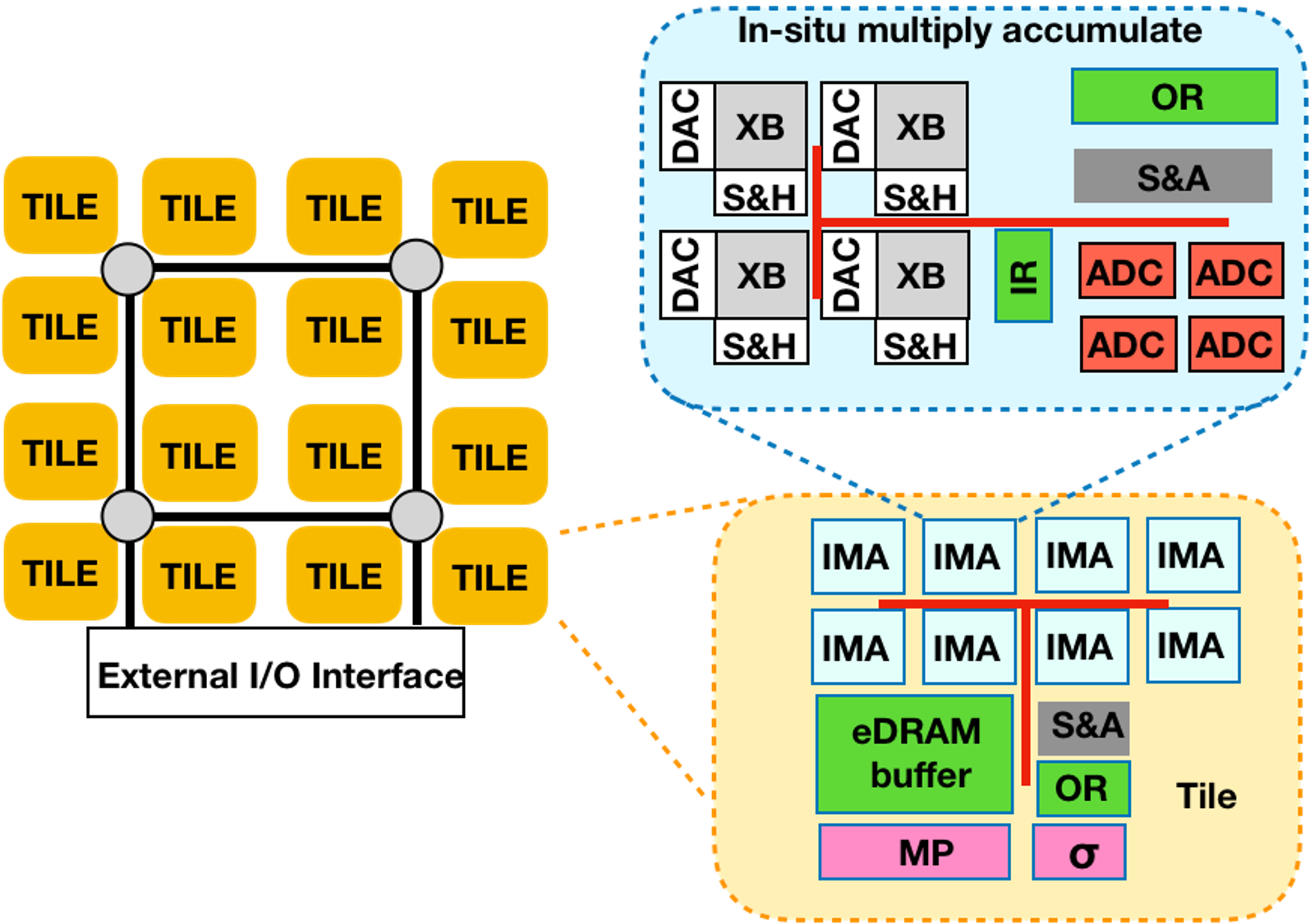}
         \caption{}
     \end{subfigure}
\hfill
     \begin{subfigure}[t]{0.45\textwidth}
         \centering
         \includegraphics[width=\textwidth]{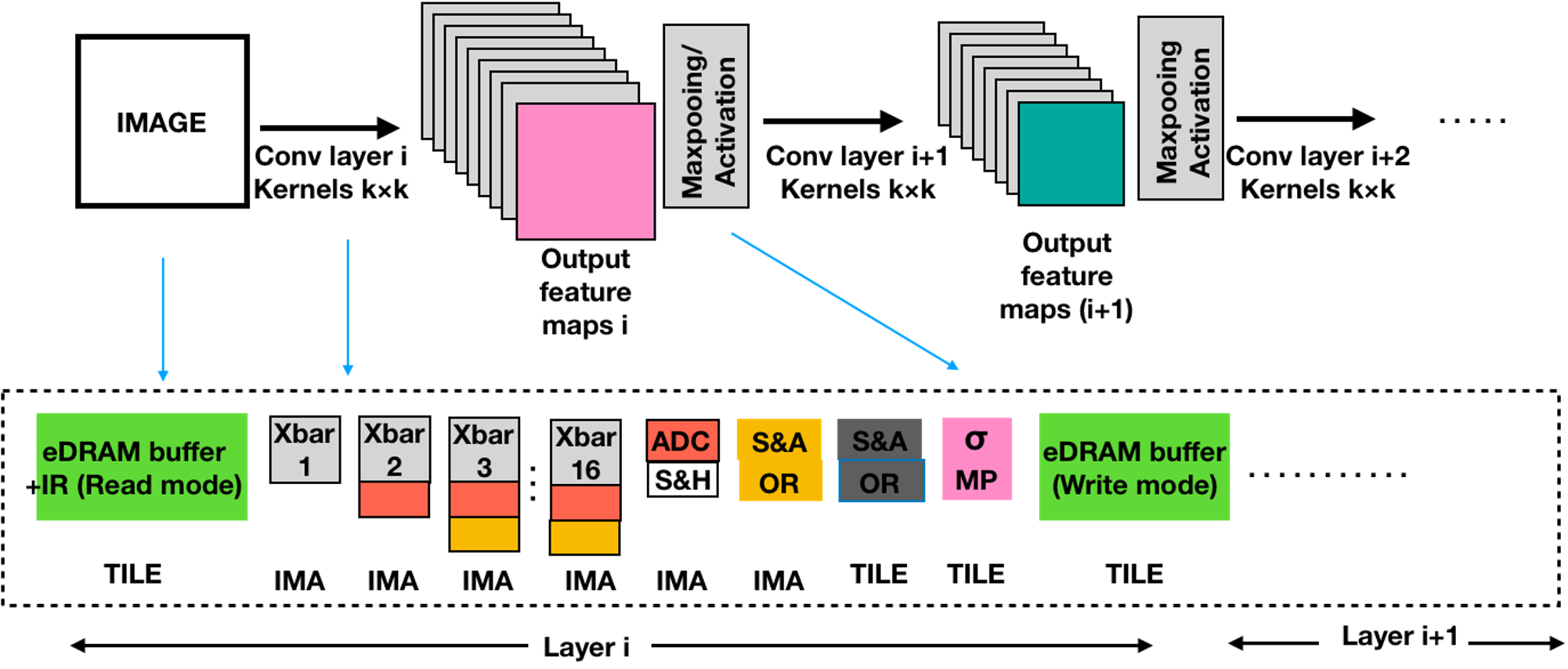}
         \caption{}
     \end{subfigure}
    \caption{a) ISAAC overview; b) pipeline of layer $i$ of the convolutional neural network with layers $i, (i+1), (i+2). $ }
    \label{fig:1}
\end{figure}

A Convolutional Neural Network Accelerator with   In-Situ Analog Arithmetic in Crossbars (ISAAC) is a ReRAM-based full-fledged accelerator for CNN \cite{shafiee2016isaac}. Its architecture is shown in Figure \ref{fig:1} and comprised of multiple tiles connected via a concentrated-mesh (c-mesh) network. In the optimal design \textit{ISAAC-CE}, each tile  includes 12 in-situ multiply-accumulates (IMAs), 1 shift-and-add (S\&A) unit, 2 sigmoid units, and 1 maxpooling unit. In addition, it has an eDRAM buffer for storage of input data and an output register (OR) used to accumulate results. A single IMA consists of its own input and output registers (IR and OR), S\&A units, and eight 128$\times$128 RCAs, referred as XB with shared analog-to-digital converters (ADCs). In addition, each RCA is accompanied with a digital-to-analog converter (DAC) and sample-and-hold (S\&H) circuit blocks.  All elements within a tile communicate via the inter-tile 32-bit link at 1.2 GHz.

\par 

The main contribution of the ISAAC design was dataflow pipelining. To understand the data flow, consider a neural network with convolutional layers \textit{i, (i+1), (i+2)} as in Figure \ref{fig:1}b. Depending on  the size of the network, each layer of  the CNN is assigned to one or several IMAs or tiles. At the initial stage, input data is fetched for the system via an I/O connection and stored in an eDRAM buffer of a tile. Before being fed into an XB in  the IMA, data are converted to analog form by DACs. After being processed by XBs, generated feature maps are converted back to digital form and passed to maxpooling and activation blocks. Computed output of the neural network layer \textit{i} is then accumulated in the S\&A and OR units and written to a new eDRAM buffer. This output also serves as the input feature map of the next layer \textit{(i+1)} and so on. The depth of the pipeline is determined by the depth of a neural network. This complicates the  training of deep neural networks. Therefore, ISAAC is designed for inference only.

\subsection{PRIME}
\begin{figure}[!h]
     \centering
     \begin{subfigure}[t]{0.4\textwidth}
         \centering
         \includegraphics[width=\textwidth]{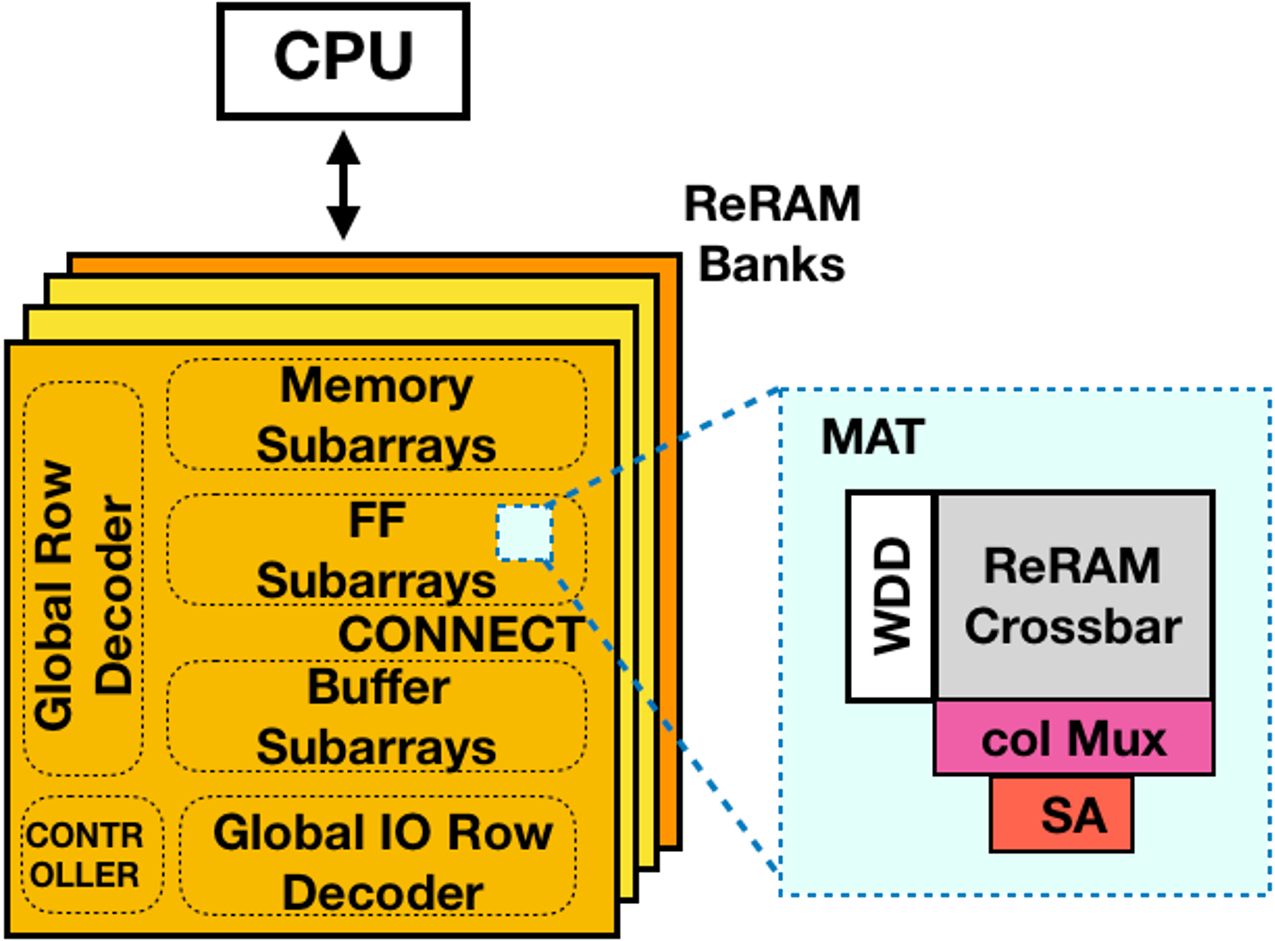}
         \caption{}
     \end{subfigure}
     \begin{subfigure}[t]{0.4\textwidth}
         \centering
         \includegraphics[width=\textwidth]{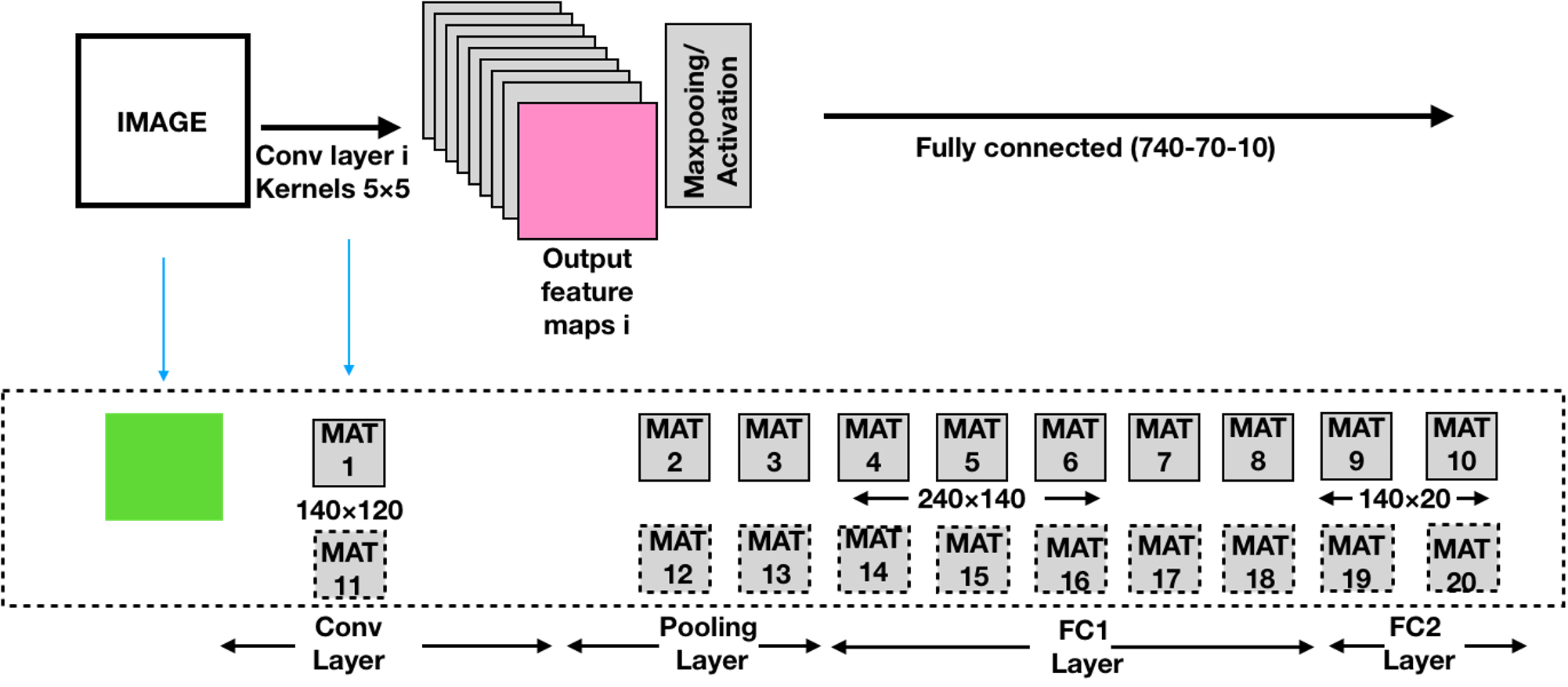}
         \caption{}
     \end{subfigure}
    \caption{ PRIME: a) Functional Architecture and b) mapping example. }
    \label{prime}
\end{figure}

PRIME consists of multiple ReRAM banks as shown in Figure \ref{prime}a \cite{chi2016prime}.  Each bank includes 64 subarrays: \textit{memory} (Mem) subarrays, two \textit{full function} (FF) subarrays, and one \textit{buffer} subarray. FF subarrays can operate as conventional memory or in an NN computation mode. The mode of operation is controlled by a PRIME controller. A typical memory matrix (mat) in  the FF subarray has 256$\times$256 ReRAM cells, whose 
  output is read by 6-bit reconfigurable local sense amplifiers (SAs). During  the computation mode,  the  resolution of ReRAM cell is 4-bit MLC, whereas at memory mode, it is SLC. Two different crossbar arrays are used to store positive and negative weights. The input of  the mat  is a 3-bit fixed point signal from a wordline decoder and driver WDD. 
Analog subtraction and sigmoid functions in  the neural network are embedded in the modified column multiplexers in  the ReRAM mats. 


Running a neural network on PRIME consists of three stages. The first stage includes mapping  the  neural network to FF subarrays and programming synaptic weights to ReRAM cells in mats.   The second stage of implementation is optimization. At this stage, depending on the size, a neural network can be mapped into a single bank or multiple banks. The first two stages happen in  the CPU. At the third stage, a set of generated instructions is passed to  the PRIME controller in ReRAM banks for  the further execution of computations. Due to  the presence of latches and output registers, computation in PRIME can be conducted in a pipelined way. The bank-level parallelism accelerates computations. Figure \ref{prime}b  shows  the acceleration of  the  \textit{CNN-1}  neural network (conv5x5-pool-720-70-10). Its layers were mapped to 10 mats of size 256$\times$256 each. In order to increase  the computational speed, the 10 mats can be duplicated (using  Mats 11-20) and configured to work in ping-pong mode: when the primary mat is loading or storing data, the duplicate mat is computing.

%

\subsection{AEPE}

\begin{figure}[!h]
\centering 
\includegraphics[width=0.5\textwidth]{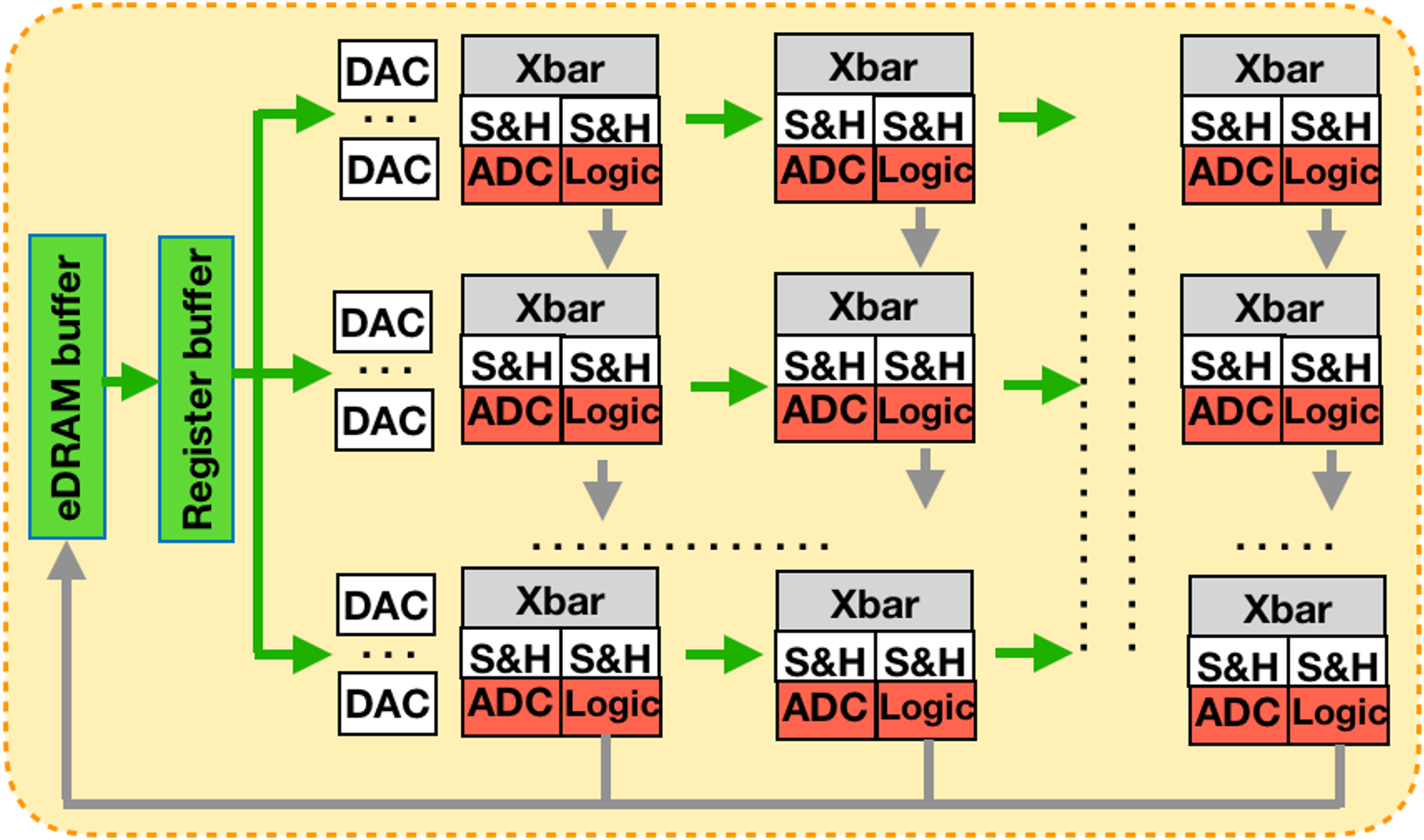}
\caption{ Overview of a tile of the AEPE.}
\label{aepe}
\end{figure}

An overview of  the Area and Power-Efficient RRAM Crossbar-based Accelerator for Deep CNNs (AEPE) tile is shown in Figure \ref{aepe}. The AEPE includes an eDRAM buffer, a ping-pong register buffer, $M\times N$ processing elements (PEs), and $M$ DACs.  Each PE unit is comprised of an $XBar$ RCA, an ADC, a logic unit, and two S\&H units. The AEPE introduced a number of changes to its configuration to improve the energy and area efficiency, compared to ISAAC, including three data reuse methods.   

In  the AEPE, the original input data is stored in eDRAM and passed to a register buffer that can reuse it without repetitive loading. This  reduced the port width of the eDRAM buffer and therefore saves area and lowers energy consumption. The ping-pong register buffer then sends data to DACs and participates in the execution of  the convolution of the input. To improve area and energy efficiency compared to ISAAC,  the AEPE uses fewer   DACs. Therefore, each column of PEs uses a signal from the same DAC.  In addition,  the AEPE utilizes lower-precision ADCs, which also contribute to the lower power consumption of the architecture.  Analog signals across PEs can be reused both horizontally and vertically. In addition,  the shared high-cost   bus connection between PEs was replaced with local connections. 


\subsection{PipeLayer}
Unlike PRIME and ISAAC, PipeLayer supports  both  the training and inference of neural networks \cite{song2017pipelayer}. This was achieved by exploiting efficient pipelines with inter- and intra-layer parallelism and 
 with a wide range of different input and kernel mapping schemes.  Similar to ISAAC, the input and output data in PipeLayer are encoded using a weighted spike-based method.

\begin{figure}[!h]
     \centering
     \begin{subfigure}[t]{0.45\textwidth}
         \centering
         \includegraphics[width=\textwidth]{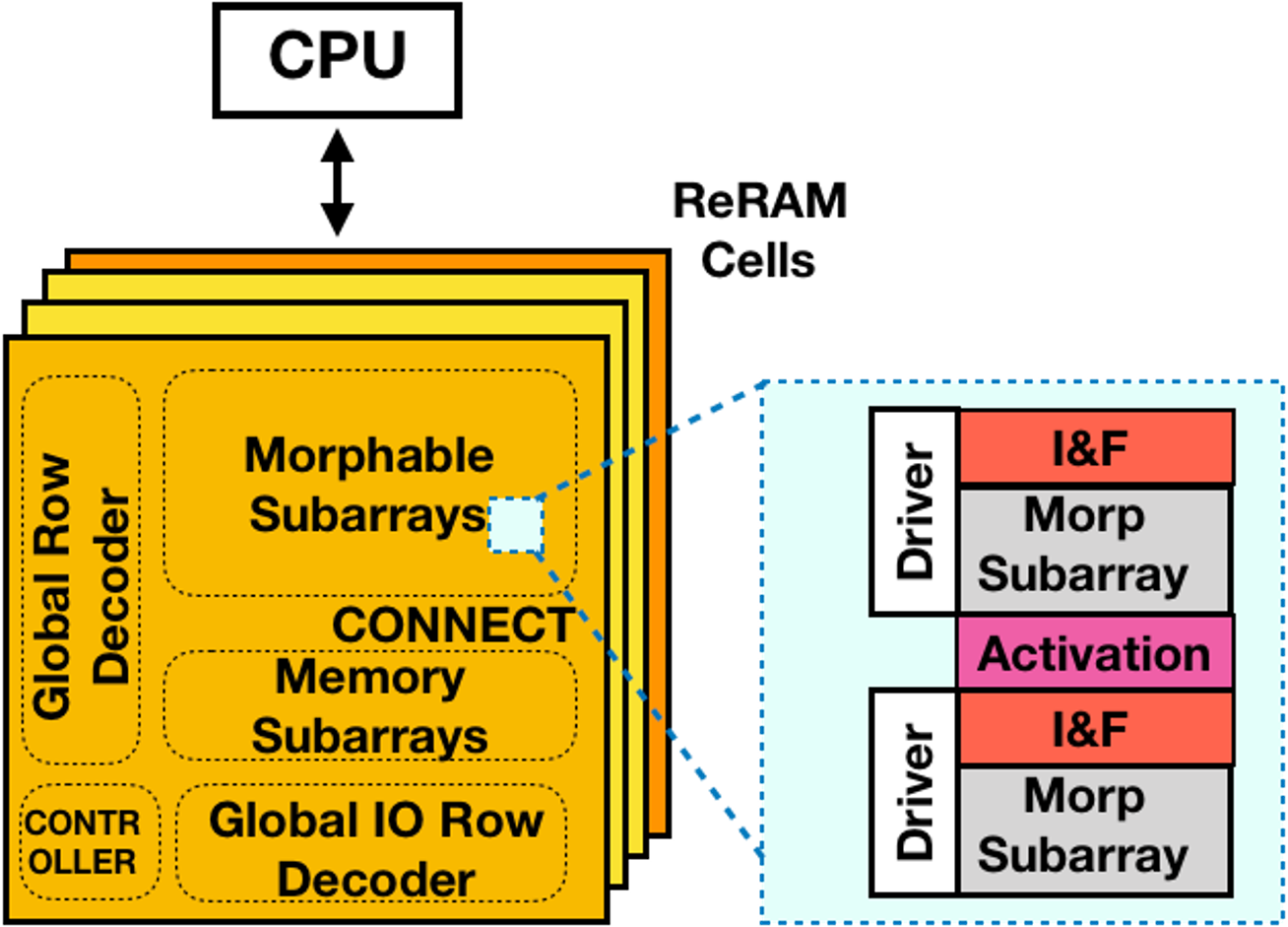}
         \caption{}
     \end{subfigure}
     \begin{subfigure}[t]{0.45\textwidth}
         \centering
         \includegraphics[width=\textwidth]{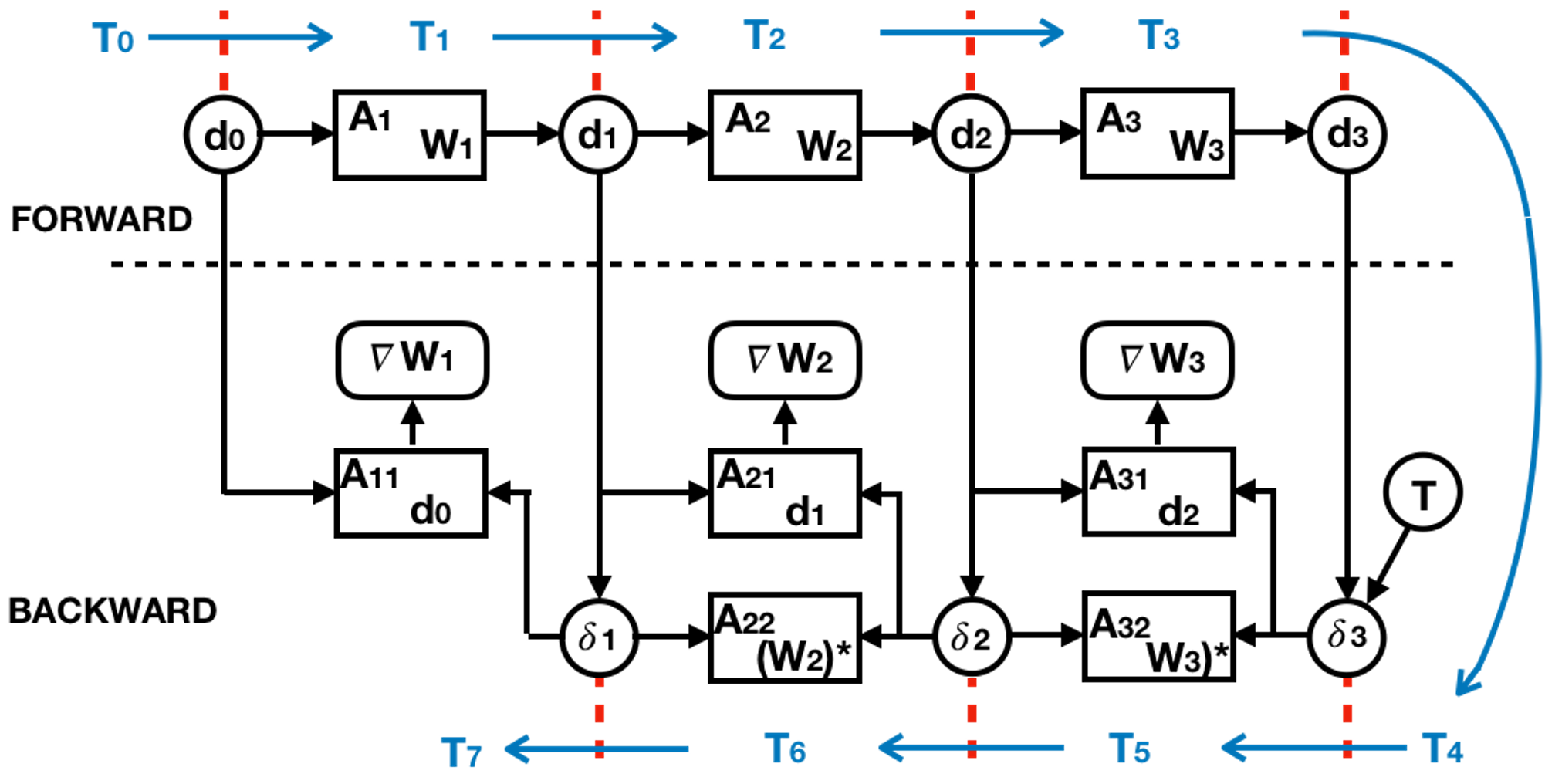}
         \caption{}
     \end{subfigure}
    \caption{a) Architecture of the PipeLayer; b) pipeline of the PipeLayer. }
    \label{pipelayer}
\end{figure}

\par
Figure \ref{pipelayer}a shows the overall architecture of the PipeLayer comprised of metal-oxide ReRAM cells. The control unit offloads instructions from  the CPU in order to supervise the processes within the cell.  To implement a spike coding scheme, the \textit{Spike Driver} block converts input data to spikes, whereas the \textit{Integration and Fire} (I\&F) component converts output spikes back to a digital format. In write mode,  the spike driver updates  the weights of  the ReRAM arrays with a 4-bit resolution. The data processing in the cell is performed by \textit{morphable} and \textit{memory subarrays} connected via the \textit{Connection} block. The morphable subarrays $A_{n}$ have two modes and are used either as a dot-product engine or as weight/data storage. In computation mode,  the output of a morphable array is passed to  the Activation block, which implements activation function determined by offloaded instructions. The memory subarrays $d_{k}$ serve as a conventional memory to store intermediate results between layers during training and testing phases. The precision of the PipeLayer computation results is 16-bit.





\subsection{AtomLayer}
The Universal ReRAM-based CNN Accelerator with Atomic Layer Computation (AtomLayer) has an architecture  similar to ISAAC and includes an array of processing elements (PEs) interconnected via three types of on-chip network:  the input network (INet), a local network (LNet), and an output network (ONet). In addition, it contains an arithmetic logic unit (ALU) tree of multiple column-ALUs (CALUs) and a global ALU (GALU). In general, ALUs are used to implement pooling and activation layers of  the neural network and to perform partial sum accumulation.
\begin{figure}[!h]
\centering
    \includegraphics[width=0.8\textwidth]{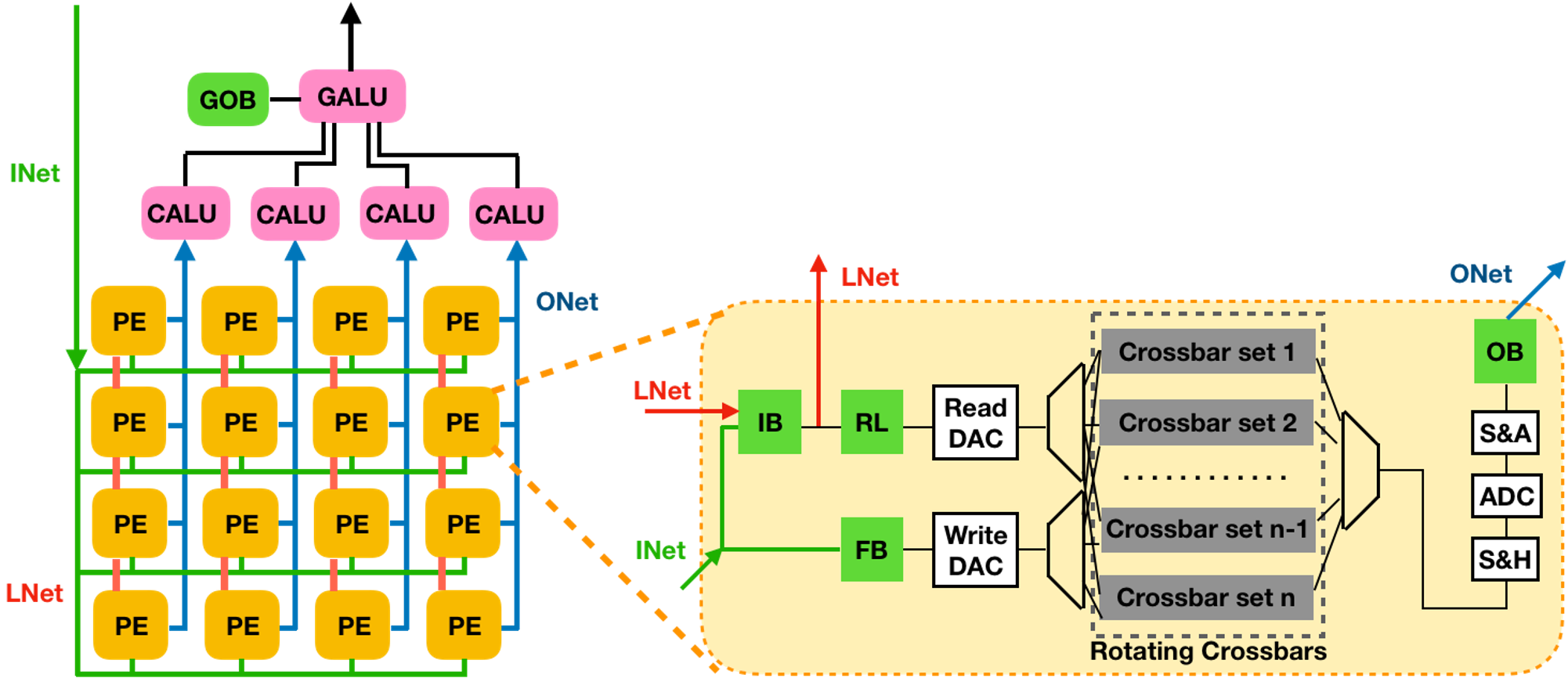}
    \caption{ The overall architecture of AtomLayer and its Processing Element. }
    \label{fig:3}
\end{figure}

The distinguishing feature of AtomLayer is rotating crossbars comprised of several 128$\times$128 crossbar sets. Each PE includes a rotating crossbars and digital computational part consisting of write and read DACs, ADCs, an S\&H unit, and an S\&A unit
. The PE also includes input, output, and filter buffers (IB, OB, and FB) and a register ladder (RL). INet transmits input data from  the DRAM and new filters from  the FB to one row of PEs. LNet ensures  the communication of two PEs from neighboring rows. ONet passes  the output of PEs to  the corresponding CALU and then from  the CALUs to  the GALU. It  processes only one layer of a neural network at a time. The FB, RL, INet, and LNet form a data reuse system.



\subsection{Newton}

A Newton architecture was proposed as an optimized version of ISAAC architecture \cite{nag2018newton}. First, it was observed that early convolutional layers have lower buffer requirements. To avoid   under-utilization, the size of the  eDRAM buffer was decreased from 64 KB to 21 KB, and the layer was spread across 10 tiles. It also introduced constraints to the workload mapping, and convolutional and fully-connected classifier layers were mapped to different tiles.  SAR ADCs were then replaced by the adaptive ADCs. In addition,  to optimize computational resources,   divide-and-conquer numeric algorithms were adopted. In particular, Karatsuba's algorithm was used to reduce power consumption by almost 25\%. However, this also leads to an increase in area by 6.4\%. Eventually,  Strassen's technique was applied to optimize  the matrix-vector-multiplication operation and to reduce computations. This helped to increase the energy efficiency by 4.5\%. The adopted measures   improved the power efficiency by 51\% and increase computation efficiency by 2.2$\times$. 

\subsection{CASCADE}

\begin{figure}[!h]
     \centering
     \begin{subfigure}[t]{0.45\textwidth}
         \centering
         \includegraphics[width=\textwidth]{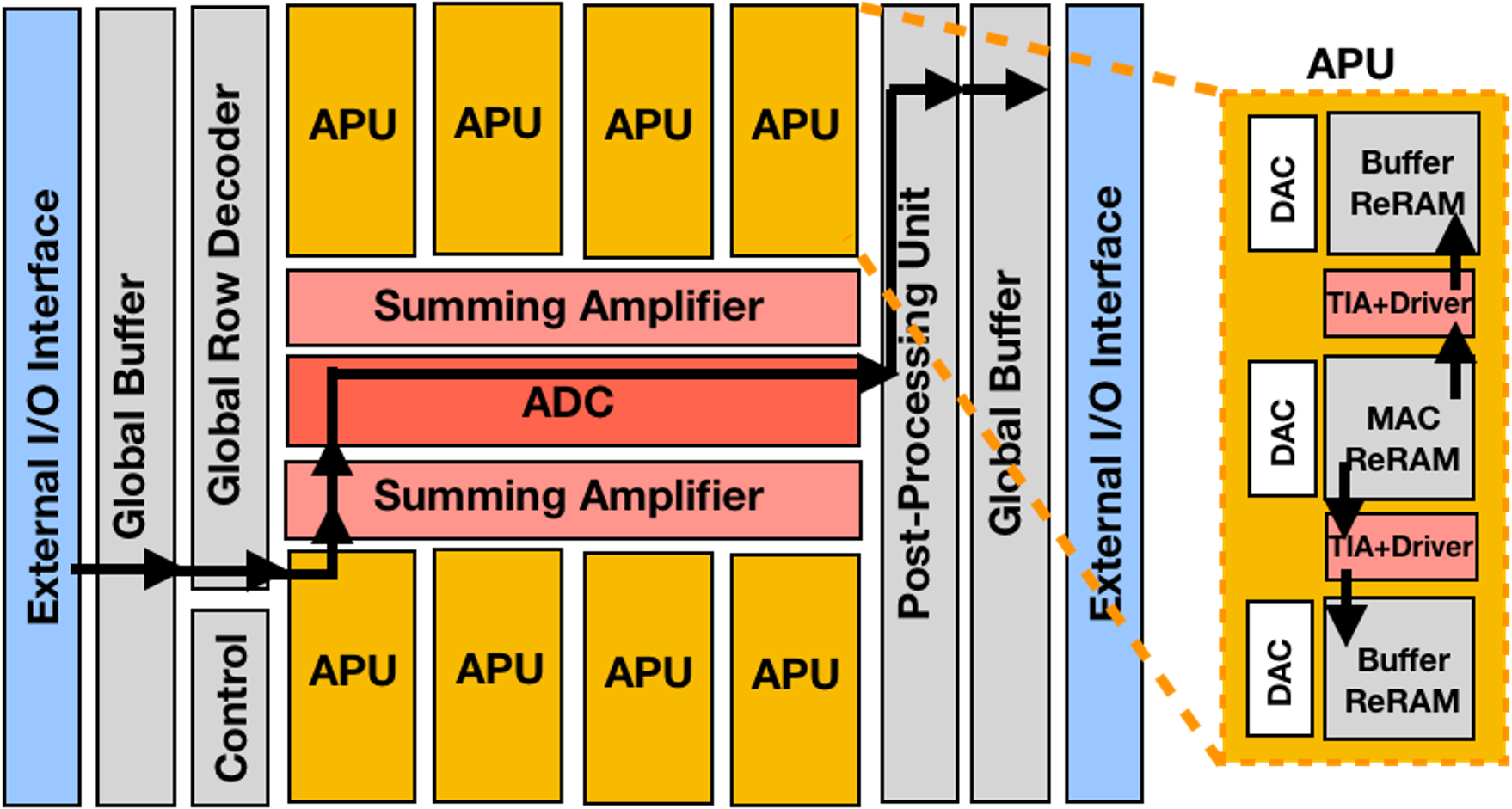}
         \caption{}
     \end{subfigure}
     \begin{subfigure}[t]{0.45\textwidth}
         \centering
         \includegraphics[width=\textwidth]{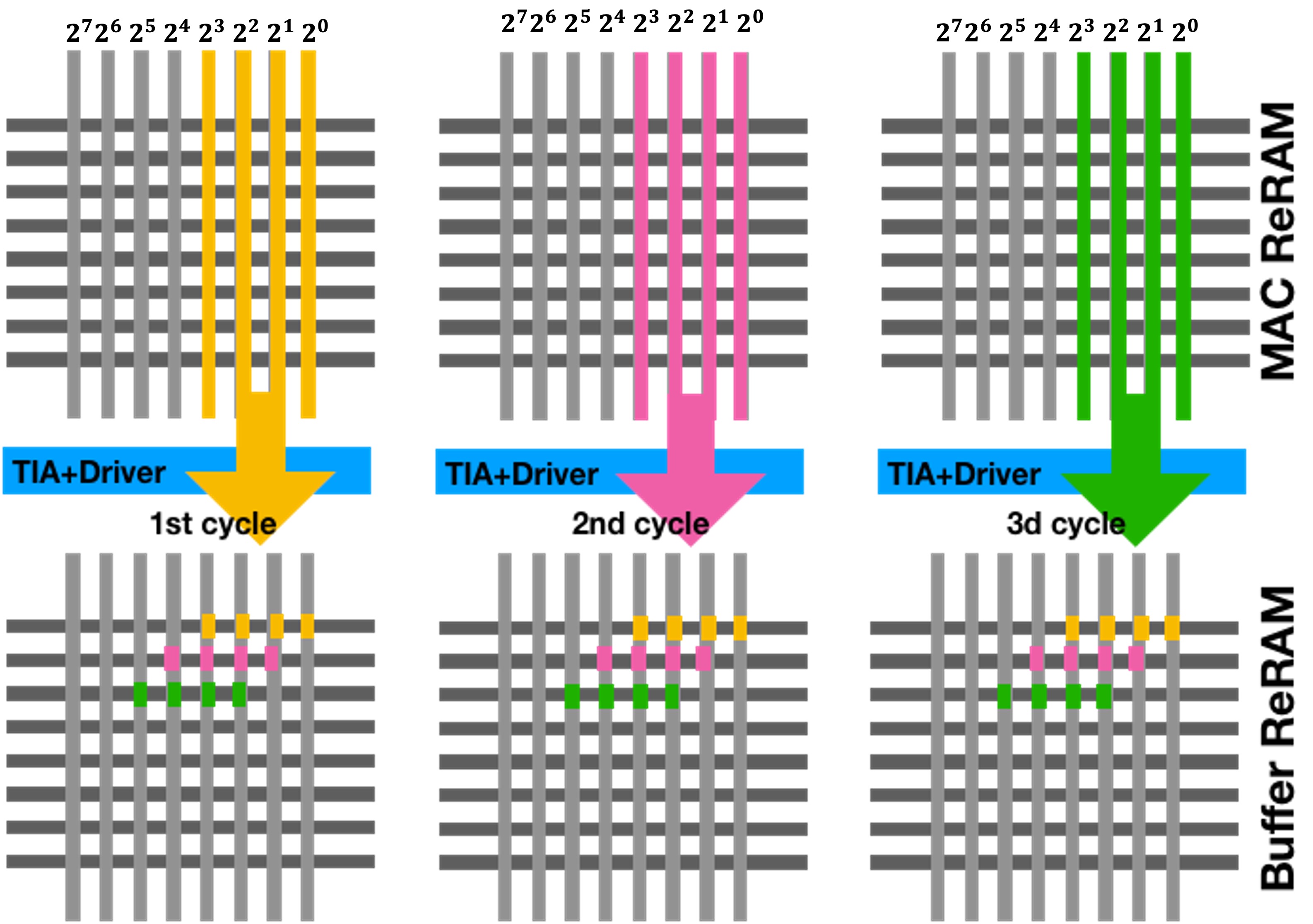}
         \caption{}
     \end{subfigure}
    \caption{a) CASCADE architecture and b) an example of partial sum accumulation.}
    \label{fig:cascade}
\end{figure}

CASCADE was proposed for DNN and RNN acceleration, which comprises several arithmetic logic units (ALUs), each consisting of MAC and Buffer arrays (Figure \ref{fig:cascade}a)  \cite{chou2019cascade}. CASCADE addresses the problem of the large overhead of A/D conversions of the previously presented designs by utilizing   a new R-Mapping scheme and transimpedance amplifiers (TIAs). In addition, CASCADE can be extended to support spiking neural networks (SNNs).

TIAs were used for sensing the output of MAC RCAs. The sensed outputs were used to produce analog partial sum accumulation in Buffer RCAs. This   eliminated   redundant conversions and kept intermediate values in the analog domain.  The utilization of Buffer arrays also reduced energy consumption by up to 7.59 × compared to digital partial-sum accumulation.  Secondly,  the CASCADE architecture considered the trade-off between resolution and inference accuracy and  adopted   lower-resolution output cascading.  CASCADE was evaluated using 65 nm CMOS technology and a 65$nm$ ReRAM model from \cite{chen201865nm}.

\subsection{PUMA/PANTHER}

A Programmable Ultra-efficient Memristor-based Accelerator for ML Inference (PUMA) is a ReRAM-based Instruction Set Architecture (ISA) designed for efficient in-memory computing. It has a three-tier spatial architecture consisting of cores, tiles, and nodes. Multiple nodes in PUMA are connected via a chip-to-chip network. Each node includes tiles connected via  an on-chip network. In turn, tiles consist of cores connected via a shared memory. During workload mapping, PUMA uses the advantage of input and weight reuse to minimize data movement and decrease  the amortization  of DRAM access. 

Due to its spatial architecture, data processing in PUMA is implemented by a runtime compiler. As the first step, the programmer creates the model and generates input/output vectors. A graph is partitioned to sub-graphs and allocated to corresponding cores. Next,   the sub-graphs execution order is scheduled. The scheduling process aims at  the efficient utilization of resources and avoiding deadlock. Due to the serial ReRAM read and write operations, PUMA can support only the  inference mode. To implement training, a Programmable Architecture for Neural Network Training Harnessing Energy-efficient ReRAM  (PANTHER) was introduced and evaluated on PUMA \cite{ankit2020panther}. PANTHER implemented a weight and gradient update scheme by using  special bit-slicing Outer Product Accumulate (OPA) operations. It also supports different training algorithms and  the transpose of weight matrix $M^{T}VM$.

\begin{figure}[!h]
     \centering
     \begin{subfigure}[t]{0.45\textwidth}
         \centering
         \includegraphics[width=\textwidth]{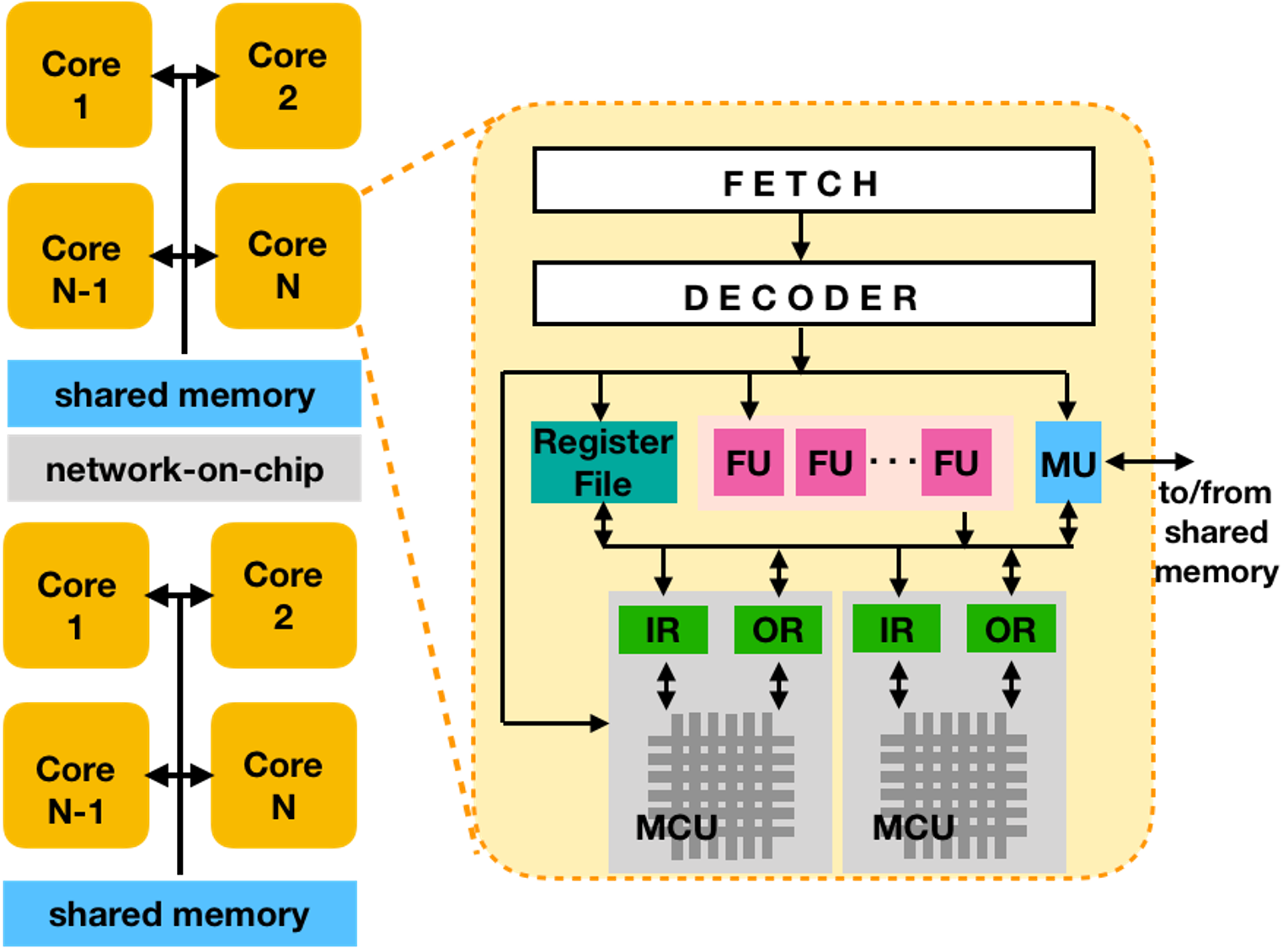}
         \caption{}
     \end{subfigure}
     \begin{subfigure}[t]{0.45\textwidth}
         \centering
         \includegraphics[width=\textwidth]{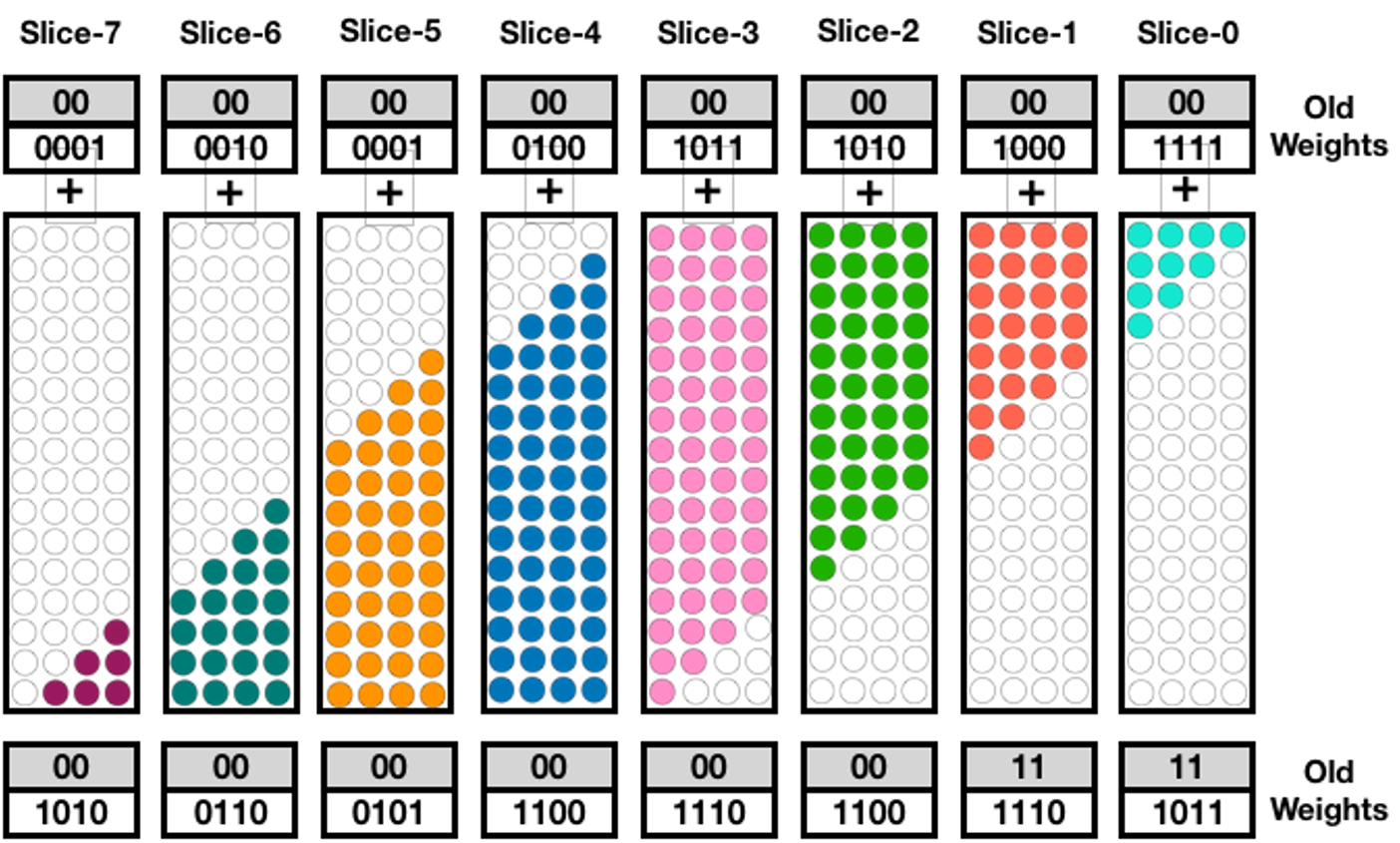}
         \caption{}
     \end{subfigure}
    \caption{a) PUMA/PANTHER architecture and b)  the bit-slicing Outer Product Accumulate (OPA) scheme. }
    \label{puma}
\end{figure}

Figure \ref{puma}a shows the PUMA architecture consisting of N cores. Unlike previously reported accelerators, the core of PUMA/PANTHER breaks the entire computational process into a set of instructions and creates corresponding Instruction Execution Pipelines. Each pipeline consists of three stages: fetch, decode, and execute. Such an approach allows PUMA to compute several different ML workloads that share low-level operations. The ISA executed by the in-order pipeline requires seven bytes and implements a wide range of operands, e.g., MVM, ALU, set, copy, load, and store. At the execution stage, the pipeline involves  the following core components: a Register File,  a Matrix-Vector Multiplication Unit (MVMU), a Vector Functional Unit (VFU), and a Memory Unit (MU).


In an MVMU, the register \textit{XBarIn} passes a digital input to DAC. The  obtained analog values are then fed to  the memristor crossbar array, which implements a 16-bit matrix-vector-multiplication operation. The precision of a device in a crossbar array is 2-bit, so eight crossbars are combined together. After MVM operation,  the computed data is converted back to digital format using ADC and stored in  the \textit{XBarOut} register. The single core may have more than one MVMU. Since there is no instruction level parallelism, multiple MVMUs are activated by a single MVM instruction using a \textit{mask} operand in  the ISA. The operand's 
  \textit{filter/stride} enable logical shuffling and avoid data movement around \textit{XBarIn}. Linear and non-linear operation in  the neural network are implemented using  the VFU.






\section{Comparison and Discussion}
\label{sec:SotA_comparison}

All the aforementioned  SotA ReRAM-based  accelerators maintain a hierarchical multi-node and multi-core architecture with a similar hierarchy. Table \ref{notations} summarizes their components that have  the same or similar functionality but different notations.  

\begin{table}[!h]
\centering
\caption{Notations.}
\resizebox{1\textwidth}{!}
{

\begin{tabular}{|l|c|c|c|c|c|c|c|c|}
\hline
Accelerator       & \textbf{ISAAC} & \textbf{PRIME}     & \textbf{AEPE}     & \textbf{PipeLayer}            & \textbf{AtomLayer}    & \textbf{Newton} & \textbf{CASCADE}   & \textbf{\begin{tabular}[c]{@{}c@{}}PUMA/\\ PANTHER\end{tabular}}      \\ \hline
\multirow{4}{*}{\textbf{\begin{tabular}[c]{@{}l@{}}Levels of\\ hierarchy\end{tabular}}}  & n/a            & n/a& n/a               & n/a           & n/a   & n/a             & n/a        & Node          \\ \cline{2-9} 
  & Chip           & Chip               & Chip              & Chip          & Chip  & Chip            & n/a        & Tile          \\ \cline{2-9} 
  & Tile           & \begin{tabular}[c]{@{}c@{}}ReRAM \\ bank\end{tabular}              & Tile              & \begin{tabular}[c]{@{}c@{}}ReRAM\\ bank\end{tabular}          & \begin{tabular}[c]{@{}c@{}}Processing\\ Element\end{tabular}          & Tile            & Chip       & Core          \\ \cline{2-9} 
  & IMA            & Mat& \begin{tabular}[c]{@{}c@{}}Processing\\ Element\end{tabular}      & n/a           & n/a   & IMA             & ALU        & MVMU          \\ \hline
\textbf{RCA}      & XB             & \begin{tabular}[c]{@{}c@{}}ReRAM \\ array\end{tabular}             & XBar              & \begin{tabular}[c]{@{}c@{}}Morphable \\ subarray\end{tabular} & \begin{tabular}[c]{@{}c@{}}ReRAM \\ crossbar\end{tabular}             & XB              & \begin{tabular}[c]{@{}c@{}}MAC\\ ReRAM\end{tabular}& MCU           \\ \hline
\multirow{9}{*}{\textbf{\begin{tabular}[c]{@{}l@{}}Peripheral \\ interface\end{tabular}}} & OR             & WDD latch          & n/a               & n/a           & OR    & OR              & \multirow{2}{*}{n/a}               & XBarOut       \\ \cline{2-7} \cline{9-9} 
  & IR             & WDD latch          & Reg buffer        & Driver        & IB    & IR              &            & XBarIn        \\ \cline{2-9} 
  & ADC            & SA & ADC               & Spike Driver  & ADC   & ADC             & \begin{tabular}[c]{@{}c@{}}TIA, ADC,\\ SA\end{tabular}             & ADC           \\ \cline{2-9} 
  & DAC            & WDD& DAC               & I\&F          & DAC   & DAC             & DAC        & DAC           \\ \cline{2-9} 
  & S\&H           & \multirow{2}{*}{n/a}               & S\&H              & \multirow{2}{*}{n/a}          & S\&H  & S\&H            & S\&H       & \multirow{2}{*}{n/a}  \\ \cline{2-2} \cline{4-4} \cline{6-8}
  & S\&A           &    & S\&A              &               & n/a   & S\&A            & \begin{tabular}[c]{@{}c@{}}Buffer\\ ReRAM\end{tabular}             &               \\ \cline{2-9} 
  & MaxPool        & SA & n/a               & \multirow{2}{*}{Activation}   & \multirow{2}{*}{\begin{tabular}[c]{@{}c@{}}CALU/\\ GALU\end{tabular}} & MaxPool         & \multirow{2}{*}{\begin{tabular}[c]{@{}c@{}}Post-\\ Processing\\ Unit\end{tabular}} & \multirow{2}{*}{VFU}  \\ \cline{2-4} \cline{7-7}
  & Sigmoid        & \begin{tabular}[c]{@{}c@{}}SA, modified \\ column MUX\end{tabular} & Logic             &               &       & Sigmoid         &            &               \\ \cline{2-9} 
  & eDRAM          & \begin{tabular}[c]{@{}c@{}}Memory \\ subarray\end{tabular}         & eDRAM             & FF subarray   & eDRAM & eDRAM           & \begin{tabular}[c]{@{}c@{}}Global\\ buffer\end{tabular}            & eDRAM         \\ \hline
\textbf{\begin{tabular}[c]{@{}l@{}}Connection\\ Link\end{tabular}}        & c-mesh         & \begin{tabular}[c]{@{}c@{}}internal \\ shared bus\end{tabular}     & \begin{tabular}[c]{@{}c@{}}local \\ connection\\ NoC\end{tabular} & connection    & \begin{tabular}[c]{@{}c@{}}INet, LNet,\\ ONet\end{tabular}            & c-mesh          & n/a        & \begin{tabular}[c]{@{}c@{}}chip2chip/\\ on-chip network,\\ shared memory\end{tabular} \\ \hline
\end{tabular}

}
\label{notations}
\end{table}

ISAAC is one of the first successful designs of ReRAM-based accelerators. It outperformed  the fully digital DaDianNao with  improvements of 14.8x, 5.5x, and 7.5x in throughput, energy, and computational density, respectively \cite{shafiee2016isaac}. Around the same time,  the architecture of PRIME was introduced \cite{chi2016prime}. Its performance depends on workload and decreases with the  increase in NN size due to the cost of the data communication between banks/chips. PRIME and ISAAC seem to have similar architectures because both of them take advantage of  the RCA for fast and efficient dot-product operation. However, they use different encoding techniques, peripheral and communication interfaces, and pipeline structures. In particular,  the output of  the RCAs in ISAAC is sensed by ADCs, whereas PRIME uses SAs.
The majority of  the subsequent works on accelerator architecture aimed to improve limitations of the design of ISAAC. Therefore,   the ISAAC-based accelerator AEPE (\textit{diff}=2) achieved a power efficiency of 2.71x and an area efficiency of 2.41x by decreasing  the number of DACs, lowering  the resolution of ADCs, and utilizing   different links for communication. The CASCADE architecture considered all of the advantages and disadvantages of  the ISAAC and PRIME architectures and proposed  the utilization of  the TIA interface. This helped to reduce energy consumption by 77.5 ×, compared to  the ADC interface, and by 325.4 × compared to  the SA interface. PUMA is a spatial processor and provides more flexibility and scalability to accelerate a wide range of workloads and different types of data \cite{ankit2019puma}. All these architectures could accelerate only  the inference of  the neural network. 

PipeLayer was one of the first ReRAM architectures that supported both inference and training phases. It achieved a speedup of 42.45x and an average energy saving of 7.17x compared with  the GPU platforms. AtomLayer also supported efficient inference and training due to a special row-disjoint kernel mapping scheme and data reuse system \cite{qiao2018atomlayer}. It provided a 1.1 × and 1.6 × higher power efficiency in  the inference and training modes compared to ISAAC and PipeLayer, respectively. PANTHER proposed the use of   bit-slicing Outer Product Accumulate (OPA) operations to achieve higher precision and to implement different training algorithms \cite{ankit2020panther}. The proposed training scheme with inter-layer parallelism and a weight reuse system was evaluated on  the PUMA inference accelerator. PANTHER achieved   up to  an 8.02 ×, 54.21 ×, and 103 × energy reduction and a 7.16 ×, 4.02 ×, and 16 × speedup compared to  the digital accelerators, ReRAM-based accelerators, and GPUs, respectively.

Table \ref{characteristics} summarizes  the basic characteristics and performance of the above-mentioned ReRAM-based accelerators. Detailed discussions are provided in  the following subsections.

\subsection{Performance}

Efficiency of  the hardware can be measured using different metrics that reflect certain characteristics. The key parameter of how  the DNN model and hardware performs on a given task is the \textbf{computational accuracy}. Its unit is usually defined by the type of  problem being solved and can be expressed in percentages, mean average precision  (MAP), root mean squared error (RMSE), and so on. When assessing the accuracy, it is also necessary to take into account the complexity of the task and the dataset. For instance, a classification problem is considered to be relatively easier than a machine translation task. Similarly,  classifications of MNIST and CIFAR-100 have different levels of difficulty. The  \textbf{scalability} metric shows the ability of a hardware to be scaled up to achieve higher performance without re-design. ISAAC, AEPE, and PUMA/PANTHER have  the  potential to be scaled further due to the  utilization of  the NoC interconnect. Bus-based PRIME, PipeLayer, and AtomLayer have limited scalability.The  \textbf{flexibility} of the ReRAM accelerator demonstrates its  ability to  support different DNN workloads and tasks. All accelerators are capable of processing CNN and DNN networks and were benchmarked on different neural network architectures, such as VGG, MSRA, AlexNet, and ResNet. Among all accelerators, PUMA has the best flexibility.

\begin{landscape}
\begin{table}[!h]
\caption{Summary of the ReRAM-based hardware accelerators.}
\resizebox{1\linewidth}{!}
{
\begin{tabular}{|l|c|c|c|c|c|c|c|c|}
\hline
\textbf{Arch. Name}& \textbf{ISAAC'16 \cite{shafiee2016isaac}}          & \textbf{PRIME'16 \cite{chi2016prime}}         & \textbf{AEPE'17 \cite{tang2017aepe}}    & \textbf{PipeLayer'17 \cite{song2017pipelayer}}  & \textbf{AtomLayer'18\cite{qiao2018atomlayer}}       & \textbf{Newton'18\cite{nag2018newton}}             & \textbf{CASCADE'19\cite{chou2019cascade}}           & \textbf{PUMA'19\cite{ankit2019puma}/PANTHER'20 \cite{ankit2020panther}}          \\ \hline
\textbf{Workload Types}  & \begin{tabular}[c]{@{}c@{}}CNN, MLP,\\ DNN\end{tabular} & CNN, MLP               & CNN, MLP         & CNN, MLP            & CNN, MLP         & \begin{tabular}[c]{@{}c@{}}CNN, MLP, \\ DNN\end{tabular}    & DNN, RNN   & \begin{tabular}[c]{@{}c@{}}MLP, CNN, LSTM, \\ GAN, RBM, RNN, \\ BM, SVM, \\ Linear  regression,\\ Logistic regression\end{tabular}      \\ \hline
\textbf{\begin{tabular}[c]{@{}l@{}}Benchmark\\ neural networks\end{tabular}}               & \begin{tabular}[c]{@{}c@{}}VGG-A/B/C/D,\\ MSRA-A/B/C,\\ DeepFace\end{tabular}   & \begin{tabular}[c]{@{}c@{}}LeNet-5, \\ CNN-1, CNN-2, \\ MLP-S/M/L,\\ VGG-D\end{tabular}        & \begin{tabular}[c]{@{}c@{}}AlexNet, \\ VGG-16,\\ ResNet50\end{tabular}   & \begin{tabular}[c]{@{}c@{}}MNIST-A-B-C-0,\\ AlexNet, \\ VGG-A/B/C/D/E\end{tabular}          & \begin{tabular}[c]{@{}c@{}}VGG-19, ResNet-152,\\ DCGAN\end{tabular}      & \begin{tabular}[c]{@{}c@{}}VGG-A/B/C/D,\\ MSRA-A/B/C,\\ AlexNet, \\ ResNet34\end{tabular}           & \begin{tabular}[c]{@{}c@{}}VGG-A/B/C,\\ MSRA-A/B/C,\\ AlexNet, DeepFace,\\ NeuralTalk\end{tabular} & \begin{tabular}[c]{@{}c@{}}BigLSTM, MLP, \\ NMT, VGG-16, \\ VGG-19, LSTM-2048\end{tabular}             \\ \hline
\textbf{Technology}& 32 nm   & 65 nm TSMC CMOS        & 32 nm            & 32 nm*              & 32 nm            & 32nm& 65nm       & 32 nm          \\ \hline
\textbf{Frequency} & 1.2 GHz & 3 GHz  & 1.2 GHz          & 1 GHz*              & 1 GHz*           & 1.2GHz      & n/a        & 1.0 GHz        \\ \hline
\textbf{\begin{tabular}[c]{@{}l@{}}Improvement\\ (baseline)\end{tabular}}  & \multicolumn{1}{c|}{\begin{tabular}[c]{@{}l@{}}Throughput: 14.8x;\\ Energy: 5.5x;\\ Comp. efficiency: 7.5x \\ (DaDianNao)\end{tabular}} & \multicolumn{1}{c|}{\begin{tabular}[c]{@{}l@{}}Performance $\sim$2360x;\\ Power efficiency  $\sim$895x  \\ (ISAAC-CE)\end{tabular}}            & \multicolumn{1}{c|}{\begin{tabular}[c]{@{}l@{}}Power efficiency \\ by 2.71x (ISAAC-CE);\\ Area efficiency \\ by 2.41x (ISAAC-CE)\end{tabular}} & \multicolumn{1}{c|}{\begin{tabular}[c]{@{}l@{}} Comp. efficiency = 3.1 x  \\
Power efficiency =0.37 x  \\(ISAAC-CE)
\end{tabular}}      & \multicolumn{1}{c|}{\begin{tabular}[c]{@{}l@{}}Power efficiency 1.1x \\ (ISAAC-CE);\\ Training efficiency 1.6x\\ (PipeLayer);\\ 15x smaller footprint\end{tabular}}&
\begin{tabular}[c]{@{}c@{}}a 77\% decrease in power, \\ 51\% decrease in \\ energy, and 2.2× increase\\  in throughput/area \\ (ISAAC-CE)\end{tabular} & \begin{tabular}[c]{@{}c@{}}consumes 3.5×\\  lower energy \\ (ISAAC-CE)\end{tabular} &
\multicolumn{1}{c|}{\begin{tabular}[c]{@{}l@{}}up to 8.02×, 54.21×, and \\ 103× energy reductions as \\ well as 7.16×, 4.02×, \\ and 16 × execution\\  time reductions\\ ( digital accelerators)\end{tabular}} 
\\ \hline
\textbf{Power, W}  & 65.8  & n/a  & 16*  & 82.6& 6.89  & n/a & n/a  & 62.5/ 105W     \\ \hline
\textbf{Area, mm2} & 85.4  & n/a  & 24*  & 168.6  & 4.80    & n/a & n/a   & 90.6/ 117 \\ \hline
\textbf{\begin{tabular}[c]{@{}l@{}}Inference \\ latency, ms\end{tabular}}  & \begin{tabular}[c]{@{}c@{}}8.00 (VGG-19) \\ 43.46 (ResNet-152)\end{tabular}     & n/a    & n/a              & \begin{tabular}[c]{@{}c@{}}2.60 (VGG-19)\\ 14.13 (ResNet-152)\end{tabular}  & \begin{tabular}[c]{@{}c@{}}6.92 (VGG-19)\\ 4.01(ResNet-152)\end{tabular} & n/a & n/a        & n/a            \\ \hline
\textbf{\begin{tabular}[c]{@{}l@{}}Peak perf., \\ GOPs\end{tabular}}       & 55 523*     & n/a    & \begin{tabular}[c]{@{}c@{}}41 943*\\ (AEPE diff =2)\end{tabular}      & 122 706* & 3 276*               & 90 006* & n/a        & 52 310          \\ \hline
\textbf{\begin{tabular}[c]{@{}l@{}}Power efficiency,\\ GOPS/W\end{tabular}}              & \begin{tabular}[c]{@{}c@{}}750 (ISAAC-CE)\\ 255.3 (ISAAC-SE)\\ 800(ISAAC-PE)\end{tabular}   & n/a    & \begin{tabular}[c]{@{}c@{}}2044.94\\ (AEPE diff=2)\end{tabular}      & \begin{tabular}[c]{@{}c@{}}142.9\\ (PipeLayer)\end{tabular} & \begin{tabular}[c]{@{}c@{}}682.5\\ (AtomLayer)\end{tabular}              & \begin{tabular}[c]{@{}c@{}}920\\ (Newton)\end{tabular} & n/a        & \begin{tabular}[c]{@{}c@{}}840\\ (PUMA)\end{tabular}           \\ \hline
\textbf{\begin{tabular}[c]{@{}l@{}}Area efficiency,\\ GOPS/mm2\end{tabular}}             & \begin{tabular}[c]{@{}c@{}}650 (ISAAC-CE)\\ 103.35 (ISAAC-SE)\\ 550(ISAAC-PE)\end{tabular} & n/a    & \begin{tabular}[c]{@{}c@{}}1789.23  \\ (AEPE diff=2)\end{tabular}              & \begin{tabular}[c]{@{}c@{}}1485\\ (PipeLayer)\end{tabular}  & \begin{tabular}[c]{@{}c@{}}615.7\\ (AtomLayer)\end{tabular}              & \begin{tabular}[c]{@{}c@{}}680\\ (Newton)\end{tabular}  &  \begin{tabular}[c]{@{}c@{}}101\\ (CASCADE)\end{tabular}         & \begin{tabular}[c]{@{}c@{}}580\\ (PUMA)\end{tabular}           \\ \hline
\textbf{\begin{tabular}[c]{@{}l@{}}Precision, \\ fixed point\end{tabular}} & 16-bit  & 16-bit & 16-bit           & 16-bit              & 16-bit           & 16-bit      & 16-bit     & \begin{tabular}[c]{@{}c@{}}16-bit (inference)\\ 16- and 32-bit (training)\end{tabular} \\ \hline
\textbf{RCA}       & 128x128 & 256x256& 128x128          & 128x128             & 128x128          & 128x128     & 64x64      & 128x128        \\ \hline
\textbf{\begin{tabular}[c]{@{}l@{}}RCA cell \\ precision\end{tabular}}     & 2-bit   & 4-bit  & 3-bit            & 4-bit               & 2-bit            & 2-bit       & 1-bit      & 2-bit          \\ \hline
\textbf{Crossbar latency}  & 100 ns  & 100 ns*& 100 ns           & 100 ns*             & 100 ns           & 100 ns*     & 100 ns*    & 100 ns*        \\ \hline
\textbf{Input encoding}    & bit-serial              & multi-level            & bit-serial       & bit-serial          & bit-serial       & bit-serial  & bit-serial & bit-serial     \\ \hline
\textbf{Input precision}   & 16-bit  & 6-bit  & 16-bit           & 16-bit              & 16-bit           & 16-bit      & 16-bit     & 16-bit         \\ \hline
\textbf{\begin{tabular}[c]{@{}l@{}}Synaptic weight\\ precision\end{tabular}}               & 16-bit  & 8-bit  & 16-bit           & 16-bit              & 16-bit           & 16-bit      & 16-bit     & \begin{tabular}[c]{@{}c@{}}16-bit (inference)\\ 16/32-bit (training)\end{tabular}      \\ \hline
\textbf{Output precision}  & 16-bit  & 6-bit  & 16-bit           & 16-bit              & 16-bit           & 16-bit      & 16-bit     & 16-bit         \\ \hline
\textbf{\begin{tabular}[c]{@{}l@{}}ReRAM materials,\\ characteristics\end{tabular}}        & \begin{tabular}[c]{@{}c@{}}TiO/HfO\\ $R_{on}/R_{off}$ = $2k\Omega/2M\Omega $\end{tabular}     & \begin{tabular}[c]{@{}c@{}}Pt/TiO2-x/Pt\\ $R_{on}/R_{off}$ = $1k\Omega/20k\Omega$\\ 2V SET/RESET\end{tabular}               & \begin{tabular}[c]{@{}c@{}}TiO/HfO\\ $R_{on}/R_{off}$ =  $2k\Omega/2M\Omega$\end{tabular}              & \begin{tabular}[c]{@{}c@{}} Pt/TiO2-x/Pt\\ $R_{on}/R_{off}$ = \\ $1k\Omega/20k\Omega$\\ 2V SET/RESET\end{tabular} & \begin{tabular}[c]{@{}c@{}}TiO/HfO\\ $R_{on}/R_{off}$ = \\ $2k\Omega/2M\Omega$\end{tabular}              & \begin{tabular}[c]{@{}c@{}}TiO/HfO\\ $R_{on}/R_{off}$ = $2k\Omega/2M\Omega$\end{tabular} & n/a        & \begin{tabular}[c]{@{}c@{}} $R_{on}/R_{off}$ = $100k\Omega/1M\Omega$\\ Read Voltage 0.5V\end{tabular}     \\ \hline
\textbf{\begin{tabular}[c]{@{}l@{}}H/W evaluation \\ models and \\ compilers\end{tabular}} & CACTI 6.5               & \begin{tabular}[c]{@{}c@{}}Synopsis Design \\ Compiler,\\ modified NVSim, \\ CACTI-3DD,\\ CACTI-IO\end{tabular}& \begin{tabular}[c]{@{}c@{}}CACTI, Orion,\\ NVsim\end{tabular}    & NVsim               & CACTI 6.5        & CACTI 6.5   & Cadence Spectre            & \begin{tabular}[c]{@{}c@{}}PUMAsim,CACTI 6.0 ,\\ Booksim 2.0 \\ Orion 3.0\end{tabular} \\ \hline
\textbf{Network-on-chip}& concentrated mesh       & memory bus             & concentrated mesh& n/a & INet, LNet, ONet & concentrated mesh           & n/a        & 2D Mesh        \\ \hline
\textbf{Off-chip links}    & HyperTransport          & HyperTransport*        & HyperTransport*  & HyperTransport*     & HyperTransport*  & HyperTransport              & HyperTransport*            & HyperTransport \\ \hline
\textbf{Largest network}   & \begin{tabular}[c]{@{}c@{}}26 layers \\ (330 million  parameters)\end{tabular}& \begin{tabular}[c]{@{}c@{}}VGG-D:\\ 16 layers;\\ 1.4*10\textasciicircum{}8 synapses\\ $\sim$1.6*10\textasciicircum{}10 operations\end{tabular} & \begin{tabular}[c]{@{}c@{}}VGG16 \\ (138 million parameters);\end{tabular}             & \begin{tabular}[c]{@{}c@{}}VGG-E\\ (141 million parameters)\end{tabular}  & \begin{tabular}[c]{@{}c@{}}VGG-19\\ (141 million parameters)\end{tabular}              & \begin{tabular}[c]{@{}c@{}}VGG-19\\ (141 million parameters)\end{tabular} & \begin{tabular}[c]{@{}c@{}}VGG-C\\ (138 million parameters)\end{tabular}         & \begin{tabular}[c]{@{}c@{}}VGG-19\\ (141 million parameters)\end{tabular}            \\ \hline
In-situ training      & \xmark      & \xmark     & \xmark     & \cmark & \cmark      & \xmark  & \xmark      & \cmark     \\ \hline
\multicolumn{9}{l}{*Expected since it is not reported in the publication}\\
\multicolumn{9}{l}{n/a: Not reported}\\
\end{tabular}
}
\label{characteristics}
\end{table}
\end{landscape}

The essential operation in NN acceleration hardware is multiply-and-accumulate (MAC) operation, which typically requires three inputs: input data, kernel, and partial sum computed from the previous layer. The output of the MAC operation is a partial sum that is stored in a separate memory unit. Another vital operation is the memory access. Normally,  the  time required for reading and writing to a memory device (e.g., eDRAM) depends on the size of the computed partial sum and memory device. Therefore,   the  efficiency of processors is traditionally measured by the  \textbf{peak performance}, determined in \textit{Giga-operations per second} (GOPS):

\begin{equation}
    \textsf{GOPS = (the number of MAC units in the chip ) x (the frequency of MAC operations x 2)}.
\end{equation}


\textbf{Throughput} shows inference efficiency  or how many data can be processed within a specific period of time. It is measured while executing convolution operations and is also expressed in GOPS. Power efficiency of the chip is defined as the   number of \textit{operations per second per watt} (e.g., GOPS/W, TOPS/W, FLOPS/W). Another performance metric is the \textbf{latency} (\textit{seconds per inference}). This is equal to the duration of time between points when input data are fetched for the model and a corresponding result is generated. However, these parameters reflect only  the maximum efficiency of  the accelerator at maximum utilization and does not fully represent average performance.

The throughput of ISAAC was increased by a very deep pipeline, which also made it vulnerable to pipeline bubbles and execution stalls. This led to the impossibility of implementing the training phase. Different ISAAC configurations were evaluated using  the following three metrics: 

\begin{itemize}
    \item Computational Efficiency (CE), represented by the number of 16-bit operations per second per unit area (GOPS/$mm^{2}$);
    \item Power Efficiency (PE), represented by the number of 16-bit operations per second per power (GOPS/W);
    \item Storage Efficiency (SE), represented by storing capacity for synaptic weights per unit area (MBytes/$mm^{2}$).

\end{itemize}

During Design Space Exploration,  the three optimal designs of ISAAC with the best computation efficiency \textit{ISAAC-CE}, power efficiency \textit{ISAAC-PE}, and storage efficiency \textit{ISAAC-SE} were identified \cite{shafiee2016isaac}.  Their peak performance was estimated assuming that all IMAs were utilized in every cycle.  The configuration of ISAAC-PE (H128-A8-C8, 8 IMAs) is close to ISAAC-CE (H128-A8-C8, 12 IMAs),  and both of them show a similar 
  performance. The architecture of ISAAC-SE (H256-A8-C512, 4 IMAs) is capable of accelerating large workloads, but it cannot achieve the performance and energy efficiency of ISAAC-CE and ISAAC-PE \cite{shafiee2016isaac}. 


\begin{figure}[!h]
     \centering
     \begin{subfigure}[t]{0.48\textwidth}
         \centering
         \includegraphics[width=\textwidth]{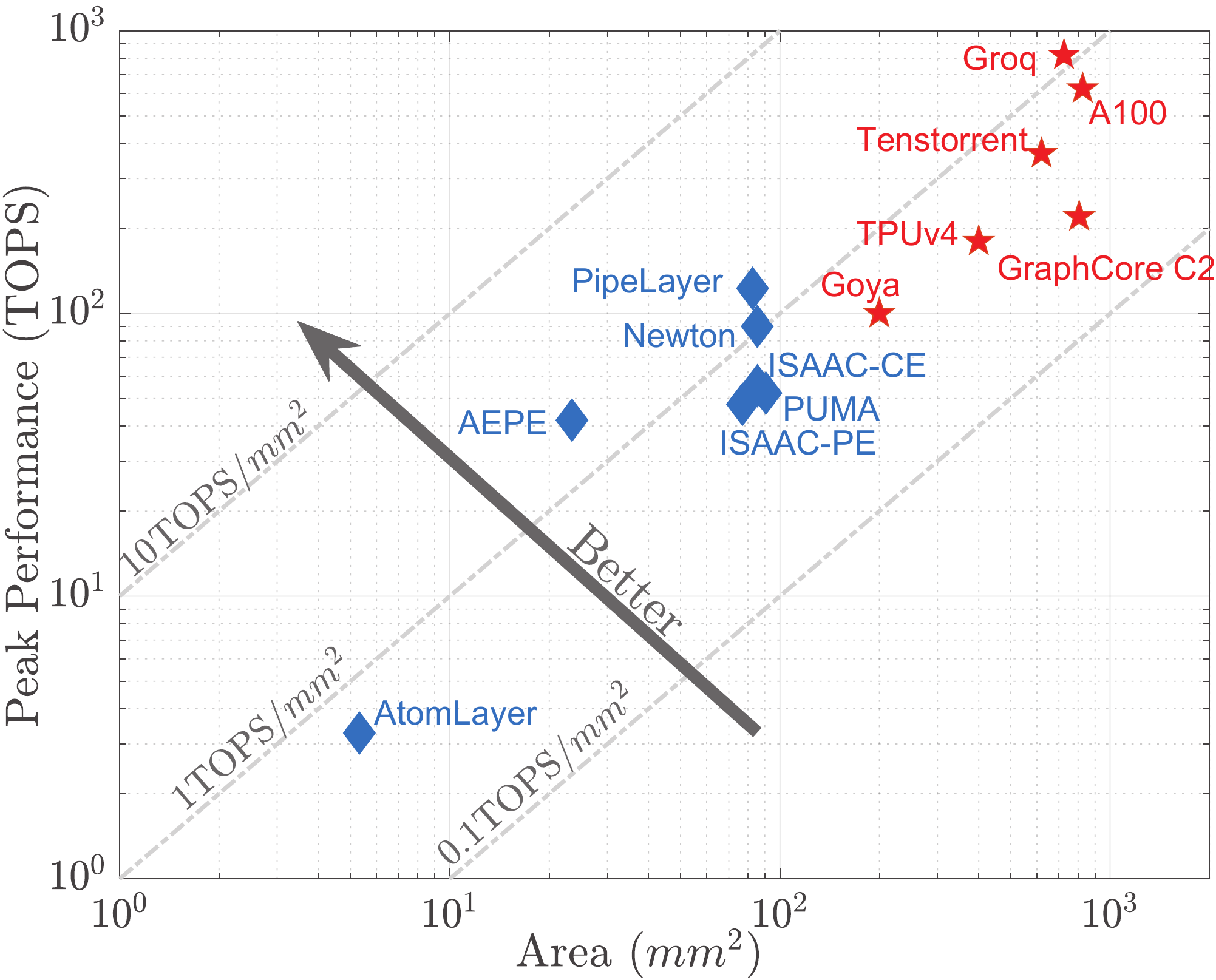}
         \caption{}
     \end{subfigure}
     \begin{subfigure}[t]{0.48\textwidth}
         \centering
         \includegraphics[width=\textwidth]{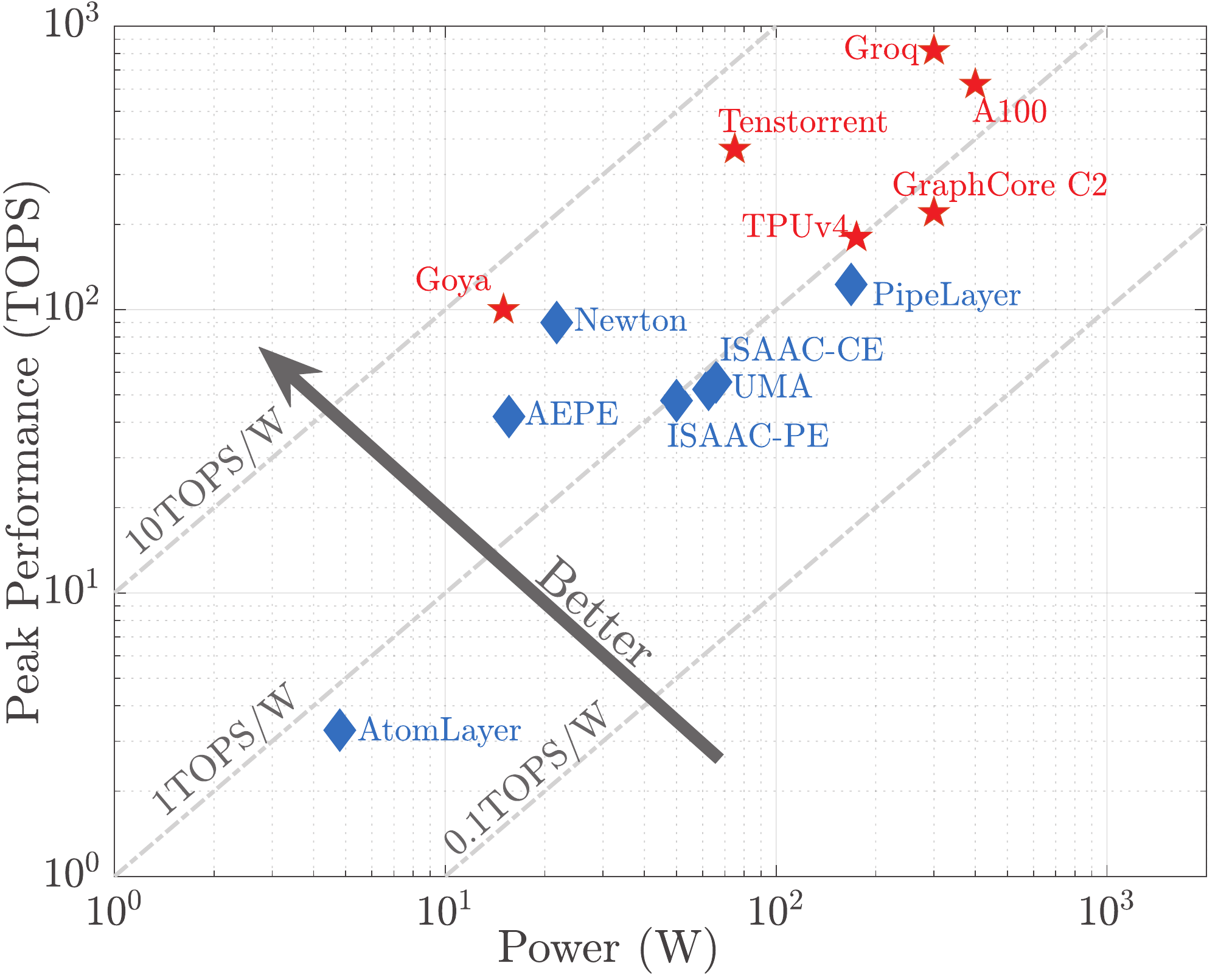}
         \caption{}
     \end{subfigure}
     \begin{subfigure}[t]{0.45\textwidth}
         \centering
         \includegraphics[width=\textwidth]{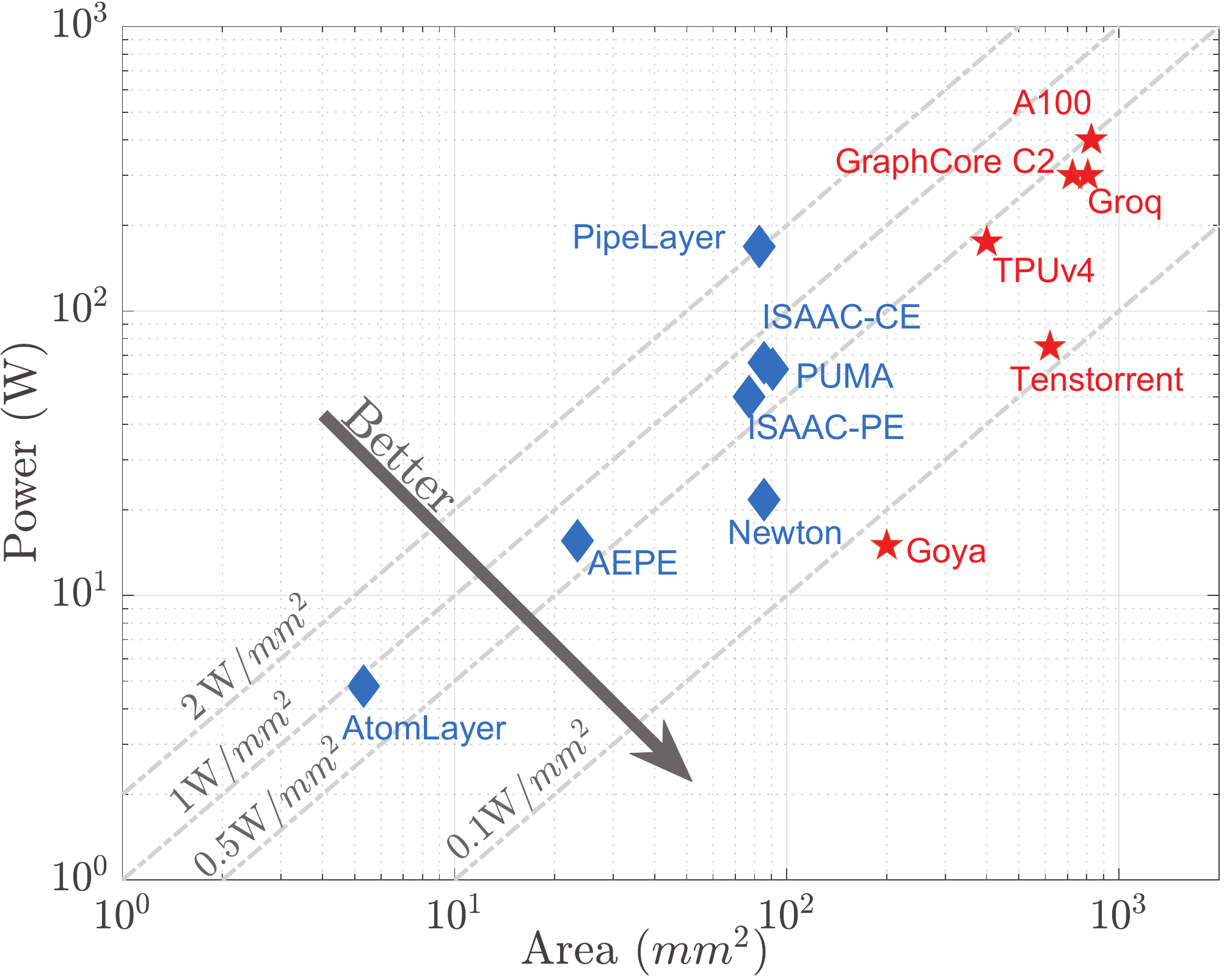}
         \caption{}
     \end{subfigure}
    
\caption{ Performance of the ReRAM accelerators against state-of-the-art commercial accelerators: a) computational efficiency;  b) power efficiency;   c) power density.
}
\label{GOPS}
\end{figure}

Figures \ref{GOPS}a-b show the relation of the estimated peak performance of ReRAM accelerators to their area and power. Figure \ref{GOPS}c reflects the corresponding power density of these accelerators. The figures depict an intra-class comparison against the performance of the state-of-the-art commercial accelerators such as Goya\cite{medina2020habana}, Google TPUv4\cite{wang_selvan}, GraphCore C2\cite{lacey}, Groq\cite{gwennap2020groq}, Nvidia A100 \cite{campa_kawalek_vo_bessoudo_2021}, and Tenstorrent \cite{Tenstorrent}. Based on the study, it is clear that the resistive accelerators have a similar performance compared to commercial accelerators. All have performances of 1 TOPS/$mm^2$ and 1 TOPS/W. However, the power density of the commercial accelerators is always less than $0.5W/mm^2$, which seems to be a bound. On the other hand, most resistive accelerators exceed this bound, showing values as high as $2W/mm^2$ for PipeLayer, except Newton, which shows a value of 0.254 $W/mm^2$.



\begin{figure}[!h]
\centering
\includegraphics[width=0.6\textwidth]{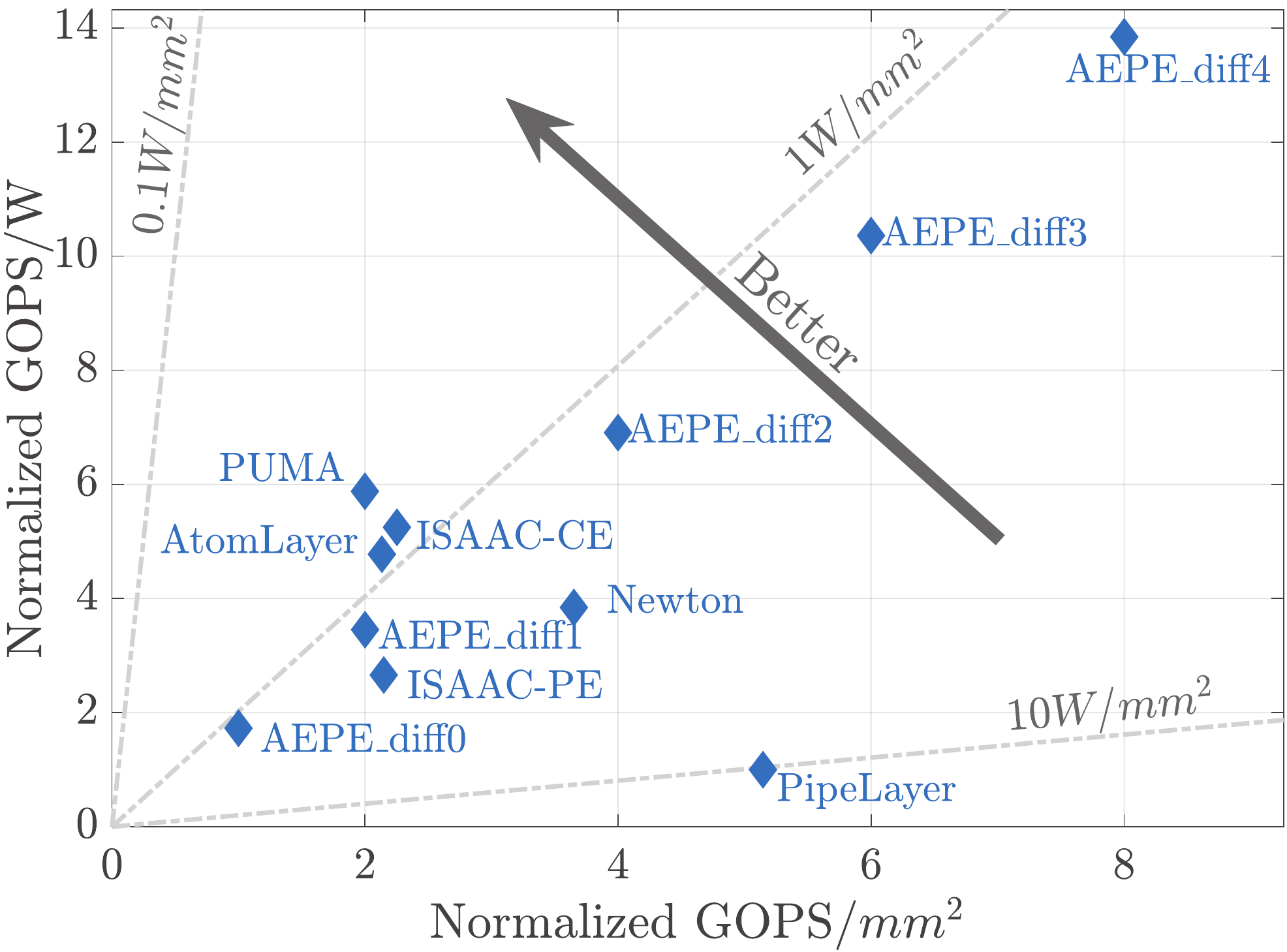}
\caption{Performance of the ReRAM accelerators: computational efficiency normalized to AEPE\_diff0 and power efficiency normalized to PipeLayer. }
\label{GOPS_reram}
\end{figure}

In order to find the optimal design,  the AEPE varied the  \textit{Diff} parameter, which is equal to a difference between the resolution of input/weight of  the NN and the resolution of  the ADCs. An increase in \textit{Diff} increases the peak performance of  the AEPE but degrades computational accuracy. The AEPE outperforms  the ISAAC-CE configuration when \textit{Diff}=2. With the same power and area,  an increase in \textit{diff} leads to an increase in the peak performance of GOPs and consequently an increase in   computational and power efficiency. However, its also leads to a decrease in   computational accuracy \cite{tang2017aepe}. The PipeLayer architecture proposed a highly parallel design that supports training. In addition, it does not use eDRAM and stores all  inputs, outputs, and filters in morphable ReRAM subarrays that can support 'storing' and 'computing' modes. This contributes to  the high computational efficiency of PipeLayer, equal to 1485 GOPS/$mm^2$. Despite   the high write energy of the RCAs, the power efficiency of  the PipeLayer is reduced to 142.9 GOPS/W \cite{song2017pipelayer}.

A special kernel mapping technique and bus-based connections helped to improve  the efficiency of AtomLayer over ISAAC in inference and that of 
 PipeLayer in training. The peak efficiency of the whole AtomLayer system was not reported, but the computational efficiency of AtomLayer for  the acceleration of VGG-19 without and with fully connected layers is 615.7 GOPS/$mm^2$ and 475.6 GOPS/$mm^2$, respectively \cite{qiao2018atomlayer}.   Newton introduced an adaptive ADC and mapping algorithms, which helped it to achieve a 2.2$\times$ higher computational efficiency and a 1.51$\times$ higher   power efficiency compared to ISAAC \cite{nag2018newton}.  The CASCADE architecture implemented an analog partial sum accumulation and achieved a peak performance of 101 GOPS/$mm2$ \cite{chou2019cascade}. PUMA has a high flexibility, scalability, and peak performance, and it is expected that its power efficiency will increase with technology scaling from 32 $nm$ to 5 $nm$ \cite{ankit2019puma}. 
The reported latency of ISAAC, PipeLayer, and AtomLayer in  the inference of VGG-19 are  8.00 ms, 2.60 ms, and 6.92 ms, respectively \cite{qiao2018atomlayer}. There are more metrics that can be used to evaluate ReRAM accelerators, including power consumption, on-chip area, and cost.

Figure \ref{GOPS_reram} shows the peak computational and power efficiency of the ReRAM accelerators normalized to  AEPE\_diff0 ( 288.61 GOPS/$mm^2$)and PipeLayer (142.9 GOPS/W), respectively. We chose to normalize these architectures      
  because they show the smallest peak performance among the others in either CE or PE.



\subsection{Power and Area Breakdown}

\begin{figure}[!h]
     \centering
     \begin{subfigure}[t]{0.45\textwidth}
         \centering
         \includegraphics[width=\textwidth]{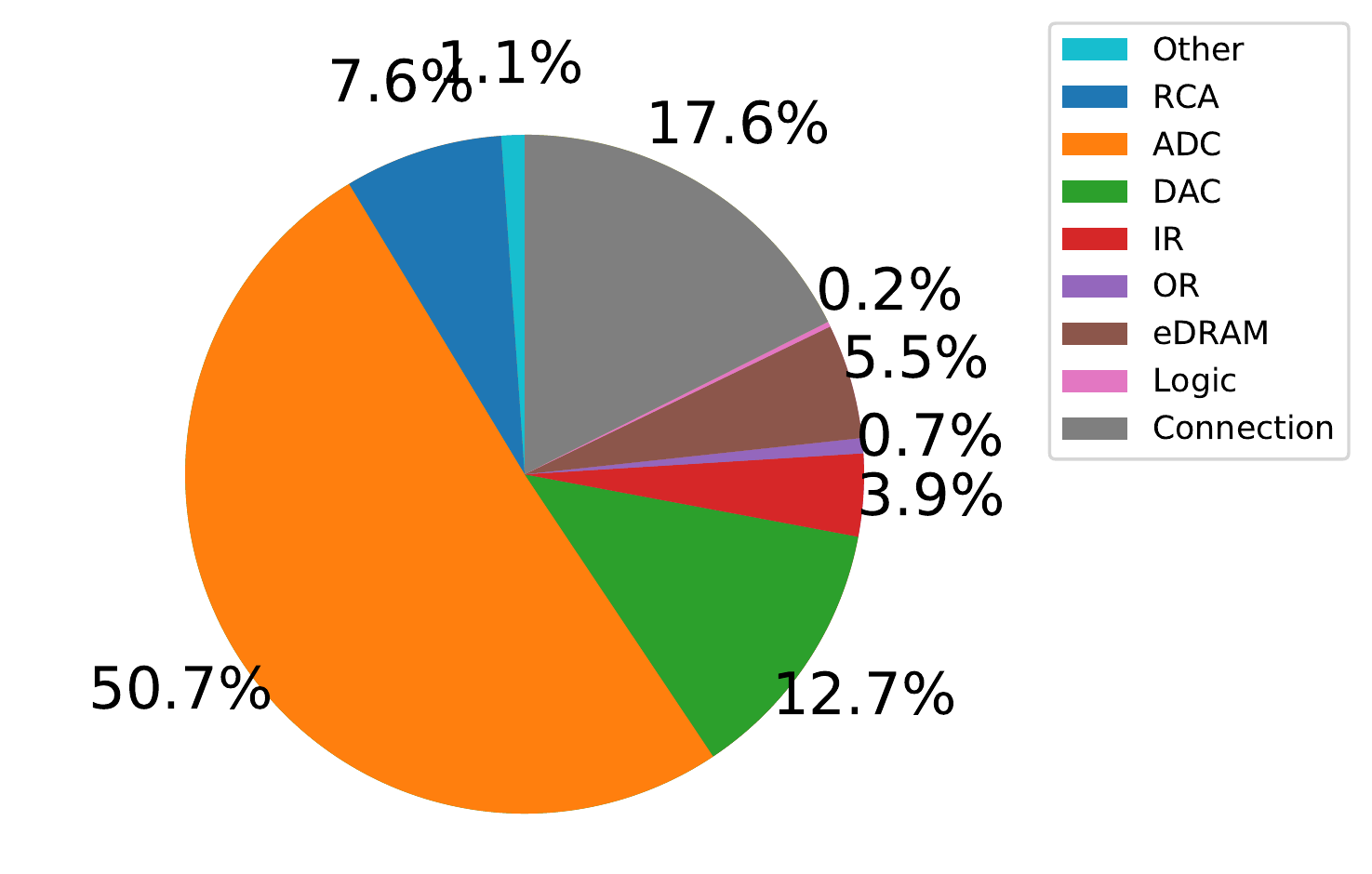}
         \caption{}
     \end{subfigure}
     \begin{subfigure}[t]{0.45\textwidth}
         \centering
         \includegraphics[width=\textwidth]{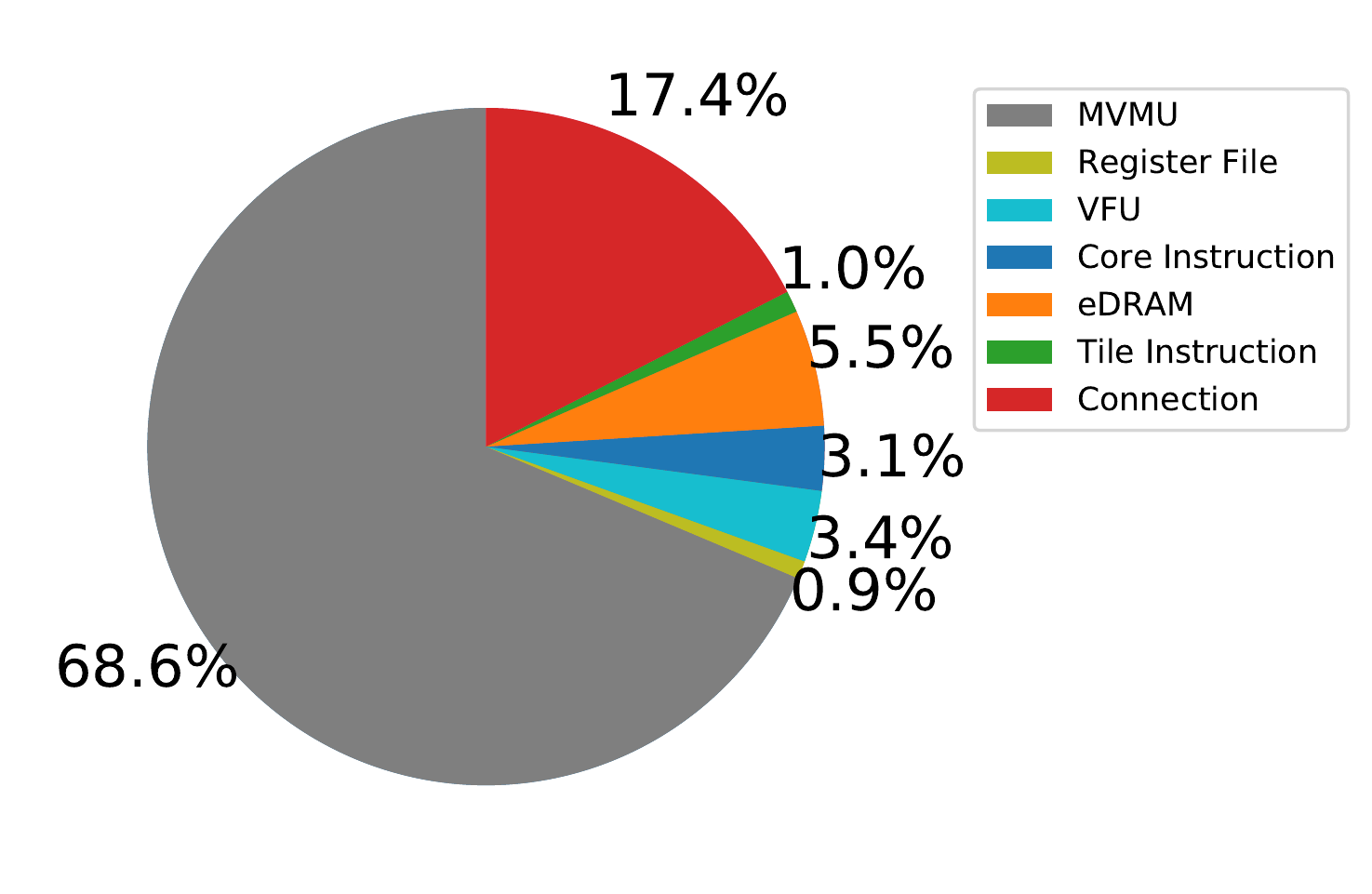}
         \caption{}
     \end{subfigure}
     \begin{subfigure}[t]{0.45\textwidth}
         \centering
         \includegraphics[width=\textwidth]{isaac_core_power.pdf}
         \caption{}
     \end{subfigure}
     \begin{subfigure}[t]{0.45\textwidth}
         \centering
         \includegraphics[width=\textwidth]{PUMA_core_power.pdf}
         \caption{}
     \end{subfigure}
\caption{ Power breakdown in the chips of a) ISAAC and b) PUMA and area breakdown in the chips of c) ISAAC and d) PUMA. }
\label{power_tiles}
\end{figure}

Figure \ref{power_tiles} depicts the power and area breakdown of components in the entire ISAAC-CE and PUMA chips. ISAAC-CE consisted of 168 tiles, each having 12 IMAs.
The PUMA node is comprised of 138 tiles. Each tile included eight cores, each having two MVMUs. The ISAAC IMA have separate units of RCA, ADC, DAC, and input and output registers. PUMA combines them in a single MVMU unit. The ISAAC tile also have independent circuit blocks for maxpooling and sigmoid functions. These operations and other types of activation functions are supported by the VFU unit of the PUMA. Both ISAAC and PUMA architectures employ a c-mesh NoC connection and a 64 KB eDRAM for storing data between NN layers. Since PUMA has an ISA-based  structure, it also includes units for storing and processing Core and Tile instruction sets. Overall, it can be concluded that ISAAC and PUMA have a similar  power distribution across their architecture.

\begin{figure}[!h]
\centering
\includegraphics[scale=0.6]{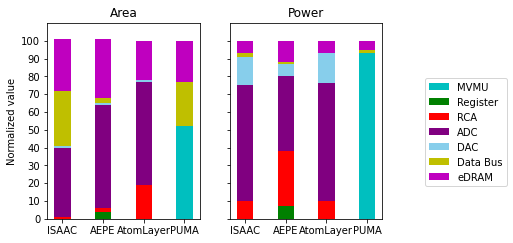}
\caption{Normalized area and power consumption of the components in  the ISAAC tile,  the AEPE tile,  the AtomLayer PE, and the  PUMA tile. }
\label{power_tiles2}
\end{figure}

Figure \ref{power_tiles2} shows the normalized power and area breakdown of the components in a single processing unit in ISAAC, AEPE, AtomLayer, and PUMA, respectively. As mentioned earlier, power and area distribution across tiles in ISAAC and PUMA are similar. AEPE is ISAAC-based configuration, but it replaced  the NoC connection by a shared bus and a smaller number of DACs. These helped to reduce  the required power and area. The overhead in  the AtomLayer architecture is caused by a number of rotating crossbars. Overall, it can be concluded that the majority of area and power in SotA ReRAM accelerators are consumed by digital peripheral circuits and communication links.

\subsection{ReRAM Technology}

MVM operation in ISAAC was implemented using 1T1R crossbar arrays. They could be built with any ReRAM technology, but to ensure a higher bit precision, the utilization of memristor devices was recommended. In particular, $TiO_{2}$ and $HfO_{x}$ memristor models satisfied the desired $R_{ON}$/$R_{OFF}$ ratio \cite{zangeneh2013design}. Moreover, transistors in  the 1T1R cell were used for access control and were not involved in dot-product operation.

The RCA in PRIME was implemented using a $Pt/TiO_{2-x}/Pt$ device model with $R_{on}/R_{off}$=1k$\Omega$ /20k$\Omega$. The device had a   resolution of up to eight bits and was tested on a small scale crossbar array in \cite{gao2013high}. According to the authors,  the ReRAM endurance in PRIME was not affected significantly. Its 
  lifetime was about $10^{12}$ in memory mode and $10^{9}$  in computational mode. In other words, over 10 years, PRIME can be reprogrammed every 300 ms, which is more than enough. In addition, the sneak-path and IR drop problems in PRIME that occur during memory mode can be overcome by available solutions, e.g., by adopting  a double-sided ground biasing technique.

Newton uses $TaO_{x}$ memristor devices based on \cite{hu2016dot}. The device was tested on different ranges of input voltages, e.g., 0$\sim$0.5V, 0$\sim$0.25V, and 0$\sim$0.125V. Good linearity was observed at maximum voltages. However, to compensate parasitic and IR-drop problems,  the DAC voltage range was limited. PUMA architecture also suggests the utilization of memristors, but other devices such as STT-MRAM and NOR Flash can also be used. Validation of PUMA/PANTHER designs were made using $TAO_{x}$ devices with stable high and low resistance states. The initial endurance was over $10^{9}$ cycles, and retention exceeded 10 years at 85$^{\circ}$C \cite{wei2008highly}. The estimated lifetime of ReRAM devices in the PUMA/PANTHER is approximately 6 years for 1000 trainings per year consisting of 100 epochs, 64 batch-size, and 1M training examples.

\subsection{Precision and Data Encoding}

Due to a limited precision of ReRAM cell values and to maintain efficient computation,   low-precision fixed-point/integer computations are preferable. Earlier work has shown that a 16-bit-wide fixed-point number representation with stochastic rounding is ample in   classification accuracy \cite{gupta2015deep}.  In order to perform an MVM operation, input data should be applied to rows of ReRAM crossbar arrays. The ISAAC accelerator performs 16-bit multiplication and therefore requires $2^{16}$ levels of input voltage. This requires 16-bit DACs, which will lead to large overheads. Therefore, in ISAAC, instead of a 16-bit fixed point number, an input is fed as multiple sequential bits and takes 16 cycles. During the first cycle, Bit 1 of the input is multiplied-and-added with synaptic weight.   It is then converted via  the ADC and stored in OR. During the second cycle, Bit 2 of the input is multiplied with synaptic weight, shifted left, added, and   stored in OR, and so on. One of the ways to reduce latency is to replicate weights on one or more IMAs. This input \textit{bit-serial method} is also used in  the AEPE, AtomLayer, PipeLayer, CASCADE, and PUMA. To enhance the weight precision, a weight-composing scheme was used in ISAAC,  the AEPE, AtomLayer, and PipeLayer. The resolution of MVM and OPA operations in PUMA/PANTHER was increased using a bit-slicing OPA technique.

\par
To improve computational accuracy, PRIME adopted an input and synapse composing scheme, which  increased the precision of input $P_{in}$ and output $P_{out}$ data from 3-bit to 6-bit and  the precision of synaptic weights $P_{\omega}$ from 4-bit to 8-bit. PRIME's FF arrays can operate in memory and computational modes. The operation of a voltage driver is controlled by multiplexer switches. During memory mode, read and write voltages have two levels, whereas, during computational mode, an  input signal is encoded to $2^{16}$ levels. \textit{Voltage-level encoding} in  the voltage driver is implemented using multilevel voltage sources. 




\subsection{Weights Mapping}

\begin{figure}[!h]
     \centering
     \begin{subfigure}[t]{0.11\textwidth}
         \centering
         \includegraphics[width=\textwidth]{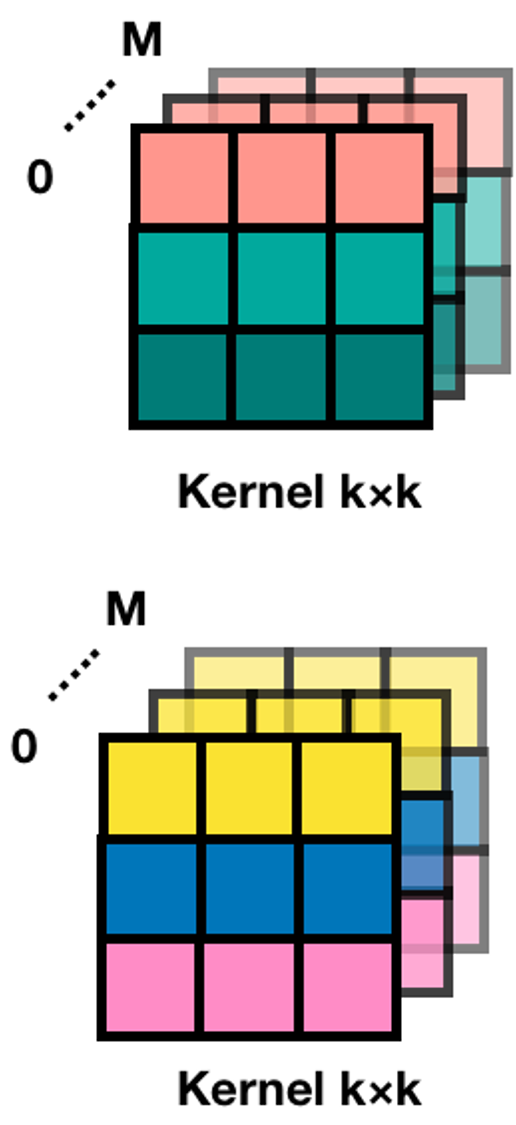}
         \caption{}
     \end{subfigure}
     \hspace{11em}
     \begin{subfigure}[t]{0.25\textwidth}
         \centering
         \includegraphics[width=\textwidth]{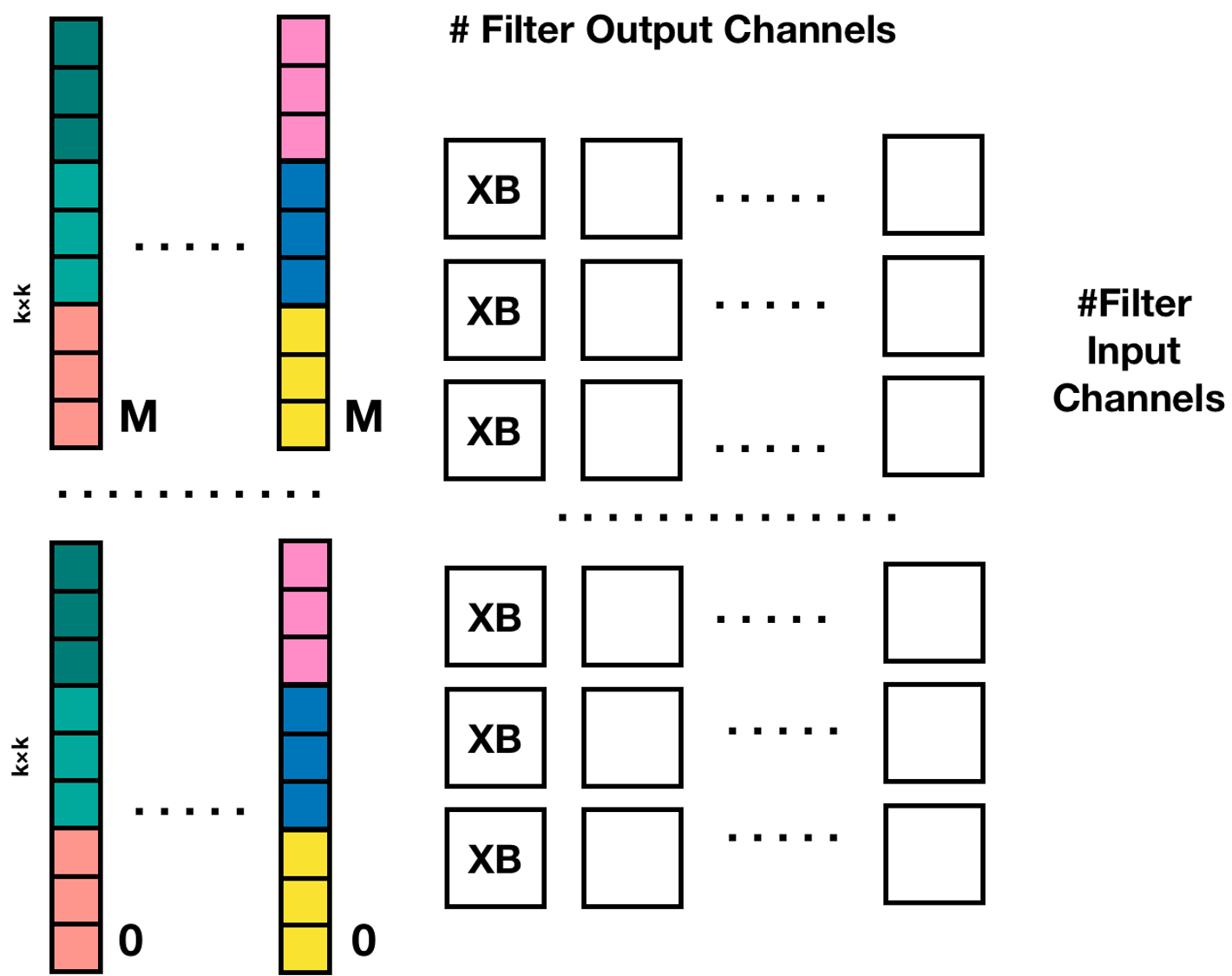}
         \caption{}
     \end{subfigure}
     \begin{subfigure}[t]{0.25\textwidth}
         \centering
         \includegraphics[width=\textwidth]{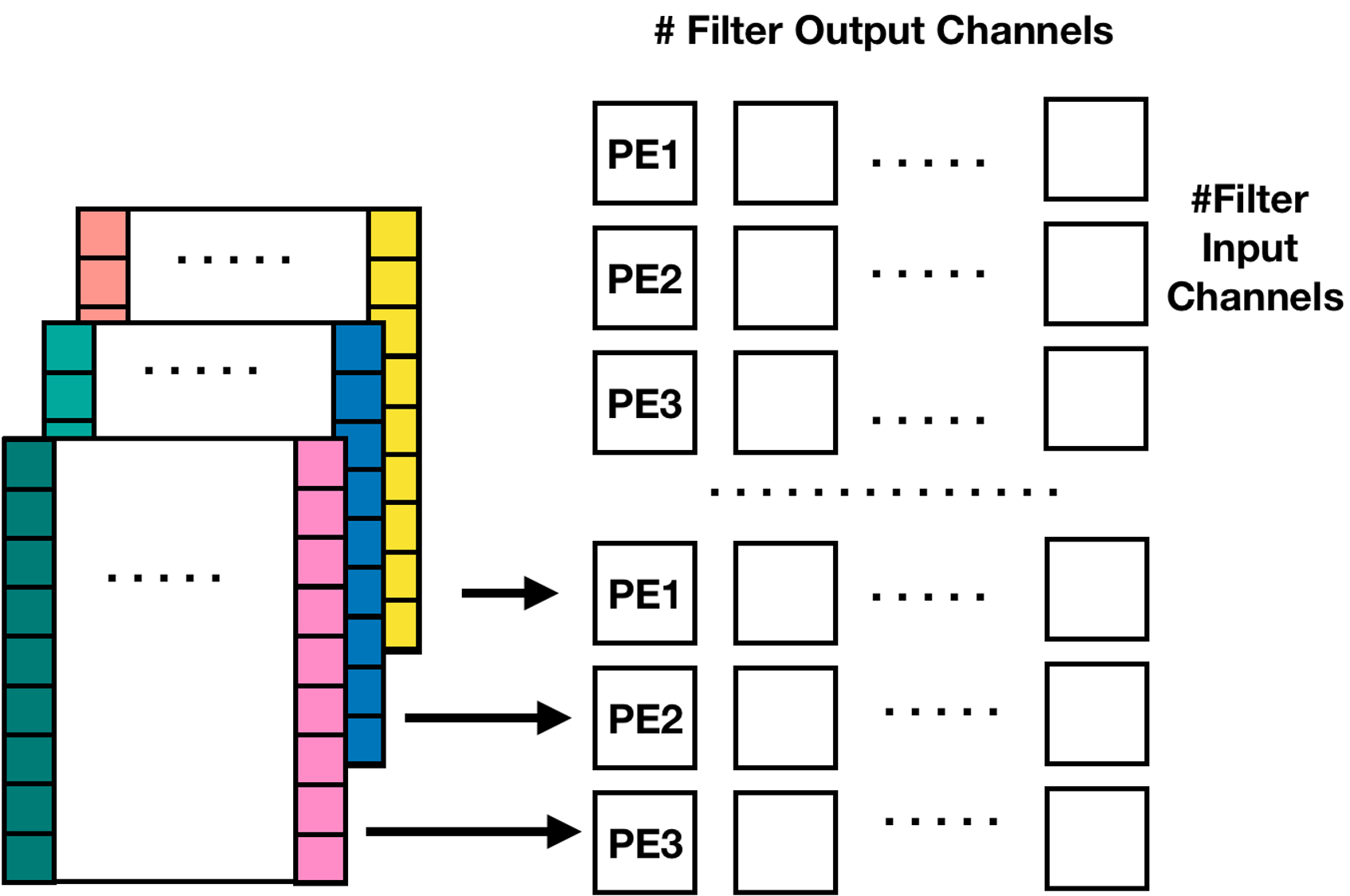}
         \caption{}
     \end{subfigure}
     \hspace{3em}
     \begin{subfigure}[t]{0.4\textwidth}
         \centering
         \includegraphics[width=\textwidth]{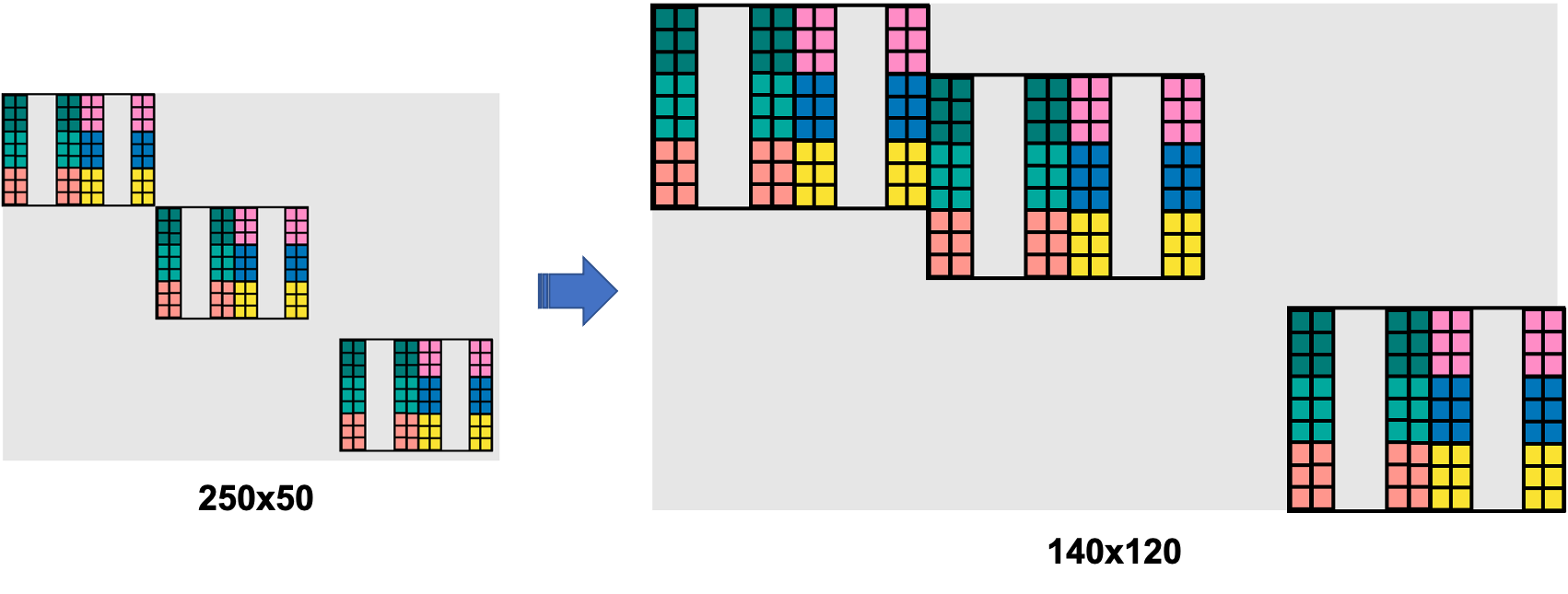}
         \caption{}
     \end{subfigure}
\caption{ a) Neural Network kernels. Kernel mapping schemes in b) ISAAC, c) AtomLayer, and d) PRIME.}
\label{kernel}
\end{figure}

The kernel to crossbar conductance mapping methods has an impact on the overall efficiency of the proposed ReRAM accelerators, including area, power consumption, and latency.  For instance, in ISAAC, each kernel of a neural network was unfolded and mapped to different columns of  the RCA, as shown in Figure \ref{kernel}b, operating on a set of inputs in parallel. To consider a sign of the synaptic weights, kernels were encoded to a 16-bit signed number. A single weight was implemented using eight 2-bit ReRAM cells. The kernel mapping of ISAAC is straightforward but    inefficient. Due to an excessive number of RCAs, further optimization is required. Newton adopts numeric algorithms to optimize MVM operation and weight mapping.


\par 
In the AEPE, kernels can be directly mapped to the tile by partitioning the weight matrix among the partitioned RCAs. AtomLayer exploits a row-disjoint filter mapping to implement intra-row and inter-row data reuse. In this approach, the same filter rows are concatenated and reshaped to fit  the crossbar  size and are mapped to the same PEs, as shown in Figure \ref{kernel}d.
PRIME utilized positive-negative-split weight mapping. In PRIME, two separate crossbar arrays were exploited for positive and negative weights, and each 8-bit weight was comprised of two 4-bit ReRAM cells. The detailed kernel mapping scheme was discussed in \cite{chi2015processing}.  Figure \ref{prime}b  illustrates how the first layer of  the \textit{CNN-1}  neural network (conv5x5-pool-720-70-10) was mapped to mats of the accelerator. The first convolutional layer of the network had five 5$\times$5 convolution kernels.  Therefore, the first layer required a weight matrix of size 25(5$\times$5)$\times$10(=5$\times$2). To improve computation efficiency, the matrix was replicated 10 times and mapped to a crossbar array mat of size 256$\times$256. Since two adjacent convolutions share inputs, the mapping scheme was further optimized. As a result, the rearranged matrix was replicated 24 times. Consequently, the number of mat inputs decreased from 250 to 140, but the number of output data increased from 50 to 120. 
CASCADE suggests using 1-bit ReRAM cells. In PUMA/PANTHER,  the  advantages of heterogeneous weight slicing were demonstrated.

\subsection{Power Management and Communication Network}

\par

The latest progress in process technologies allowed for the improvement of computational efficiency via  the placement of multiple cores on a single chip. As a result, the increase of the number of heterogeneous components on a chip raised the issue of effective communication. The most common and simplest type of inter-core communication is a bus interconnection. It has a relatively simple topology and  a low area cost, and it is easy to build and efficient to implement \cite{bjerregaard2006survey}. For these reasons, bus lines were used in the design of PRIME and AtomLayer accelerators.  For inter-bank communication, PRIME uses an internal shared bus.  Nevertheless, traditional bus-based communication systems suffer from a long propagation delay for data transfer, a larger load per data bus line, a lack of bandwidth, and high power consumption. \cite{mitic2006overview}.

\begin{figure}[!h]
     \centering
     \hspace{3em}
     \begin{subfigure}[t]{0.15\textwidth}
         \centering
         \includegraphics[width=\textwidth]{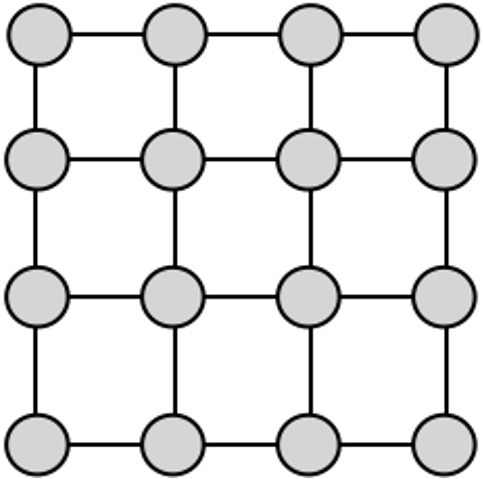}
         \caption{}
     \end{subfigure}
     \hspace{4em}
     \begin{subfigure}[t]{0.15\textwidth}
         \centering
         \includegraphics[width=\textwidth]{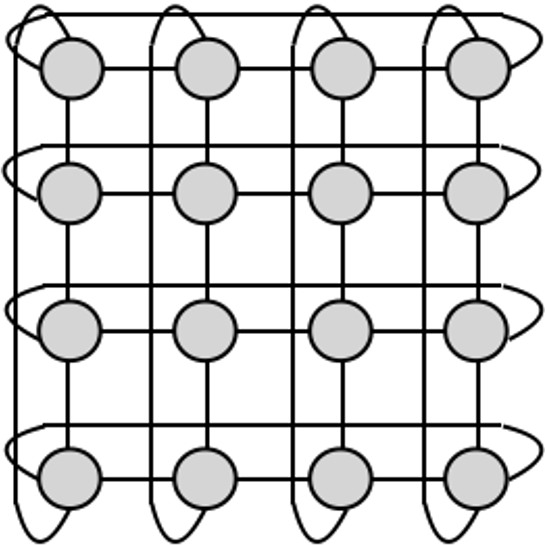}
         \caption{}
     \end{subfigure}
     \hspace{3em}
     \begin{subfigure}[t]{0.15\textwidth}
         \centering
         \includegraphics[width=\textwidth]{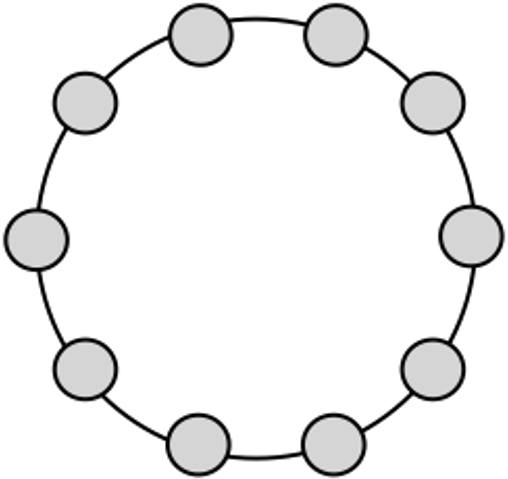}
         \caption{}
     \end{subfigure}
     \\
     \begin{subfigure}[t]{0.3\textwidth}
         \centering
         \includegraphics[width=\textwidth]{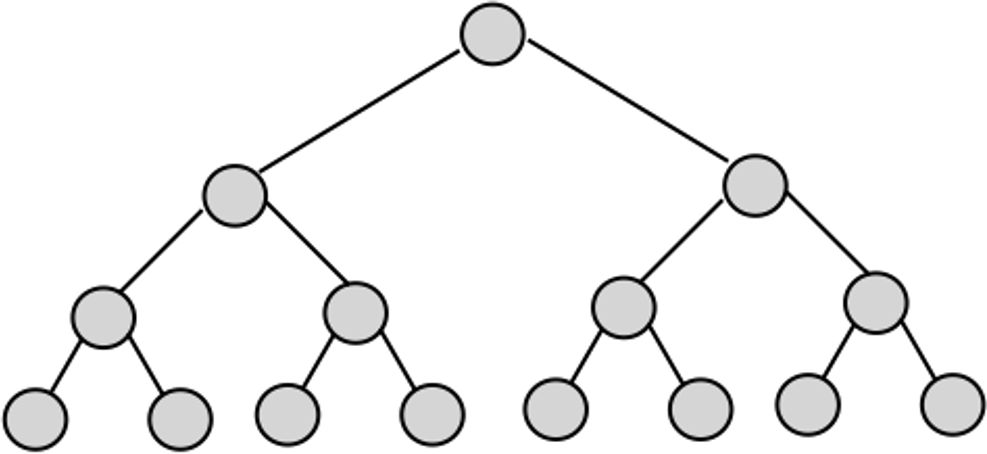}
         \caption{}
     \end{subfigure}
     \hspace{1em}
     \begin{subfigure}[t]{0.15\textwidth}
         \centering
         \includegraphics[width=\textwidth]{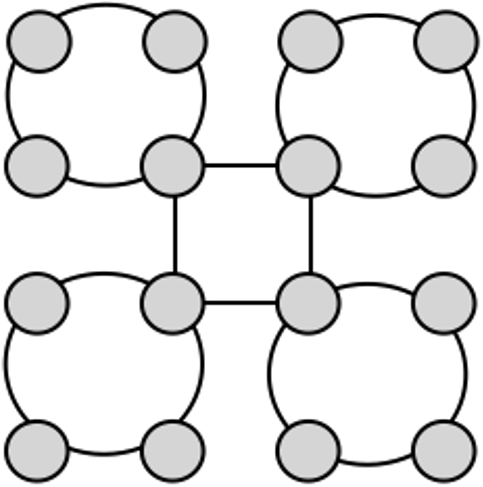}
         \caption{}
     \end{subfigure}
     \hspace{3em}
     \begin{subfigure}[t]{0.15\textwidth}
         \centering
         \includegraphics[width=\textwidth]{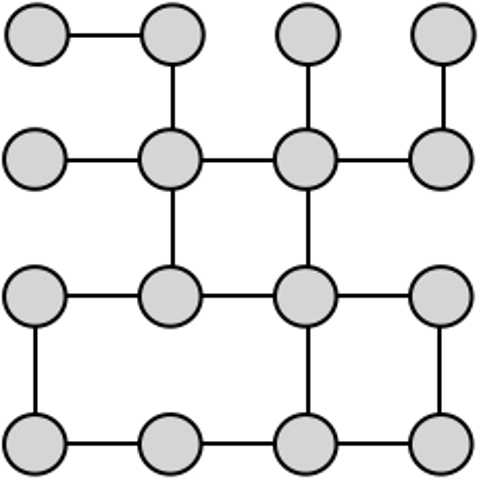}
         \caption{}
     \end{subfigure}     
\caption{ Regular NoC topologies: a) mesh; b) torus; c) ring; d) tree. Irregular NoC topologies: e) mixed mesh with ring; f) mesh with removed links. }
\label{NoCtypes}
\end{figure}

An alternative architecture of bus interconnection,  Network-on-chip (NoC) is gaining more   attention as a promising intrachip communication infrastructure in multi-core processors  \cite{tsai2012networks}. The concept of NoC aims to improve the power efficiency, fault tolerance,  parallelism, and scalability of the SoC. This is especially beneficial for  the deployment of DNN models that imply massive parallel computations. Generally, NoC has regular and irregular types of topology, and Figure \ref{NoCtypes} shows  the most common of these. The first class consists of basic topologies such as   the  2D mesh, torus, ring, and tree. The 2D mesh in Figure \ref{NoCtypes}a is considered to be   a common router-based NoC topology in the  layout of multi-core architectures   due to its high scalability and high throughput \cite{lin2019optimizing}. A torus interconnection (Figure \ref{NoCtypes}b) is similar to a mesh with additional wrap-around connections between edge switches. These wrap-around links increase the speed of processing due to additional routes, but also complicate the  physical design and  the cost of the chip. The next popular configurations are  the ring and  the tree. In a single core \cite{lin2019optimizing}, the data are injected into one of the nodes and passed to a destination node along the ring. The irregular connections as in Figures \ref{NoCtypes}e-f are mainly comprised of modified versions of regular connections, e.g., a mixed mesh with a ring and a mesh with removed links.
The tiles in ISAAC were connected via NoC with concentrated-mesh (c-mesh) topology \cite{shafiee2016isaac}. Its effectiveness over  the 2D mesh has been shown in a number of conducted studies  \cite{kumar2009exploring, kalimuthu2012comparative, chmaj2015concentrated}. Energy and area  
 nodes of PUMA were connected via a chip-to-chip 2D-mesh NoC interconnect for large-scale execution. 

\subsection{Modeling Tools}

Power and area estimation of ISAAC and AtomLayer was achieved using CACTI 6.5. Accelerators such as PRIME, AEPE, and PipeLayer used NVsim and extracted some of the parameters from CACTI and Orion. NVsim allows for more flexibility in array organization and peripheral device  modeling \cite{dong2012nvsim}. PUMA and PANTHER used a specially designed PUMAsim simulator, which also utilized models extracted from CACTI, Orion, and NVsim.
All accelerators apart from PRIME were designed and scaled to  the 32 nm technology at a 1 GHz frequency. Connections in all models were estimated using HyperTransport. The link bandwidth of PUMA and ISAAC was 6.4 GB/s.

\section{Fabricated ReRAM hardware}
\label{sec:SotA_RRAM_macro}

Non-volatile computing-in-memory (nvCIM) macros aim to realize  the high-speed and energy-efficient acceleration of artificial intelligence applications in   edge devices. Table \ref{macros} summarizes  the performance of   recent ReRAM-based macros. A 65 nm ReRAm macro was designed to accelerate a binary convolution neural network (CNN) \cite{chen201865nm}. In this design, positive and negative weights were implemented using two separate nvCIM-P and nvCIM-N subarrays.  The increase in  the  precision of the readout circuit  increased  the inference accuracy but also lowered energy efficiency.  The next design was a 55 nm  1 Mb Multibit ReRAM CIM Macro \cite{xue2019embedded}, which allowed for the implementation of a CNN on macros using multibit input/weight/output. The  achieved accuracy in  the CIFAR10 classification task was 88.52\%. In a fabricated 22 nm 2 Mb ReRAM-CIM macro  \cite{xue202015},  the  precision of input data was increased from binary to 4-bit. It also eliminated the previous positive-negative-split weight mapping and decreased area. The device was proved to be competitive with only a 0.93\% degradation compared to software simulation results.  A Fully Integrated Analog ReRAM-based 130 nm macro used a 2T2R cell, which  decreased the effect of the IR drop by decreasing the accumulative SL current \cite{liu202033}. It also introduced a low-power resolution-adjustable ADC with a precision of up to 8 bits. The obtained accuracy of the accelerated 784-100-10 fully connected neural network in the  classification of    MNIST data was 94.4\%. The design utilized a signed ternary weights. A Reconfigurable 4T2R ReRAM Computing In-Memory Macro on a 40 nm process  \cite{chen2021reconfigurable} utilized a 4T2R cell, which allowed for   row-wise memory access. Therefore, it can either function  as a reliable non-volatile memory or perform fast CIM. A recently reported 40 nm ReRAM macro supported the highest resolution and allowed for only positive weights and inputs \cite{yoon202129}.  It also provided read-disturb tolerance due to voltage-mode sensing. The first ReRAM-based system for edge AI inference and incremental training with no off-chip memory was recently presented and called CHIMERA  \cite{giordano2021chimera}.   Its special incremental training scheme    achieved 283x fewer RRAM weight updates and a 340x efficient energy-delay product. CHIMERA utilizes the \textit{illusion} mapping schemes to combine several CHIMERA chips in order to implement  a neural network of different scale.


\begin{table}[]
\caption{Recent Fabricated ReRAM macros for ML/DL Acceleration.}
\centering
\resizebox{\linewidth}{!}{
\begin{tabular}{|l|c|c|c|c|c|c|c|}
\hline
   & \textbf{ISSCC'18}\cite{chen201865nm}   & \textbf{ISSCC'19}\cite{xue2019embedded}        & \textbf{ISSCC'20}\cite{xue202015}            & \textbf{ISSCC'20}\cite{liu202033}     & \textbf{ OJCAS'21} \cite{chen2021reconfigurable}
    & \textbf{ ISSCC'21} \cite{yoon202129}
    & \textbf{ VLSIC'21} \cite{giordano2021chimera}\\
    \hline
\textbf{Technology}   & 65nm& 55nm     & 22nm        & 130nm & 40nm   &40nm &  40nm \\ \hline
\textbf{Capacity}     & 1Mb & 1Mb      & 2Mb         & 158.8Kb            & 2Mb    & 64Kb    & 2Mb  \\ \hline
\textbf{Subarray Size}     &8x128K& 256x512  & 512x512     & N/A& 128x128 & N/A & 16x16 \\ \hline
\textbf{Cell}         & 1T1R& 1T1R     & 1T1R        & 2T2R  & 4T2R    & 1T1R  & Systolic Array (SA) \\ \hline
\textbf{ReRAM size}   & 0.25$\mu m^2$          & 0.2025$\mu m^2$& N/A       & 1.69$\mu m^2$            & 0.55$\mu m^2$   & N/A  & N/A \\ \hline
\textbf{Cell size}    & N/A& N/A    & N/A       & 3.38$\mu m^2$            & N/A     & N/A  & N/A\\ \hline
\textbf{Die area}     & N/A& N/A   & 6$mm^2$     & 21.82$mm^2$           & N/A    & 0.437$mm^2$    & 29.16$mm^2$ \\ \hline
\textbf{ReRAM mode}   & Memory/ CIM     & Memory/ CIM & Memory/ CIM& Memory/ CIM& Memory/ CIM & Memory/ CIM  & Memory/ CIM \\ \hline
\textbf{RRAM precision}  & binary           & ternary  & 4bit        & Analog (0.4-4uA)   & ternary   & N/A & N/A\\ \hline
\textbf{ADC} & 3b  & 3b & N/A & 1b-8b & N/A & 4b& N/A\\ \hline
\multicolumn{1}{|l|}{\textbf{Precision (I,W,O)}}       & 1,T,3 &2,3,3 & 2,4,10& 1,T,8 & N/A & 1-8b/1-8b/20b & INT8, FP16 \\ \hline
\textbf{Energy efficiency, (TOPS/W)} & 25.42 @ 1V  & 21.9 @1V     & 45.52@0.8V & 78.4 @4.2V & 223.6 @ 0.7V  & 56.67 @ 0.9V  & 2.2@1.1V\\ \hline
\textbf{Read Delay (ns)} & 14.8& 14.6&13.1&51.1& 0.92 & N/A & N/A\\ \hline
\textbf{Inference Speed}           & N/A& N/A    & N/A       & 77 $\mu$ s/Image & N/A    & N/A & N/A \\ \hline
\textbf{Accuracy}     &  MNIST: N/A       & CIFAR10: 88.52\% & \begin{tabular}[c]{@{}c@{}}CIFAR10: 90.18\%  \\ CIFAR100: 64.15\%\end{tabular} & MNIST: 94.4\%  & \begin{tabular}[c]{@{}c@{}} MNIST: 95.7\% \\ CIFAR10: 81.7\% \end{tabular} & N/A  & \begin{tabular}[c]{@{}c@{}} ImageNet: 69.35\% \\Flowers102: 39.98\% \end{tabular} \\ \hline
\end{tabular}
}
\label{macros}
\end{table}

The overview of the presented architectures shows that an improvement of inference accuracy was typically achieved by increasing the precision of input/weight/output in the design of ReRAM macros. In particular, there is a rise in the number of ReRAM devices and transistors in a cell of crossbar arrays. This requires a trade-off between precision and energy efficiency.

\section{Potential Directions}
\label{sec:SotA_potential}


Chip design requires  the consideration of several aspects such as an I/O interface, communication, computation accuracy, and memory capacity. At  a high level, a typical neural network accelerator consists of multiple processing units (PUs) and an inter-PU communication network. Depending on  the DNN model, the design of  the hardware varies in  the number of PUs,  the technology used to implement MAC operation,  the amount of memory, the  connection network, and other characteristics. Potential research  on the  improvement of ReRAM multi-node and multi-core accelerators include several directions. First of all, this includes  the  revision of   system evaluation metrics. Hardware advancements are also required at different levels of the system hierarchy:  the ReRAM array level, the  peripheral interfacing level, and  the inter-chip communication level.  One of the essential aspects that affect inference and training latency and accuracy is pipelining   RCAs and PUs. More details on potential research directions are provided below.

\subsection{New Performance Metrics}

Recently, many hardware architectures for DNN acceleration have been proposed. There is a need for new performance metrics that could fairly compare among them,  especially for CIM-based architectures. Simple metrics such as GOPs, GOPS/W, or GOPS/mm$^2$ are not sufficient and do not reflect the limitations of   architectures. A new performance evaluation framework such as \textit{Eyexam} is needed, which extends the conventional roofline model for DNN accelerators and provides a full performance profile considering both hardware and architecture characteristics such as  the neural network design,   the total number of PUs,  the number of utilized and active PUs,  the physical dimensions of PUs, workload and dataflow parallelism, and  the amount of precision  \cite{sze2020evaluate}. It relates bandwidth and  the computational roof with   peak performance. The evaluation of DNN of certain hardware using Eyexam  helps to determine its memory  or computation bounds. The identified performance bound is used to improve the computation or memory capabilities of the application design.



The throughput of the hardware system is determined by  the  \textit{operations per second} (which depends on DNN hardware and  the DNN model) and  \textit{operations per inference} (which depends on  the DNN model only) metrics. 
In the case of the system with multiple PUs, they can be further decomposed in order to reflect the peak throughput of a single PU  as well as the utilized and the active PUs. 
In such systems, the throughput can be increased by increasing the number of PUs or  the utilization of the reduced precision. In addition, the running time required for MAC and the memory access operations can vary significantly. Therefore, defining them in \textit{operations per second} does not fully reflect the efficiency. In addition, 
not all operations are created equal. Therefore, the total operations can be further decomposed to \textit{effectual} (i.e., non-zero) and \textit{ineffectual} (i.e., multiplication by zero) operations. The number of effectual operations decreases with the increase in the amount of sparsity in  the DNN model. The ineffectual operations are operations that do not change the accumulated value.  
It is preferable that all ineffectual operations would be skipped by hardware, which is difficult to implement. Therefore, for evaluation purposes, they are further decomposed to \textit{exploited} (i.e., skipped) and \textit{unexploited} (i.e., not skipped) operations. The analysis of other metrics and their decomposition showed a close connection between them. In particular, batching   input data  increases throughput as well as  the latency of the system. A decreasing number of PUs saves area and cost but degrades overall latency and throughput. Increasing resolution can lead to an increase in accuracy, but also   large overheads. Therefore, there are always design trade-offs between  the desired metrics, which are summarized in Table \ref{PU_perf}. Unfortunately, due to the lack of data, the performance of the aforementioned accelerators has not been evaluated accordingly.

Eyexam provides a full performance profile of DNN hardware as a function of its characteristics. The analysis consists of seven steps, and each step puts on a certain constraint. Extending such framework to consider the nature of CIM architectures would be of interest.  For a fair comparison, new performance metrics should be introduced. For instance, \textit{operations per second} should be normalized to 1 bit. Moreover, through a  layer-wise performance comparison of the DNN using \textit{inferences per second}, \textit{energy per inference} metrics  can be conducted. 
Generally, the evaluation metrics of DNN accelerators should not be limited only to the peak performance and is 
 in need of adjustment and revision. 

\begin{table}[!h]
\centering
\caption{Evaluation metrics and performance improvement methods in the ReRAM-based accelerators.}
\resizebox{\textwidth}{!}{
\begin{tabular}{|l|l|l|l|}
\hline
\multicolumn{1}{|c|}{\textbf{\begin{tabular}[c]{@{}c@{}}Performance\\ Metric\end{tabular}}} & \multicolumn{1}{c|}{\textbf{Measurement}}        & \multicolumn{1}{c|}{\textbf{Note}}& \multicolumn{1}{c|}{\textbf{\begin{tabular}[c]{@{}c@{}}Improvement\\ Technique\end{tabular}}}  \\ \hline
\textbf{\begin{tabular}[c]{@{}l@{}}Computation \\ accuracy\end{tabular}}       & \% , MAP, RMSE , etc.  & \begin{tabular}[c]{@{}l@{}}The unit is defined by \\ a performed task\end{tabular}    & \begin{tabular}[c]{@{}l@{}}Efficient neural network architecture;\\ Increasing bit precision;\\ Increasing reliability of the hardware; \\ High-endurance and long-term retention ReRAM;\\ Accurate weight update scheme;\\ Decreasing latency.\end{tabular}         \\ \hline
\textbf{Throughput}       & \begin{tabular}[c]{@{}l@{}}Generically: \\ Number of operations/\\ second\\ \\ For inference:\\ Seconds/inference\end{tabular} & \begin{tabular}[c]{@{}l@{}}The throughput reflects \\ the data volume that\\ can be processed/  \\ a number of tasks that \\ can be completed within \\ a specified time period.\end{tabular} & \begin{tabular}[c]{@{}l@{}}Efficient neural network architecture;\\ Batching input data;\\ Efficient kernel mapping;\\ Efficient data encoding;\\ Efficient communication network;\\ Sufficient storage capacity\\ Decreasing power consumption;\\ Increasing energy efficiency\\ Increasing number of PUs (=maximum number \\ of MAC operations performed in parallel)\\ Increasing the utilization of PUs;\\ Weight reuse system.\end{tabular} \\ \hline
\textbf{Latency}           & Inferences/second      & \begin{tabular}[c]{@{}l@{}}The period of time between\\  points when input data\\  is fetched to a system and\\ corresponding results is \\ generated\end{tabular}  & \begin{tabular}[c]{@{}l@{}}Efficient neural network architecture;\\ Parallel data fetching;\\ Data reuse;\\ Parallel processing;\\ Reducing precision;\\ Short interconnects;\\ Efficient communication network topology.\end{tabular}        \\ \hline
\textbf{\begin{tabular}[c]{@{}l@{}}Power \\ consumption\end{tabular}}          & Watt or Joules/second  & \begin{tabular}[c]{@{}l@{}}The amount of energy \\ consumed per unit time.\end{tabular}            & \begin{tabular}[c]{@{}l@{}}Reducing number of PUs;\\ Lowering precision;\\ Lowering ReRAM conductance;\\ Reducing number of multi-hop links;\\ Reducing number of peripheral devices.\end{tabular}     \\ \hline
\textbf{\begin{tabular}[c]{@{}l@{}}Energy \\ efficiency\end{tabular}}          & \begin{tabular}[c]{@{}l@{}}Generically: \\ Number of operations/\\ joule\\ \\ For inference:\\ Seconds/joule\end{tabular}      & \begin{tabular}[c]{@{}l@{}}The data volume that\\ can be processed/a number \\ of tasks that can \\ be completed for a given \\ unit of energy\end{tabular}        & \begin{tabular}[c]{@{}l@{}}Data reuse;Reducing number of PUs;\\ Reducing precision.\end{tabular}            \\ \hline
\textbf{\begin{tabular}[c]{@{}l@{}}Energy \\ consumption\end{tabular}}         & Joules/inference       & \begin{tabular}[c]{@{}l@{}}Total energy consumption include \\ energy consumed during data \\ movement and arithmetic/ MVM\\ operations\end{tabular}   & \begin{tabular}[c]{@{}l@{}}Data reuse;\\ Short interconnects;\\ Utilization of smaller technology; \\ Energy-efficient implementation of arithmetic\\ operations;\\ Reducing precision.\end{tabular}    \\ \hline
\textbf{Area} & $mm^2$& On-chip area         & \begin{tabular}[c]{@{}l@{}}Reducing number of PUs;\\ Reducing resolution of ADC/DAC;\\ Reducing number of peripheral devices;\\ Utilization of smaller technology;\\ Utilization of MLC ReRAM.\end{tabular}         \\ \hline
\textbf{Cost} & Monetary expression    & \begin{tabular}[c]{@{}l@{}}Financial metric of the system \\ or/and operational expenses \\ and design\end{tabular}          & \begin{tabular}[c]{@{}l@{}}Reducing number of PUs;\\ Reducing chip area;\\ Reducing cost of technology;\\ Lowering power consumption;\\ Decreasing flexibility;\\ Decreasing scalability.\end{tabular} \\ \hline
\textbf{Flexibility}       & The degree of flexibility           & The range of supported workloads  & \begin{tabular}[c]{@{}l@{}}Increasing number of supported hyper-parameters;\\ Support of different levels of precision;\\ Increasing number of PUs;\\ Increasing storage capacity.\end{tabular}        \\ \hline
\textbf{Scalability}       & The degree of scalability           & insignificant re-design           & \begin{tabular}[c]{@{}l@{}}Support of different types of data;\\ Reconfigurable interconnects;\\ Increasing number of PUs;\\ Increasing storage capacity.\end{tabular}       \\ \hline
\end{tabular}
}
\label{PU_perf}
\end{table}



\subsection{ReRAM Array Level}

The most popular types of ReRAM based on the filament composition can be classified into the following two types: (a) oxygen vacancy filament-based ReRAM (OxReRAM) and (b) conductive bridge random access memory (CBRAM). The very first step of filament creation is referred as a 'FORMING'. A 'SET' operation is a switching of ReRAM state from a high-resistance state (HRS) to a low-resistance state (LRS), whereas a 'RESET' operation is a switching from LRS to HRS. The filament switching can be unipolar (SET and RESET voltages have the same polarity) and bipolar (SET and RESET voltages have an opposite polarity).

The wide adoption of ReRAM devices is inhibited by a number of manufacturing and operation constraints.  Overcoming the limitations and reliability issues creates new areas of study, such as weight update linearity,  the symmetry and stability of resistance levels, and  increasing endurance and long-term retention. In addition, extensive research is still   concentrated at a single device level. However, the organizing ReRAM devices in high density crossbar arrays have greater potential for application purposes. Therefore, a comprehensive investigation of arrays' switching time, voltage drop, endurance, retention, wordline current, and other parameters is required. Research possibilities in the area of  the search of high-performance ReRAM architecture for building, both standalone and integrated onto CMOS chips, have  great potential. Currently, the choice of a ReRAM device depends on the application and includes a consideration of the  following  features: weight precision, energy consumption, scalability, and latency.

\subsubsection{Limited Device Precision}

One of the parameters that contributes to the computational accuracy of the neural network accelerator is   weight precision. The reduced bit precision can be acceptable during  the inference phase, whereas the \textit{in situ} training phase requires higher bit representation. Currently only binary and ternary ReRAM devices are prevailing in the fabricated nvCIM platforms. In high-density applications, multi-bit precision is achieved by combining multiple ReRAM cells using various \textbf{weight-composing schemes}. Since this approach contributes to the increases in   chip area,  the majority of current research is concentrated on scaling down the size of arrays and stacking cells in multi-layer 3D crossbars (e.g., 3D-VRAM and 3D-HRAM \cite{xu2014modeling}). Alternatively,  the density and cost reduction of  the crossbar array can be achieved by the utilization of  the  \textbf{multilevel per cell} (MLC) behavior of ReRAM. In other words, ReRAM devices should have a wide window margin $R_{ON}$/$R_{OFF}$ ratio with multiple distinct states \cite{zhu2015overview}. Moreover, to perform energy-efficient MVM operation, the states should be programmed to very low conductance levels.
\par

\par
MLC in CBRAM and OxReRAM can be achieved in the    following three ways: (a) by changing  the compliance current during  the 'SET' operation; (b) by controlling the  'RESET' voltage;  (c) by changing  the program/erase pulse width \cite{prakash2016multilevel, zahoor2020resistive}. The increase in compliance current $I_{cc}$ applied to a ReRAM cell is accompanied by widening   its conductive filament (CF) and  increasing $I_{reset}$. As a result, LRS resistance levels are lowered, and several new $R_{LRS}$ can be set. 
Recent individual multi-level devices have been reported to be programmed up to 3-bit with seven low-resistance-state levels \cite{prakash2014demonstration}. The application of this method is constrained due to  the  difficulty of limiting the current in a passive cross-point array.
A less complex circuit is required for  the reset voltage-controlled mode. The increase in the amplitude of reset voltage $V_{reset}$ leads to a decrease in  the  current $I_{HRS}$ and a corresponding change in resistance states $R_{HRS}$. In addition, $V_{reset}$ also increases $V_{set}$, but $I_{reset}$ remains constant. In the third scheme, only  the pulse duration of the applied voltage is varied, and the amplitude remains the same. A frequency-dependent scheme to implement analog-valued weights from a single-bit memristor is discussed in \cite{eshraghian2019analog}. Due to a utilization of comparator circuits, the  implementation of this approach requires a higher energy consumption.

The performance of the device is also affected by the physical size and  the material of conductive filament and electrodes. Their non-uniformity and a lack of precise control during the  fabrication process can cause (spatial) device-to-device and (temporal) cycle-to-cycle variability issues. In addition, resistive switching variability in a single device might occur during read/write phases. Moreover, device variability starts to have a significant impact with the increase in the number of ReRAM cells and therefore determines the maximum density of the RCA.  Programming  all elements in an array to multiple states is an even more sophisticated task. The latter was achieved in \cite{le2018resistive}, where each cell in a crossbar array was programmed to three bits. 













\subsubsection{Limited endurance and retention}

Each transition of a ReRAM device from HRS and LRS, and vice versa, causes damage in the  filament.  The forming/set operations are known as a reversible \textit{soft breakdown} (SB). However, after a certain number of cycles, they are permanently stuck and cannot be RESET anymore. This problem is known as a \textit{hard breakdown}  \cite{guan2012switching, carta2016x}.  
\textit{Endurance} represents a maximum number of potentiation and depression cycles that ReRAM can   toggle
. Typically its value is between $10^{6}$  and $10^{12}$ \cite{zahoor2020resistive, cai2018long}, which is still not enough to replace DRAM with endurance  > $10^{15}$ \cite{chiu2015low}. 
The lifetime of a ReRAM cell depends on the initial conditions of a device,  the switching and electrode materials, and the value of  the applied switching current. Although the dynamics of endurance is not well studied, current methods on extending  the ReRAM lifetime include  a reduction of the  switching current \cite{butcher2011high} and a decrease in the voltage amplitude and pulse width \cite{chen2014understanding}.  In addition, the OxReRAM has a better endurance than CBRAM \cite{zahoor2020resistive}. In particular, the utilization of tantalum-oxide-based materials can help to increase endurance above $10^{12}$ cycles \cite{prakash2015resistance}.  Better endurance characteristics can be achieved with    increases in the device size and filament geometry \cite{fantini2014lateral}. \textit{Retention} is another aspect of ReRAM reliability. It reflects the capability of ReRAM cells to keep its programmed state with no degradation in time. For example,   $\mathrm{HfO_x}$ devices have   an approximately $10^4$ s (2.78 h) retention \cite{azzaz2016endurance}. Moreover, thermal and electrical noise make ReRAM cells susceptible to retention failure and require   reprogramming. A retention model of $Ir/Ta_{2}O_{5-\delta}/TaO_{x}/TaN$ maintaining a retention window of no less than 10 years was presented in \cite{wei2011demonstration}. To save testing time, the retention of  the ReRAM cell is usually examined at high-temperature conditions.

The need of a trade-off between endurance, writing speeds, window margin, and data retention  in OxReRAMs was validated in \cite{azzaz2016endurance, nail2016understanding}. High-endurance devices are especially important during  the deployment of gradient-based online learning, which requires  the frequent update of devices. In particular, classification benchmarks such as  the MNIST handwritten digit recognition dataset require writing approximately $10^4$ cycles. The implementation of off-chip training and semi-online training would relax the ReRAM endurance requirements. Fully online on-chip learning requires high endurance and moderate retention, whereas semi-online requires moderate endurance and retention. 
Typically, the reported endurance and retention values  provide information on  the operation of a single device rather than the whole crossbar array.

Reducing the number of writes helps to improve the lifetime of devices and to enable   online training. Thus, new  endurance-aware training algorithms are of great importance. For instance, the authors of \cite{payvand2020chip,payvand2020error} proposed an error-triggered training scheme where the weights are only updated when the update value is greater than an adaptive threshold. Utilization of such a technique reduces the number of writes by two orders of magnitudes and therefore improves the lifetime of the devices.

\subsection{Thermal Density}

ReRAM characteristics such as  the reset current, switching speed, retention, and endurance are highly dependent on temperature. 
In particular, temperature fluctuations affect the readout margin of a device and cause  reliability issues. The study of  the conduction mechanism in $HfO_{2}$-based  
ReRAM has shown that   stable switching behavior is observed within a temperature range of 213–413$K$. Further temperature increases cause a decrease in the resistance of the OFF-state and an increase in the resistance of the ON-state \cite{walczyk2011impact}. In addition, an increase in the  write operation  rate leads to an increase in Joule heating. A higher temperature results in reduced retention and endurance.
Moreover, the device geometry influences its susceptibility to the heat. Scaling down the feature size ($F$) in devices such as $NiO$ ReRAM from a 100 $nm$ to a 30 $nm$ node can lead to an increase in temperature from approximately  400   to 1800 K \cite{lohn2014analytical}. 


 ReRAM thermal models have been comprehensively reviewed   in \cite{roldan2021thermal}. However, the individual ReRAM device models may not be applicable in RCAs, and further research is required. A passive temperature increase in neighbour ReRAM cells can be induced by neighboring thermal crosstalk and along  the word-line/bit-line \cite{al2020reliability}. The study has shown that, within the same layer in a 3D array, the thermal crosstalk from adjacent devices is stronger along  the vertical direction than the  horizontal one \cite{sun2015thermal}. A die stacking design also plays a crucial role in temperature distribution. A stacking 3D ReRAM crossbar array next to a digital processor showed a heat increase in ReRAM banks of up to 70.8 $^\circ$C and therefore led to a decrease in their lifetime close to or below ESL. Stacking 3D RCAs on top of the digital processor led to a temperature rise of up to 107.1 $^\circ$C, which consequently reduced  the ReRAM bank lifetime to less than 2.6 years \cite{beigi2019thermal}.

\begin{figure}[!h]
\centering
\includegraphics[scale=0.3]{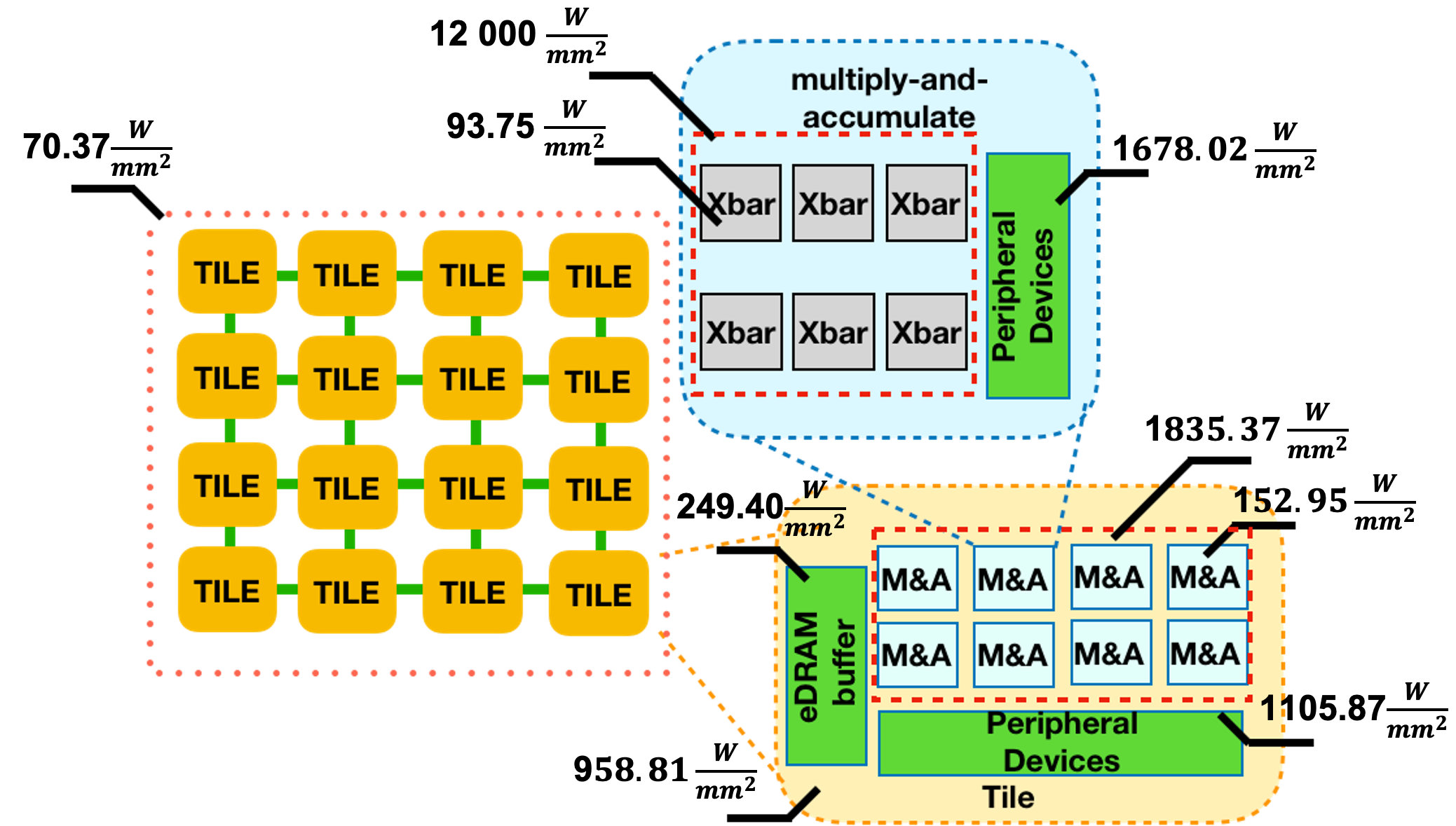}
\caption{Power density of ISAAC-CE.}
\label{isaac_power_density}
\end{figure}

The existing ReRAM-based many-core accelerators comprise multiple heterogeneous systems. Figure \ref{isaac_power_density} illustrates the average power density in   ISAAC-CE. Such non-uniformity in the thermal density distribution can severely affect the performance of the accelerators and require a study of the thermal density. Several solutions for thermal-aware optimization in ReRAM-based architectures have been proposed. One of them is a thermal-aware optimization for extending ReRAM lifetime (THOR). THOR senses the temperature of ReRAM banks and delays access to the hot banks until they are cooled down. Such an approach provides the lifetime enhancement of ReRAM-based memory by 1.36$\times$ and a power reduction by 4.6\%. DeepSwapper was designed for hybrid DRAM/ReRAM memory and enhanced the lifetime of ReRAM by 1.87$\times$. A thermal-aware optimization framework for accelerating  the DNN on ReRAM (TOPAR) was proposed to  reduce   the average  temperature  and  the  temperature  variance between  ReRAM  arrays in  DNN  accelerators. It involves three online and offline stages that   improve   the ReRAM  endurance  up  to  2.39 $\times$ and   the  overall inference accuracy. 



\subsubsection{Stuck-At-Fault}

Stuck-at-fault (SAF) is a common problem of  the ReRAM crossbar array that occurs due to temporary writing failures (TWFs) \cite{chen2014rram}, \cite{xia2017stuck}.  When programming current becomes very high during a RESET operation, ReRAM is stuck in HRS. In the case of SET operation, due to the high programming current, ReRAM becomes stuck in LRS. In some cases, ReRAM is stuck between LRS and HRS, which is known as the \textit{resistance drift} failure mechanism \cite{barci2015bilayer}. ReRAM cells exposed to stuck-at-fault tend to have limited endurance \cite{grossi2019resistive}.

The proposed solutions aimed to overcome SAF can be divided into three groups. The first one is \textbf{retraining} \cite{date17afn,dac17rescuing,charan20joint, dac17ftt}. It can be implemented using a fault map that contains information on SAF weights  \cite{chen14modeling,aspdac19ft}. Although this method requires high computational resources, the updated weights provide a recovery of accuracy without hardware modification. In a knowledge distillation (KD)-based retraining,  the  teacher network transfers 'knowledge' to a student network \cite{charan20joint}. As a result, the student network can outperform the  teacher network. The second group  proposed post-processing \textbf{correction} \cite{dac19nia,aspdac20defects, charan20joint}. In \cite{dac19nia}, the  authors estimated  the error contributed by SAF cells and recovered accuracy by additional CMOS circuits. This method requires a high cost due to the utilization of digital hardware. The overhead can be reduced by the  introduction of   the weight importance technique \cite{aspdac20defects,charan20joint}. The third group implies a \textbf{remapping/reshaping} of  the matrix.  Matrix permutation \cite{date17afn,aspdac20defects,aspdac19handling,dac17ftt} can be based on row permutation \cite{date17afn,aspdac20defects} and neuron permutation \cite{date17afn,aspdac20defects}. 
Matrix permutation requires a considerable amount of time for preprocessing, whereas  node permutation is restricted only to fully connected dense layers and cannot be applied to convolutional layers.  


In \cite{Jung2021Cost}, a dataset-free and cost-free method to mitigate the impact of stuck-at-faults by exploiting 
  the statistical properties of deep learning applications. The method relies on only changing the statistical parameters of   batch normalization while keeping the trainable parameters without change.





\subsubsection{IR Drop and Sneak-path Effect}

Depending on the presence of access selectors, ReRAM crossbar arrays can be classified as (a) active  (e.g.,  the 1T1R array)  or  (b) passive  (e.g.,  the selector-less 0T1R arrays). All cells in an array can be categorized as \textit{selected} (on the selected wordline and bitline), \textit{half-selected} (either on the selected wordline or bitline), or \textit{not-selected}  (no voltage applied). A major challenge in  the operation of large crossbar arrays occurs due to a sneak-path current flowing via half-selected cells causing a significant voltage drop. Inevitable parasitic wire resistance at ReRAM interconnects also contribute to the IR-drop issue. According to International Technology Roadmap of Semiconductors (ITRS), the expected feature size for crossbar technology  is  $5nm$ \cite{wilson2013international}, and the anticipated wire resistance for this dimension is approximately $92\Omega$ \cite{fouda2018modeling}.
In the ideal case, the measured weights should be similar to the ideal weight values. However, even the small value of the wire resistance has a significant effect on the weights stored in RCA\cite{lee2020learning,fouda2019mask,fouda2020ir}. In particular, the weight values exponentially decay across the diagonal of  the RCA. Subsequently, the opposite corner cells witness the least and worst sneak path effects \cite{fouda2020spiking}.


Previous work on  the alleviation of sneak current error in MLC ReRAM includes exploiting diode gating (e.g.,1D1M) \cite{gul2019addressing}, transistor gating (e.g., 1/2/4T1R), multistage reading, unfolded crossbar architectures,  and including non-linear selector (e.g.,1S1R) devices. However, they can drastically increase area, power, and latency overhead. In \cite{ramadan2019adaptive}, the  authors addressed  the sneak-path problem in their adaptive programming scheme. In addition, the  IR-drop-aware 
training framework was introduced in \cite{lee2020learning,fouda2019mask}.
The conductance of some devices   is exponential or a quadratic function of the applied voltage \cite{fouda2018modeling} and can help reduce the sneak path problem in resistive memories on RCAs \cite{fouda2019resistive} due to single cell reading.  
However, in neuromorphic applications, this adds  exponential behavior to vector-matrix multiplication (VMM). This exponential nonlinearity makes the VMM operation inaccurate, which deteriorates the training performance \cite{kim2018deep}. Some algorithms were developed to take the effect of the device's voltage dependency into consideration, while training ANNs such as \cite{kim2018deep}.

The performance of the devices are also affected by remoteness from    write drivers. The multiple paths created by undesired currents  disturb  the reading and writing of the weights along  the sneak path. The problem worsens with the  increase in the size of the array mat. Therefore, the implementation of large matrices using a single large crossbar array is prohibited. One possible solution is    \textbf{partitioning} layer matrices into small, realizable RCAs with the interconnect fabric between them. The horizontal and vertical interconnects are used to connect RCAs with the same array rows and columns, respectively. The typical ReRAM mats size are 128$\times$128, 512$\times$512, and 1024$\times$1024 \cite{xu2015overcoming}. Although partitioning the array diminishes the sneak path problem, it might degrade  the performance of  the RCA due to routing problems and non-idealities of the routing fabric. In the case of full-offline learning, the residual sneak path problem after partitioning and routing  the non-idealities   needs 
 to be taken into account. Such algorithmic work is performed in the masking technique \cite{fouda2018overcoming}.

\subsubsection{ReRAM Asymmetric Programming Nonlinearity }

ReRAM's potentiation and depression behaviors demonstrate exponential dynamics versus the programming time or the number of pulses. Moreover, the depression curve has a higher slope compared to the potentiation curve and leads to asymmetric programming. Nonlinear and asymmetric update dynamics in some   RRAM devices hinder large-scale deployment in neural networks \cite{yu2018neuro}. Thus, to ensure convergence of the network, the vanilla backpropagation algorithms should take into consideration the device non-idealities. The device's asymmetric nonlinearity (ANL) can be derived from a closed-form model and further added to the neural network training algorithm.


The common and accurate way to update the ReRAM conductance is applying   program/erase pulses with a constant voltage $V_p$ and a varying width $T$. 
The potentiation and depression conductances as a function of the number of pulses can be written as \cite{fouda2018independent,fouda2019effect}

\begin{eqnarray}
G_{LTP}=G_{max}-\beta_P e^{-\alpha_P V_P T n  },\,\, \text{and}\\
G_{LTD}=G_{min}+\beta_D e^{-\alpha_D V_D T n},
\end{eqnarray}
\noindent respectively, where $n$ is the number of pulses, $G_{max}$ and $G_{min}$ are the maximum and minimum conductance values, respectively, and $\alpha_P, \alpha_D, \beta_P$, and $\beta_D$ are fitting coefficients. $\beta_P$ and $\beta_D$ are related to the difference between $G_{max}$ and $G_{min}$.

The device potentiation and depression asymmetry and linearity can be quantified using the asymmetric nonlinearity factors \cite{woo2018resistive}.
The potentiation asymmetric nonlinearity (PANL) factor and depression asymmetric nonlinearity (DANL) are defined as
$PANL={G_{LTP}\left(N/2\right)}/{\Delta G}-0.5$ and $DANL=0.5-{G_{LTD}\left(N/2\right)}/{\Delta G}$, respectively, where $N$ is the total number of pulses to fully potentiate the device. The impact of these factors is reflected in the coefficients $\alpha_P, \alpha_D, \beta_P$, and $\beta_D$.  $PANL$ and $DANL$ are between $[0, 0.5]$. Their sum represents is equal to the total asymmetric nonlinearity (ANL):
\begin{equation}
ANL=1-\frac{\beta_P e^{-0.5\alpha_P N}+\beta_D e^{-0.5\alpha_D N}}{\Delta G}. 
\end{equation}

Such a problem causes unsuccessful training with a non-negligible drop in   network performance. An asymmetric ternary stochastic update method is proposed in \cite{fouda2019effect} to overcome   the asymmetric nonlinear problem. The results shows that the network was able to train well with less than a 2\% drop in   accuracy. 









\subsubsection{Manufacturing}
\begin{figure}[!h]
\centering 
\includegraphics[scale=0.35]{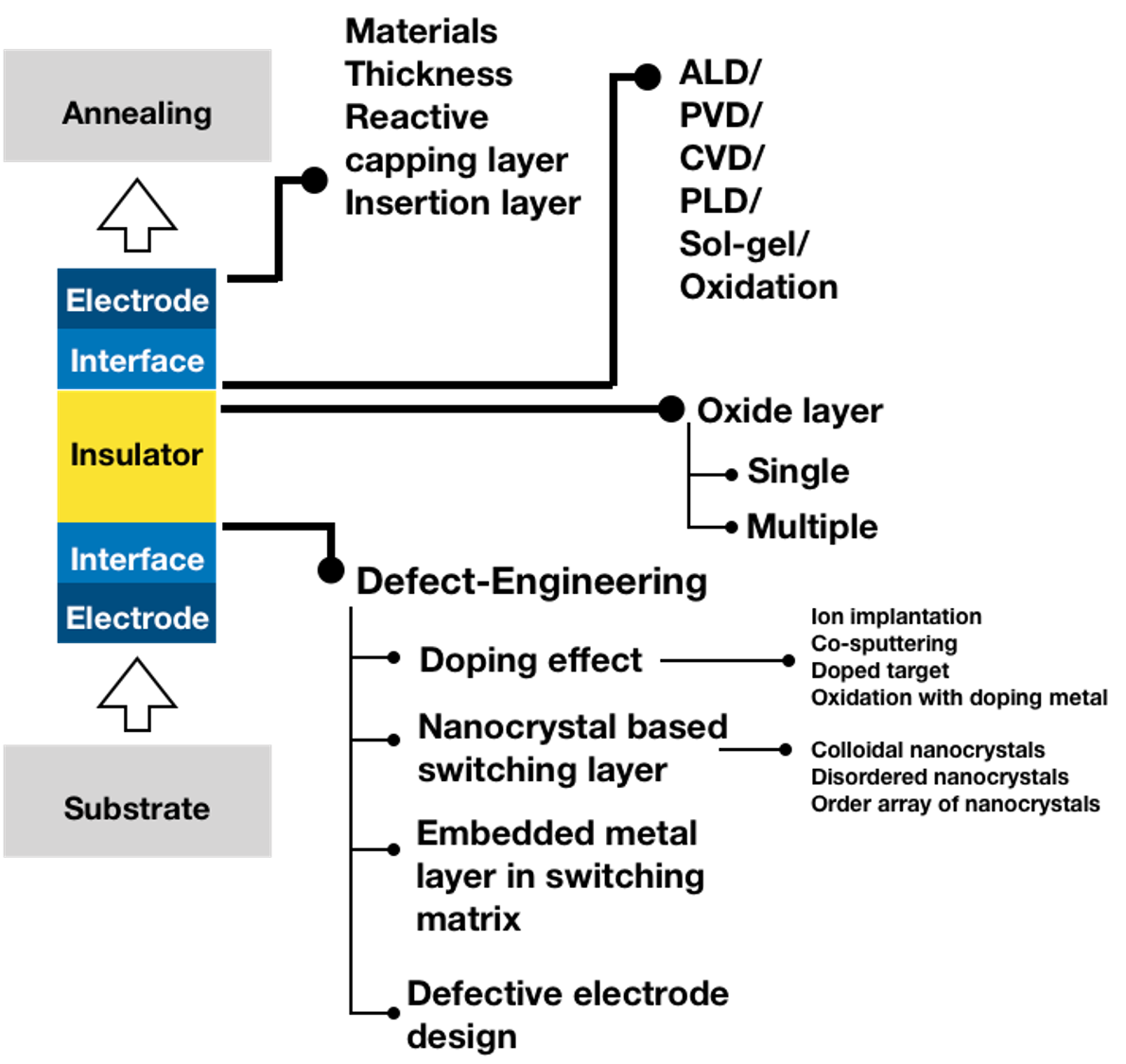}
\caption{ Material engineering of ReRAM (adapted from \cite{banerjee2020challenges}).}
\label{rram_eng}
\end{figure}

The structure of ReRAM is based on metal-insulator-metal. Its manufacturing involves different techniques, such as atomic layer deposition (ALD), physical vapor deposition (PVD), chemical vapor deposition (CVD), pulsed laser deposition (PLD), and sol-gel synthesis and oxidation. Fabrication of ReRAM is accompanied by controlled and uncontrolled defects. These defects are essential for switching ReRAM's resistance states.  Defect engineering can be implemented at the electrode, the  interface of insulating  layers \cite{banerjee2020engineering}. The choice of electrode material is conditioned by the metal work function and  the free energy of  the oxidation parameter. The insulation material can be a single layer or a compound of multiple layers. Stacking several oxide layers contribute to a better performance, such as a high endurance, a wide $R_{ON}/R_{OFF}$ ratio, and low power. The key points of ReRAM material engineering are summarized in Figure \ref{rram_eng}.

One of the advantages of ReRAM technology is its compatibility with CMOS BEOL processing. During  the fabrication of OxReRAM, preference is given to atomic layer deposition (ALD) \cite{wong2012metal, beckmann2020towards}.  Research on interfacing   ReRAM with CMOS technology was reported in   \cite{lee2012integration, shahrabi2016chip}. The $25nm$ tantalum oxide memristors  integrated on CMOS transistor circuits was presented in \cite{sheng2019low}. In addition, there is   active research in the field of selector device requirements, e.g.,  insulator-metal-transition selectors \cite{cha2016comprehensive, zhang2014high} and novel Ovonic Threshold Switch (OTS) selectors  \cite{yasuda2017cross}. Nevertheless, there is still no industry standard for   hybrid CMOS-ReRAM fabrication 
 devices.

\subsection{Processing Unit level}

Peripheral circuits of processing elements play a crucial role in  the overall performance of the system. It is well known that ADC and DAC contribute to most  
  power and area overheads. 
To reduce their impact,   certain design configurations utilizing shared ADCs and DACs were proposed 
  \cite{tang2017aepe}. Another way is reducing  the resolution of  the input/output. Therefore, there is a trade-off between accuracy and  the resolution of peripheral devices. including  the  design of low-power and small ADC/DAC or  the exploration of new designs of driving/sensing circuits. Alternative solutions include ADC/DAC-free architectures. One of the potential directions is a shift towards fully analog circuit designs.

Processing elements function in close connection with memory architectures. The majority of the proposed accelerator architectures used volatile memory devices such as eDRAM and SRAM. PRIME used its ReRAM crossbar array in memory mode as an embedded cache. The improvement of non-volatile ReRAM memory capacities for storing the  input and output of neural network layers is an essential research direction.

High accuracy requires an accurate programming of ReRAM cells to conductance levels. This opens up possibilities in search of programming architectures that include a number of tradeoffs.   For instance, long write latency degrades  the endurance of the device. Architecture performance estimation also requires a consideration of write
switching speed and energy.

Chip cost is one of the driving factors of the architecture design.
The corresponding analysis has shown that the  main contributors of  the cost-per-bit of ReRAM cells are sensing and control circuits. Multi-layer stacking also significantly affects   cost efficiency due to fabrication size and  the associated challenges  \cite{xu2013understanding}, \cite{xu2015overcoming}.



\subsection{Efficient Pipelining}

In multi/many-core systems-on-chip (SoC) platforms,  communication between on-chip components takes place via interconnection backbones, interfaces, and special protocols. Their aim is to provide energy-efficient, low-latency, and high-throughput data transmission between chip components. The design of  the communication network of ReRAM-based accelerators requires  the support of different traffic types and parallel processing.  Moreover, they can operate in either training or inference phases.   Although  most functions between them are shared, each phase has different requirements for memory and  performance metrics. In particular, training requires high accuracy and high throughput, whereas  the inference phase has lower requirements for accuracy but prioritizes low latency and high throughput 
\cite{nabavinejad2020overview}.

\par 
The reviewed accelerators have a multicore design with different types of communication networks and topologies. Although accelerators with processing elements connected via a shared bus consume less area and energy, systems based on various NoC interconnects have greater potential for scalability  \cite{bjerregaard2006survey}. NoC consists of network interfaces (NIs) and routers (also called switches). Several routers are interconnected with each other via physical links  and form different network topologies. Every PE is connected to its own router via NI. When a source PE finishes processing, data packages are passed to a destination PE via routers. The increase in the number of PEs also increases  the potential of for deadlock or livelock problems due to   
   data packages stuck between PEs.


Learning and inference capabilities of the chip  directly depend on pipelining. The utilization of SMART flow control in \cite{ko2020smart}   decreased the number of hop counts in NoC interconnects. In addition, large parallelism was exploited using intra-layer, inter-layer, batch pipelining,  and weight replication techniques.

\par
Nevertheless, neither a wire-based bus nor NoC systems can fully satisfy the requirements of ReRAM-based DNN accelerators \cite{nabavinejad2020overview}. Their major challenges include   the high latency and high power consumption of multi-hop links.  The evolution of new platforms have also driven  the development of new interconnection technologies beyond the conventional metal/dielectric systems such as  wireless and optical interconnects. In addition, the  failure   of a single interconnect in NoC can cause a permanent fault in  the  entire communication medium and cause the  system to fail \cite{deorio2012reliable}. This requires more attention on reliability.

The optimization of NoC topologies implies a consideration of meaningful operational metrics such as speed, throughput, area, and energy. In particular, NoC power saving techniques were revised in \cite{ofori2017survey}. The fault tolerance of NoC was studied in \cite{hemmati2018increasing}. 
The proposed solutions  should be deadlock- and livelock-free. The continued development of research in NoC can significantly boost the computing performance of ReRAM accelerators. Moreover,  the coexistence of two or more communication standards and interconnects and their multi-mode operation can become the next trend in SoC communication. Therefore,   research on their  compatibility is required.

  




\subsection{CAD Tools}


The increasing functional complexity increases the design complexity of the system. The proposed ReRAM accelerators are comprised of heterogeneous circuits that are simulated independently.
Simulation tools allow one to estimate  the performance of the hardware at four abstraction levels, including  the system, architectural, circuit, and device levels.

For instance,  the power and area of the peripheral devices of SotA ReRAM accelerators were estimated in CACTI and   Orion. The performance of the interconnect was evaluated using HyperTransport. The assessment of the ReRAM crossbar was conducted in  the NVSim tool, which was developed and validated by data from fabricated chips.
There are other available crossbar simulation frameworks such as MNSIM, NeuroSim, and XB-sim. They also provide a simplified estimation of power, area, and latency \cite{zhang2020neuro}, \cite{fei2020xb}. 
The \textit{MNSIM 2.0} modeling tool includes three stages: CNN optimization, performance modeling, and evaluation. During the first stage, the mixed-precision CNN quantization and the non-uniform activation quantization based on PIM computing performance is conducted. The reference data of various memory technologies used in MNISM was extracted from CACTI and NVsim. 
\textit{XB-sim} supports  the ReRAM-aware training algorithm and resource allocation for kernel mapping. The end-to-end evaluation tool  was verified by  the circuit simulation of  the fabricated chip. The integration of NoC estimation and optimization in XB-sim are planned as future work. \textit{NeuroSim 2.0} is an end-to-end  framework that also supports algorithm-to-hardware mapping and  the evaluation of hardware performance. Other simulators used for  the modeling and assessment of neuromorphic computing hardware are reviewed in \cite{staudigl2021survey}.


Generally, the existing simulators are still lacking in their performance. At device-level simulation, there is a need to decrease the gap between devices and their simulation models. At  the system level, the design tools should consider  the non-idealities of all components. Other aspects include decreasing the simulation time and providing more flexibility in the design. Compatibility of the simulator results is the matter of a communication between existing EDA vendors. Overall, the diversity and disunity of design tools used in the modeling of certain parts of the ReRAM accelerators showed the need for an integrated universal simulation tool.

\par

\par


\subsection{Benchmarking}

Due to a vast range of network configurations, finding a neural network topology with the best performance can be  challenging. The Neural Architecture Search (NAS) aims to automate that search based on three components: a search space, a search algorithm, and a child model evolution method \cite{elsken2019neural}. It is a challenging procedure that requires an iteration of thousands of different models. 
The state-of-the-art NAS tools use evolutionary algorithms and reinforcement learning \cite{liashchynskyi2019grid}, \cite{gao2019graphnas}.
After a successful search, the deployment of a neural network on a hardware can be constrained by its resources \cite{ding2020hardware}.
An increase in the number of hardware and software systems for  the acceleration of ML inspired industry and academia to create \textit{application-to-hardware} performance benchmarks for DNN. 
A fair evaluation of  the  new neural network architecture/system requires a balanced and detailed benchmarking methodology. Moreover, it should be reproducible and  affordable. The recently designed benchmark suit called \textit{MLPerf} provides up-to-date publicly available benchmarking results for both the  inference and training of DNN software and hardware  architectures  \cite{mattson2020mlperf}.  The benchmark suite includes five high-level benchmark  tasks:    image classification, object detection, translation, recommendation, and reinforcement learning. Each task comes with a canonical reference model in different frameworks. The reported results are labelled according to the division (open or closed), category (available, preview, or research), and system type (on-Premises or cloud). The Closed division addresses the lack of a standard inference 
  benchmarking workflow and has strict rules, whereas  the Open division gives more flexibility and allows one to change the model and other parameters.

Another publicly available industry-standard AI benchmark suite is the \textit{AIBench}. It has been designed in joint partnership with 17 industry partners and identified 17  prominent AI problem domains (the component benchmarks) and 12 computation units (the micro benchmarks). In addition, it includes 17 representative datasets in the form of text, image, audio, video, and 3D data. It has a separate set of evaluation metrics for online inference and offline training. AIBench has a wider coverage than MLPerf. In general, AIBench distinguishes three levels of benchmarking:  the hardware level,  the system level, and  the free level. These levels are comprised of six 
  benchmarking layers, e.g.,  the type of hardware, compiler, framework, workload, hyperparameters, dataset,  target, and metrics. The end-to-end benchmarking by AIBench is done on CPU and GPU platforms.

Tables \ref{aibench_reram} and \ref{microbench} list AIBench component benchmarks and micro benchmarks, respectively. The tables also reflect which benchmarks are supported by ReRAM accelerators such as ISAAC, PRIME, AEPE, PipeLayer, AtomLayer, Newton, CASCADE, and PUMA+PANTHER. Currently, the primary benchmark application for  the evaluation of ReRAM accelerators is image classification and is  the most widely used benchmarking DNN models are MLP and CNN. In comparison to other accelerators, PUMA supports a wider range of component benchmarks such as text-to-text, image-to-text, and speech-to-text translation and object detection. All accelerators were reported to support VGG-19 workload. All of them support fully connected, convolution, pooling, and activation layers. They also perform element-wise multiplication.  The support of  the softmax function was mentioned only in PUMA/PANTHER. However, they do not support normalization and dropout operations. The popular DNN frameworks include PyTorch, Caffe, TensorFlow, MXNet, and CNTK. However,  most ReRAM accelerators provide no or little details on software components. Therefore,  the application of ReRAM accelerators requires an extension of the list of component benchmarks and micro benchmarks.




\begin{table}[]

\caption{Seventeen prominent AI domains and their support by ReRAM accelerators. }
\resizebox{\textwidth}{!}{
\begin{tabular}{|l|l|c|c|c|c|c|c|c|c|}
\hline
\multicolumn{2}{|l|}{\textbf{Component Benchmark}}               & \textbf{ISAAC} & \textbf{PRIME} & \textbf{AEPE} & \textbf{PipeLayer} & \textbf{AtomLayer} & \textbf{Newton} & \textbf{CASCADE} & \textbf{\begin{tabular}[c]{@{}c@{}}PUMA+\\ PANTHER\end{tabular}} \\ \hline
\multirow{2}{*}{\textbf{Image classification}}       & inference & \cmark              & \cmark              & \cmark             & \cmark                  & \cmark                  & \cmark               & \cmark                & \cmark                                                                \\ \cline{2-10} 
                                                     & train     & \xmark              & \xmark               & \xmark              & \cmark                  & \cmark                  & \xmark               & \xmark                & \cmark                                                                \\ \hline
\multirow{2}{*}{\textbf{Image generation}}           & inference & \xmark               & \xmark               & \xmark              & \xmark                   & \xmark                   & \xmark                & \xmark                 & \xmark                                                                 \\ \cline{2-10} 
                                                     & train     &\xmark             &\xmark             &\xmark            &\xmark                 &\xmark                 &\xmark              &\xmark               &\xmark                                                               \\ \hline
\multirow{2}{*}{\textbf{Text-to-Text translation}}   & inference &\xmark             &\xmark             &\xmark            &\xmark                 &\xmark                 &\xmark              &\xmark               &\cmark                                                               \\ \cline{2-10} 
                                                     & train     &\xmark             &\xmark             &\xmark            &\xmark                 &\xmark                 &\xmark              &\xmark               &\xmark                                                               \\ \hline
\multirow{2}{*}{\textbf{Image-to-Text}}              & inference &\xmark             &\xmark             &\xmark            &\xmark                 &\xmark                 &\xmark              &\xmark               &\cmark                                                               \\ \cline{2-10} 
                                                     & train     &\xmark             &\xmark             &\xmark            &\xmark                 &\xmark                 &\xmark              &\xmark               &\xmark                                                               \\ \hline
\multirow{2}{*}{\textbf{Image-to-Image}}             & inference &\xmark             &\xmark             &\xmark            &\xmark                 &\xmark                 &\xmark              &\xmark               &\xmark                                                               \\ \cline{2-10} 
                                                     & train     &\xmark             &\xmark             &\xmark            &\xmark                 &\xmark                 &\xmark              &\xmark               &\xmark                                                               \\ \hline
\multirow{2}{*}{\textbf{Speech-to-Text}}             & inference &\xmark             &\xmark             &\xmark            &\xmark                 &\xmark                 &\xmark              &\xmark               &\cmark                                                               \\ \cline{2-10} 
                                                     & train     &\xmark             &\xmark             &\xmark            &\xmark                 &\xmark                 &\xmark              &\xmark               &\xmark                                                               \\ \hline
\multirow{2}{*}{\textbf{Face embedding}}             & inference &\xmark             &\xmark             &\xmark            &\xmark                 &\xmark                 &\xmark              &\xmark               &\xmark                                                               \\ \cline{2-10} 
                                                     & train     &\xmark             &\xmark             &\xmark            &\xmark                 &\xmark                 &\xmark              &\xmark               &\xmark                                                               \\ \hline
\multirow{2}{*}{\textbf{3D Face Recognition}}        & inference &\xmark             &\xmark             &\xmark            &\xmark                 &\xmark                 &\xmark              &\xmark               &\xmark                                                               \\ \cline{2-10} 
                                                     & train     &\xmark             &\xmark             &\xmark            &\xmark                 &\xmark                 &\xmark              &\xmark               &\xmark                                                               \\ \hline
\multirow{2}{*}{\textbf{Object detection}}           & inference &\xmark             &\xmark             &\xmark            &\xmark                 &\xmark                 &\xmark              &\xmark               &\cmark                                                               \\ \cline{2-10} 
                                                     & train     &\xmark             &\xmark             &\xmark            &\xmark                 &\xmark                 &\xmark              &\xmark               &\xmark                                                               \\ \hline
\multirow{2}{*}{\textbf{Recommendation}}             & inference &\xmark             &\xmark             &\xmark            &\xmark                 &\xmark                 &\xmark              &\xmark               &\xmark                                                               \\ \cline{2-10} 
                                                     & train     &\xmark             &\xmark             &\xmark            &\xmark                 &\xmark                 &\xmark              &\xmark               &\xmark                                                               \\ \hline
\multirow{2}{*}{\textbf{Video prediction}}           & inference &\xmark             &\xmark             &\xmark            &\xmark                 &\xmark                 &\xmark              &\xmark               &\xmark                                                               \\ \cline{2-10} 
                                                     & train     &\xmark             &\xmark             &\xmark            &\xmark                 &\xmark                 &\xmark              &\xmark               &\xmark                                                               \\ \hline
\multirow{2}{*}{\textbf{Image compression}}          & inference &\xmark             &\xmark             &\xmark            &\xmark                 &\xmark                 &\xmark              &\xmark               &\xmark                                                               \\ \cline{2-10} 
                                                     & train     &\xmark             &\xmark             &\xmark            &\xmark                 &\xmark                 &\xmark              &\xmark               &\xmark                                                               \\ \hline
\multirow{2}{*}{\textbf{3D object reconstruction}}   & inference &\xmark             &\xmark             &\xmark            &\xmark                 &\xmark                 &\xmark              &\xmark               &\xmark                                                               \\ \cline{2-10} 
                                                     & train     &\xmark             &\xmark             &\xmark            &\xmark                 &\xmark                 &\xmark              &\xmark               &\xmark                                                               \\ \hline
\multirow{2}{*}{\textbf{Text summarization}}         & inference &\xmark             &\xmark             &\xmark            &\xmark                 &\xmark                 &\xmark              &\xmark               &\xmark                                                               \\ \cline{2-10} 
                                                     & train     &\xmark             &\xmark             &\xmark            &\xmark                 &\xmark                 &\xmark              &\xmark               &\xmark                                                               \\ \hline
\multirow{2}{*}{\textbf{Spatial transformer}}        & inference &\xmark             &\xmark             &\xmark            &\xmark                 &\xmark                 &\xmark              &\xmark               &\xmark                                                               \\ \cline{2-10} 
                                                     & train     &\xmark             &\xmark             &\xmark            &\xmark                 &\xmark                 &\xmark              &\xmark               &\xmark                                                               \\ \hline
\multirow{2}{*}{\textbf{Learning to rank}}           & inference &\xmark             &\xmark             &\xmark            &\xmark                 &\xmark                 &\xmark              &\xmark               &\xmark                                                               \\ \cline{2-10} 
                                                     & train     &\xmark             &\xmark             &\xmark            &\xmark                 &\xmark                 &\xmark              &\xmark               &\xmark                                                               \\ \hline
\multirow{2}{*}{\textbf{Neural architecture search}} & inference &\xmark             &\xmark             &\xmark            &\xmark                 &\xmark                 &\xmark              &\xmark               &\xmark                                                               \\ \cline{2-10} 
                                                     & train     &\xmark             &\xmark             &\xmark            &\xmark                 &\xmark                 &\xmark              &\xmark               &\xmark                                                               \\ \hline
\end{tabular}
}
\label{aibench_reram}
\end{table}

\begin{table}[]
\caption{Micro benchmarks and their support by ReRAM accelerators. }

\resizebox{\textwidth}{!}{
\begin{tabular}{|l|c|c|c|c|c|c|c|c|}
\hline
\textbf{Micro benchmark}  & \textbf{ISAAC} & \textbf{PRIME} & \textbf{AEPE} & \textbf{PipeLayer} & \textbf{AtomLayer} & \textbf{Newton} & \textbf{CASCADE} & \textbf{\begin{tabular}[c]{@{}c@{}}PUMA/\\ PANTHER\end{tabular}} \\ \hline
\textbf{Convolution}      &\cmark         &\cmark         &\cmark        &\cmark             &\cmark             &\cmark          &\cmark           &\cmark                                                           \\ \hline
\textbf{Fully-connected}  &\cmark         &\cmark         &\cmark        &\cmark             &\cmark             &\cmark          &\cmark           &\cmark                                                           \\ \hline
\textbf{Sigmoid}          &\cmark         &\cmark         &\cmark        &\cmark             &\cmark             &\cmark          &\cmark           &\cmark                                                           \\ \hline
\textbf{Hyperbolic tangent}          &\cmark         &\cmark         &\cmark        &\cmark             &\cmark             &\cmark          &\cmark           &\cmark                                                           \\ \hline
\textbf{ReLU}             &\cmark         &\cmark         &\cmark        &\cmark             &\cmark             &\cmark          &\cmark           &\cmark                                                           \\ \hline
\textbf{Pooling}          &\cmark         &\cmark         &\cmark        &\cmark             &\cmark             &\cmark          &\cmark           &\cmark                                                           \\ \hline
\textbf{Normalization}    &\xmark         &\xmark         &\xmark        &\xmark             &\xmark             &\xmark          &\cmark           &\xmark                                                           \\ \hline
\textbf{Dropout}          &\xmark         &\xmark         &\xmark        &\xmark             &\xmark             &\xmark          &\xmark           &\xmark                                                           \\ \hline
\textbf{Softmax}          & \xmark         &\ \xmark         & \xmark       & \xmark           & \xmark             & \xmark          & \xmark          & \xmark                                                           \\ \hline
\textbf{Element-wise mul} &\cmark         &\cmark         &\cmark        &\cmark             &\cmark             &\cmark          &\cmark          &\cmark                                                           \\ \hline
\textbf{AllReduced}       &\xmark         &\xmark         &\xmark        &\xmark             &\xmark             &\xmark          &\xmark            &\xmark                                                           \\ \hline
\end{tabular}
}
\label{microbench}
\end{table}

\subsection{Software and Hardware Co-design Tool}



Unlike a pure hardware benchmarking of a certain set of applications, a software--hardware co-design benchmarking methodology can benefit from a 
  more accurate evaluation. 
The Open Division of MLPerf provides more flexibility. 
In contrast, NNBench-X proposes concrete end-to-end methodology that consists of three stages:  the selection of representative application, its optimization, and  an evaluation of  the hardware \cite{xie2020nnbench}. 
The first stage involves operator-level and application-level  analyses. During operator level analysis, all possible operators are extracted from  the application pool and studied for parallelism and locality. Afterwards, they are clustered into three,   $R_{1}, R_{2}, and R_{3}$, using a k-means algorithm because clustering operators, by  a similar or  the same functionality, may cause  the bottleneck problem. Application-level analysis is conducted using identified clusters. In particular, if an application has a large number of $R_{2}$ (typically Conv and MatMul) operators, it requires a computation-centric accelerator. In the case of a large number of $R_{3}$ (typically element-wise) operators, a memory-centric accelerator is required. The advantage of NNBench-X is that it can be applied for benchmarking   new NN workloads and different types of hardware such as ASIC and FPGA, which includes   
  emerging  technologies, e.g., neuromorphic chips, PIM accelerators, and near-data processing accelerators. In the next stage,  the benchmark suite is generated with the support of a wide range of compression methods. In the final stage,  the hardware system is evaluated on different types of accelerators. The demonstrated results are as follows: 
   DianNao, Neurocube, and Cambricon-X.

Generally, neural network systems can be implemented with comprehensive or affordable characteristics. To implement  the software--hardware co-design of a DNN with the desired requirements, various Design Space Exploration (DSE) frameworks have   emerged. Their design spaces include the algorithm space, hardware space, and algorithm-to-hardware mapping space. The majority of DNN-DSE concentrates on finding the optimal points in the latter two spaces. DSEs also provide  a hardware cost estimation. Its accuracy depends on  the speed of evaluation. ZigZag is a rapid DSE for DNNs that have a fully flexible hardware design space with even and uneven mapping.  In comparison to the other state-of-the-art (SotA) DSEs, it is  able to provide up to 64\% more energy-efficient solutions \cite{mei2021zigzag}. To evaluate  the efficiency of both digital and analog accelerators, ZigZag was extended to the analog in-memory (AiMAC) solutions. Importantly, AiMAC also considers the recent emerging technologies as an alternative to SRAM \cite{houshmand2020opportunities}. 


Most of the proposed ReRAM accelerators provide no or few details on neural network programming, compilation, and optimization. The design of ReRAM-based accelerators using  the software and hardware co-design search tool should also discover an efficient kernel-to-crossbar 
 conversion algorithm with consideration of data/kernel patterns, data reuse, and hardware non-idealities. It should be able to implement hardware-friendly optimization and ensure efficiency to stay constant across performance. In addition, the systems should be free of human bias.

\section{Conclusion}
\label{sec:conclusion}

The inference and training of neural networks using  ReRAM-based accelerators are   evolving fields of study. The reviewed accelerators demonstrated better efficiency and performance compared with their fully CMOS digital counterparts. However, their widespread use is hindered by a number of reasons. First, ReRAM devices have reliability issues.   Secondly, significant work is required to improve  the performance and design of peripheral devices such as ADC/DACs, buffers, and interconnects.  
Deep research on efficient kernel mapping and data routing is essential for efficient inference. The development of computer-aided tools for identifying hardware-aware neural network models and circuit-level simulation with realistic device models is a vital direction of   future work.  In addition, accurate training requires  the development of techniques to increase  the precision of the system or  the fabrication of multilevel cell arrays. In addition,  the large area overhead of peripheral circuits also limits the  precision of existing ReRAM accelerators to 16-bit. Interconnect topologies and technologies should be revised, and more study is required to implement scalable and flexible hardware and balanced pipelining.  In addition, a trade-off between performance and manufacturing cost is required. Mitigating and eliminating   the  limitations and non-idealities of ReRAM crossbar arrays can expedite the process of their broad use in commercial applications.

There is a limitation in  the  accurate evaluation of the performance of   ReRAM accelerators due to their multi-core nature. The reported parameters typically do not fully reflect their performance, and the reported numbers are given at a maximum rate of utilization. For a fair evaluation, parameters such as throughput and power efficiency should be reported at average utilization. Moreover, the results should cover the utilization of  the entire accelerator and its single processing unit. The number of MAC operations and the sparsity depend on the  DNN workload, but the hardware should be able to recognize them. Therefore, software--hardware co-design is required.  In addition, there are no industry-standard benchmarks for ReRAM accelerators. Most of them mainly cover only one application from AIBench and MLPerf benchmarking standards. The reported ReRAM accelerators have been evaluated with 1 or 2 workloads. Future work could aim at extending  the  coverage of benchmarking tasks, workload models, and operations by ReRAM hardware. In addition, data storage and movement are the greatest challenges in the design of ReRAM accelerators and 
  significantly affect their overall performance. The state-of-the-art DSE frameworks are designed for computation- and memory-centric architectures, whereas ReRAM crossbar arrays function as memory and computational units. 
   The research gap in designing     ReRAM accelerators  also includes the absence of CAD tools that take into consideration  the physical characteristics of ReRAM arrays.  To sum up, ReRAM-based multi-core parallel and spatial processing is an inevitable trend of future electronics and requires  the interdisciplinary research of material, software and hardware  engineers, and scientists.








\footnotesize
\bibliographystyle{unsrt}  
\bibliography{references}

\end{document}